\documentclass[twocolumn,twocolappendix]{aastex63}
\usepackage{graphicx}
\usepackage{threeparttable}
\usepackage{amsmath}
\definecolor{citecolor}{cmyk}{1,1,0,0}

\newcommand{\tauboo}{$\tau$~Boo~b }
\newcommand{\tauboos}{$\tau$~Boo~b's }

\received{}
\revised{}
\accepted{}
\submitjournal{ApJ} 

\shorttitle{Jupiter-like C/H on $\tau$~Boo}
\shortauthors{Pelletier et al.}

\begin{document}

\title{Where is the Water?  Jupiter-like C/H ratio but strong H$_2$O depletion found on $\tau$~Bo\"otis~b using SPIRou}

\correspondingauthor{Stefan Pelletier}
\email{stefan.pelletier@umontreal.ca}

\affiliation{Institut de Recherche sur les Exoplan\`etes, D\'epartement de Physique, Universit\'e de Montr\'eal,\\
1375 Avenue Th\'er\`ese-Lavoie-Roux, Montreal, H2V 0B3, Canada}


\affiliation{Universit\'e de Toulouse, CNRS, IRAP, 
14 avenue Belin, 
31400 Toulouse, France}

\affiliation{Canada-France-Hawaii Telescope, CNRS, 96743 Kamuela, Hawaii, USA}
 
 \affiliation{Universit\'e Grenoble-Alpes, CNRS, IPAG, 38000 Grenoble, France}

\affiliation{Institut d'Astrophysique de Paris, UMR7095 CNRS, Universit\'e Pierre \& Marie Curie, 
98bis boulevard Arago,
 75014 Paris, France}
 
\affiliation{Instituto de Astrof\'isica e Ci\^encias do Espa\c{c}o, Universidade do Porto, CAUP, Rua das Estrelas, P-4150-762 Porto, Portugal}

\affiliation{Observatoire de Haute Provence, 
 St Michel l’Observatoire, France}

\affiliation{Laborat\'orio Nacional de Astrof\'isica, Rua Estados Unidos 154,
37504-364, Itajub\'a - MG, Brazil}


\author[0000-0002-8573-805X]{Stefan Pelletier}
\affiliation{Institut de Recherche sur les Exoplan\`etes, D\'epartement de Physique, Universit\'e de Montr\'eal,\\
1375 Avenue Th\'er\`ese-Lavoie-Roux, Montreal, H2V 0B3, Canada}

\author[0000-0001-5578-1498]{Bj\"orn Benneke}
\affiliation{Institut de Recherche sur les Exoplan\`etes, D\'epartement de Physique, Universit\'e de Montr\'eal,\\
1375 Avenue Th\'er\`ese-Lavoie-Roux, Montreal, H2V 0B3, Canada}

\author[0000-0002-7786-0661]{Antoine Darveau-Bernier}
\affiliation{Institut de Recherche sur les Exoplan\`etes, D\'epartement de Physique, Universit\'e de Montr\'eal,\\
1375 Avenue Th\'er\`ese-Lavoie-Roux, Montreal, H2V 0B3, Canada}

\author[0000-0001-9427-1642]{Anne Boucher}
\affiliation{Institut de Recherche sur les Exoplan\`etes, D\'epartement de Physique, Universit\'e de Montr\'eal,\\
1375 Avenue Th\'er\`ese-Lavoie-Roux, Montreal, H2V 0B3, Canada}

\author[0000-0003-4166-4121]{Neil J. Cook}
\affiliation{Institut de Recherche sur les Exoplan\`etes, D\'epartement de Physique, Universit\'e de Montr\'eal,\\
1375 Avenue Th\'er\`ese-Lavoie-Roux, Montreal, H2V 0B3, Canada}

\author[0000-0002-2875-917X]{Caroline Piaulet}
\affiliation{Institut de Recherche sur les Exoplan\`etes, D\'epartement de Physique, Universit\'e de Montr\'eal,\\
1375 Avenue Th\'er\`ese-Lavoie-Roux, Montreal, H2V 0B3, Canada}

\author[0000-0002-2195-735X]{Louis-Philippe Coulombe}
\affiliation{Institut de Recherche sur les Exoplan\`etes, D\'epartement de Physique, Universit\'e de Montr\'eal,\\
1375 Avenue Th\'er\`ese-Lavoie-Roux, Montreal, H2V 0B3, Canada}

\author[0000-0003-3506-5667]{\'Etienne Artigau}
\affiliation{Institut de Recherche sur les Exoplan\`etes, D\'epartement de Physique, Universit\'e de Montr\'eal,\\
1375 Avenue Th\'er\`ese-Lavoie-Roux, Montreal, H2V 0B3, Canada}

\author[0000-0002-6780-4252]{David Lafreni\`ere}
\affiliation{Institut de Recherche sur les Exoplan\`etes, D\'epartement de Physique, Universit\'e de Montr\'eal,\\
1375 Avenue Th\'er\`ese-Lavoie-Roux, Montreal, H2V 0B3, Canada}

\author[0000-0002-4771-0312]{Simon Delisle}
\affiliation{Institut de Recherche sur les Exoplan\`etes, D\'epartement de Physique, Universit\'e de Montr\'eal,\\
1375 Avenue Th\'er\`ese-Lavoie-Roux, Montreal, H2V 0B3, Canada}

\author[0000-0002-1199-9759]{Romain Allart}
\altaffiliation{Trottier Fellow}
\affiliation{Institut de Recherche sur les Exoplan\`etes, D\'epartement de Physique, Universit\'e de Montr\'eal,\\
1375 Avenue Th\'er\`ese-Lavoie-Roux, Montreal, H2V 0B3, Canada}

\author[0000-0001-5485-4675]{Ren\'e Doyon}
\affiliation{Institut de Recherche sur les Exoplan\`etes, D\'epartement de Physique, Universit\'e de Montr\'eal,\\
1375 Avenue Th\'er\`ese-Lavoie-Roux, Montreal, H2V 0B3, Canada}

\author[0000-0001-5541-2887]{Jean-Fran\c{c}ois Donati}
\affiliation{Universit\'e de Toulouse, CNRS, IRAP, 
14 avenue Belin, 
31400 Toulouse, France}

\author[0000-0002-1436-7351]{Pascal Fouqu\'e}
\affiliation{Canada-France-Hawaii Telescope, CNRS, 96743 Kamuela, Hawaii, USA}
\affiliation{Universit\'e de Toulouse, CNRS, IRAP, 
14 avenue Belin, 
31400 Toulouse, France}

\author[0000-0002-2842-3924]{Claire Moutou}
\affiliation{Universit\'e de Toulouse, CNRS, IRAP, 
14 avenue Belin, 
31400 Toulouse, France}

\author[0000-0001-9291-5555]{Charles Cadieux}
\affiliation{Institut de Recherche sur les Exoplan\`etes, D\'epartement de Physique, Universit\'e de Montr\'eal,\\
1375 Avenue Th\'er\`ese-Lavoie-Roux, Montreal, H2V 0B3, Canada}

\author[0000-0001-5099-7978]{Xavier Delfosse}
\affiliation{Universit\'e Grenoble-Alpes, CNRS, IPAG, 38000 Grenoble, France}

\author[0000-0001-5450-7067]{Guillaume H\'ebrard}
\affiliation{Institut d'Astrophysique de Paris, UMR7095 CNRS, Universit\'e Pierre \& Marie Curie, 
98bis boulevard Arago,
 75014 Paris, France}
\affiliation{Observatoire de Haute Provence, 
 St Michel l’Observatoire, France}

 
\author[0000-0002-1532-9082]{Jorge H. C. Martins}
\affiliation{Instituto de Astrof\'isica e Ci\^encias do Espa\c{c}o, Universidade do Porto, CAUP, Rua das Estrelas, P-4150-762 Porto, Portugal}


\author[0000-0002-5084-168X]{Eder Martioli}
\affiliation{Institut d'Astrophysique de Paris, UMR7095 CNRS, Universit\'e Pierre \& Marie Curie, 
98bis boulevard Arago, 75014 Paris, France}
\affiliation{Laborat\'orio Nacional de Astrof\'isica, Rua Estados Unidos 154,
37504-364, Itajub\'a - MG, Brazil}

\author[0000-0002-5922-8267]{Thomas Vandal}
\affiliation{Institut de Recherche sur les Exoplan\`etes, D\'epartement de Physique, Universit\'e de Montr\'eal,\\
1375 Avenue Th\'er\`ese-Lavoie-Roux, Montreal, H2V 0B3, Canada}

\begin{abstract}
The present-day envelope of gaseous planets is a relic of how these giant planets originated and evolved.  Measuring their elemental composition therefore presents a powerful opportunity to answer long-standing questions regarding planet formation. Obtaining precise observational constraints on the elemental inventory of giant exoplanets has, however, remained challenging due to the limited simultaneous wavelength coverage of current space-based instruments. Here, we present thermal emission observations of the non-transiting hot Jupiter $\tau$~Boo~b using the new wide wavelength coverage (0.95--2.50\,$\mu$m) and high spectral resolution ($R=70\,000$) SPIRou spectrograph. By combining a total of 20 hours of SPIRou data obtained over five nights in a full atmospheric retrieval framework designed for high-resolution data, we constrain the abundances of all the major oxygen- and carbon-bearing molecules and recover a non-inverted temperature structure using a new free-shape, nonparametric TP profile retrieval approach.  We find a volume mixing ratio of log(CO)$\,\,=-2.46_{-0.29}^{+0.25}$ and a highly depleted water abundance of less than $0.0072$ times the value expected for a solar composition envelope. Combined with upper limits on the abundances of CH$_4$, CO$_2$, HCN, TiO, and C$_2$H$_2$, this results in a gas-phase C/H ratio of 5.85$_{-2.82}^{+4.44}$\,$\times$\,solar, consistent with the value of Jupiter, and an envelope C/O ratio robustly greater than 0.60, even when taking into account the oxygen that may be sequestered out of the gas-phase. Combined, the inferred super-solar C/H, O/H, and C/O ratios on $\tau$~Boo~b support a formation scenario beyond the water snowline in a disk enriched in CO due to pebble drift.

\end{abstract}

\keywords{planets and satellites: atmospheres --- planets and satellites: individual: $\tau$~Bo\"otis~b --- methods: data analysis --- techniques: spectroscopic}


\section{Introduction} \label{sec:introduction}



Hot Jupiters offer an unprecedented opportunity to probe planet formation by measuring the contents of their atmospheres.  
Not only are they an intriguing population without analogs in our Solar System, their extreme temperatures allow for conditions where the main carbon- and oxygen-bearing molecular species such as H$_2$O, CO, and CH$_4$ are expected to be in gaseous form accessible to remote sensing~\citep{burrows_chemical_1999, lodders_atmospheric_2002}.  This makes hot Jupiters opportune targets for estimating atmospheric elemental abundance ratios such as C/H, O/H, and C/O, quantities that can provide key insight into a giant planet's formation history~\citep[e.g.,][]{oberg_effects_2011, madhusudhan_toward_2014} but are hard to measure for Solar System giants.  

\subsection{The Challenge of Measuring the C/O Ratio of Solar System Giant Planets}

The high temperature that enables water to exist in vapor form in the upper atmosphere of hot Jupiters is in sharp contrast to the Solar System gas giants.  Jupiter and Saturn are comparatively much colder and therefore have some of their atmospheric constituents (such as H$_2$O) condensed out to deep layers beneath the cloud deck where they cannot easily be probed.
For example, Jupiter was found to have a factor of $\sim$3 enhancement of C, N,  S, P, Ar, Kr, and Xe compared to solar by the Galileo mission but its H$_2$O content, and hence the O/H ratio, remains poorly constrained~\citep{niemann_composition_1998, owen_low-temperature_1999, atreya_origin_2018}.   The O abundance was only recorded as a lower limit of $0.3\times$\,solar because the Galileo probe entered an anomalously dry region of Jupiter's atmosphere and still recorded rising abundances before it stopped transmitting~\citep{orton_characteristics_1998,atreya_coupled_2005}.  More recently, the Juno satellite was used to probe deeper layers and infer an oxygen abundance of $2.7_{-1.7}^{+2.4}$ relative to solar in the 0$^{\circ}$ to $+4^{\circ}$ latitude equatorial regions of the Jovian giant, although a value of zero cannot be ruled out at the 2$\sigma$ level~\citep{li_water_2020}.  Moreover, the authors caution that Jupiter exhibits large latitudinal variations and so its equatorial regions may not be representative of the average composition.  Obtaining precise constraints on the bulk oxygen content is crucial for comparing Jupiter's C/O ratio relative to the solar value of 0.54~\citep{asplund_chemical_2009}.






\subsection{The Challenge of Measuring the C/O Ratio of Exoplanets from Space}

Outside of our Solar System, H$_2$O absorption features have been detected on a large sample of hot Jupiters, although often at weaker levels than would be expected from cloud-free solar compositions~\citep[e.g.,][]{madhusudhan_h2o_2014, sing_continuum_2016}. 
While clouds can play a significant role in damping the observed signals, current retrieval studies suggest that sub-solar H$_2$O abundances remain necessary to fully explain the suite of observed water features in the majority of cases~\citep{barstow_consistent_2017, pinhas_h2o_2019, welbanks_massmetallicity_2019}.  It is, however, harder to place these H$_2$O measurements in context from a chemical standpoint without accompanying abundances of other molecules, such as CO or CH$_4$.  Unfortunately, these are more difficult to constrain with the limited wavelength coverage of the \textit{Hubble Space Telescope} Wide Field Camera~3 (\textit{HST}/WFC3) data sets from which almost all current hot Jupiters H$_2$O measurements are made.
This then raises the following question: is the trend of sub-solar water abundances observed on hot Jupiters inherently due to low overall metallicities, or rather the results of high C/O ratios in which case oxygen would preferentially be bound in CO rather than H$_2$O~\citep{madhusudhan_co_2012}?  

The possibility of high C/O (carbon-rich) planets was first explored by \cite{madhusudhan_carbon-rich_2011}, who reported a C/O~$\geq$~1 by analyzing the dayside spectrum of the hot Jupiter WASP-12b.  However, the high C/O claim rests primarily on the single datapoint that is the photometric \textit{Spitzer Space Telescope} secondary eclipse depth at 4.5\,$\mu$m and has widely been both contested~\citep{crossfield_re-evaluating_2012, line_systematic_2014} and re-affirmed~\citep{fohring_ultracam_2013, stevenson_deciphering_2014}.  Meanwhile, the analysis of follow-up transit observations using 
\textit{HST}/WFC3 favor C/O values closer to solar~\citep{swain_probing_2013, benneke_strict_2015, kreidberg_detection_2015}, although carbon-rich scenarios are still plausible under different model parameterizations in some cases~\citep{kreidberg_detection_2015}.  

Comparing instead planetary measurements to that of their host stars, \cite{brewer_co_2017} found that hot Jupiters generally have super-stellar C/O ratios, although most cases remain consistent with the C/O value of their host star to  1$\sigma$ due to large measurement uncertainties.
In a separate study, \cite{welbanks_massmetallicity_2019} obtained constraints on the water and alkali metal abundances of 19 exoplanets and found prevalent super-solar Na and K abundances in conjunction to low H$_2$O concentrations on hot Jupiters.  This combination would suggest that superstellar C/O, Na/O, and K/O ratios 
exist, and that oxygen is depleted relative to other species.  While large error bars prevent definitive conclusions, the general trend found in these studies favor high C/O ratios as being common in hot Jupiter atmospheres.

The difficulty in obtaining precise C/O measurements for exoplanets with current space telescopes lies primarily in their limited wavelength coverage, which currently does not enable us to obtain spectroscopy past 1.7\,$\mu$m to directly probe the prevalent bands of the main carbon-bearing molecules.  This limits us to broadband photometric measurements beyond 1.7\,$\mu$m, which are inherently difficult to interpret rendering conclusions often sensitive to model assumptions~\citep[e.g.,][]{kreidberg_detection_2015, spake_abundance_2020}.

Circumventing the need for space-based low-resolution instruments and partially overcoming these limitations has proven to be possible in some cases for directly imaged planets, providing C/O measurements ranging from being sub- or near-solar~\citep{konopacky_detection_2013, barman_simultaneous_2015, todorov_water_2016, lavie_heliosretrievalopen-source_2017, nowak_peering_2020, molliere_retrieving_2020, wang_chemical_2020}, or being super-solar close to unity in the case of HR~8799b~\citep{lee_atmospheric_2013, lavie_heliosretrievalopen-source_2017}.  However, direct imaging is currently only applicable to a small sample of nearby systems with young giant planets that orbit at very wide separations.

\subsection{High-Dispersion Cross-Correlation Spectroscopy}
In the last decade, 
high-dispersion cross-correlation spectroscopy \hbox{(HDCCS)} has 
emerged as an extremely powerful alternative method for identifying molecular species in hot Jupiter atmospheres.  Core to the technique is the fact that short period planets experience large line-of-sight velocity changes that induce Doppler shifts corresponding to multiple resolution elements per hour on a high-resolution spectrograph.  
This can allow for atmospheric signatures of exoplanets to be disentangled from contributions of both their host star and the Earth's transmittance which, in contrast, are essentially stationary in wavelength~\citep{birkby_spectroscopic_2018}. 
\hbox{HDCCS} has successfully been used to produce a wealth of unambiguous molecular detections in both transiting and non-transiting exoplanet atmospheres using CRIRES~\citep{snellen_exoplanet_2010, brogi_signature_2012, rodler_weighing_2012, birkby_detection_2013, de_kok_detection_2013, brogi_detection_2013,brogi_carbon_2014, schwarz_evidence_2015, brogi_rotation_2016, brogi_framework_2017, birkby_discovery_2017,hawker_evidence_2018, cabot_robustness_2019, webb_weak_2020}, Keck NIRSPEC~\citep{rodler_detection_2013, lockwood_near-ir_2014, piskorz_evidence_2016, piskorz_detection_2017, piskorz_ground-_2018, buzard_simulating_2020}, Subaru HDS~\citep{nugroho_high-resolution_2017}, GIANO~\citep{brogi_exoplanet_2018, guilluy_exoplanet_2019, giacobbe_five_2021}, CARMENES~\citep{alonso-floriano_multiple_2019,sanchez-lopez_water_2019, sanchez-lopez_discriminating_2020}, 
IGRINS~\citep{flagg_co_2019}, as well as Subaru IRD~\citep{nugroho_first_2021}.

A cross-correlation analysis can be insightful in determining whether an exoplanet atmosphere contains a certain species or a thermal inversion.  However, one area in which HDCCS has shown more limited success in is retrieving molecular abundances and temperature structures in a robust Bayesian fashion.  
This is a consequence of observing through Earth's atmosphere and the detrending procedure necessary for analysing such data sets which removes all continuum information, framing all time varying effects of the planet's atmosphere in relative terms with respect to the stellar flux.
Such a treatment of the data effectively prevents the use of standard `data minus model' likelihood calculations used in conventional atmospheric retrieval frameworks~\citep{brogi_retrieving_2019}.  As a chi-squared based approach is also non-ideal for model comparison to data that may have leftover broadband variations or correlated noise from poorly removed telluric lines, the challenge is then to obtain a robust goodness-of-fit estimator from the more outlier resistant cross-correlation function (CCF).  Frameworks to perform retrievals on high-resolution data sets have since been proposed and implemented, making use of CCF-to-likelihood mappings~\citep{brogi_retrieving_2019, gandhi_hydra-h_2019, gibson_detection_2020}.  
Applied on CRIRES observations, this has led to more robust atmospheric constraints on H$_2$O and CO abundances, even with only the $\leq 0.1$\,$\mu$m span of wavelength covered by such data sets.  Results from these earlier Bayesian inference analyses have shown evidence for water abundances that are underabundant relative to carbon monoxide~\citep{brogi_retrieving_2019, gandhi_hydra-h_2019}.
For time series observations taken with wider spectral range instruments with hundreds of thousands of pixels per exposure, performing full scale high-resolution atmospheric retrievals remains a computationally challenging feat. 

\subsection{Outline of this Work}
In this work, we present the detection and characterization of \tauboos dayside atmosphere using the new SPIRou high-resolution spectrograph and the \hbox{SCARLET} atmospheric retrieval framework adapted for \hbox{HDCCS} data sets.  In Section~\ref{sec:obs+data_processing} we describe the observations as well as the data reduction and detrending procedures.  We discuss the modeling and adopted likelihood prescription in Section~\ref{sec:modeling} as well as the cross-correlation and retrieval analyses in Section~\ref{sec:analysis}.  In Section~\ref{sec:lowwater} we explore possible non-astrophysical causes for the inferred water abundance upper limit.  Finally, we discuss the implications of our results for planet formation in Section~\ref{sec:implications} and conclude in Section~\ref{sec:conclusion}.


\section{Observations and Data Processing}
\label{sec:obs+data_processing}


We used the SpectroPolarim\`etre InfraRouge (SPIRou) spectrograph~\citep{donati_spirou_2018, donati_spirou_2020} to observe the thermal emission of the non-transiting hot Jupiter \tauboo for a total of 20 hours spanning five different nights.  
SPIRou is a high-resolution cryogenic echelle spectrograph newly installed at the Canada-France-Hawaii Telescope that is fiber-fed from a Cassegrain unit feeding an achromatic polarimeter with twin fluoride fibers linking to two science channels and one fiber connecting a reference lamp to a calibration channel~\citep{artigau_spirou_2014}.  The infrared detector utilizes a 4k$\times$4k H4RG detector that allows for a simultaneous wavelength coverage between 0.95--2.50\,$\mu$m continuously spanning across 50 spectral orders at a resolution of $R = 70\,000$.  This combination of high resolving power and large spectral grasp across the near-infrared makes SPIRou an ideal instrument for targeting absorption bands of water, carbon monoxide, and other molecules simultaneously.  

\begin{table}
\centering
\caption{\label{tab:tauboo_params} $\tau$~Boo System Properties}
\vspace{-3mm}
\renewcommand{\arraystretch}{1.1}
\begin{tabular}{lc}
\hline
\hline
Stellar parameters               &  Value           \\
\hline
Stellar type\textsuperscript{a}
&  
F7V
\\
$K$ magnitude\textsuperscript{b}                                             &  3.36   \\
Stellar mass ($M_*$)\textsuperscript{c}                                            &  $1.38\pm0.05\,M_{\odot}$     \\
Stellar radius ($R_*$)\textsuperscript{c}                                         &  $1.42\pm0.08\,R_{\odot}$     \\
Luminosity ($L_*$)\textsuperscript{c}         &  $3.06\pm0.16\,L_{\odot}$    \\
Temperature ($T_*$)\textsuperscript{c}                 &  $6399\pm45$\,K    \\
Metallicity (Fe/H)\textsuperscript{c}                 &  $0.26\pm0.03$    \\
Age\textsuperscript{c}         &  $0.9\pm0.5\,$Gyr   \\
Distance\textsuperscript{d}           &   $15.66\pm0.08$\,pc     \\
Systemic velocity ($V_{\mathrm{sys}}$)\textsuperscript{e}              &  $-16.4\pm0.1$\,km\,s$^{-1}$\\
Velocity semi-amplitude ($K_*$)\textsuperscript{c}                &  $471.73\pm2.97$\,m\,s$^{-1}$\\ 
\hline
Planet parameters               &      Value       \\
\hline
Orbital period ($P$)\textsuperscript{c}            &   3.3124568(69)\,days     \\
Semi-major axis ($a$)\textsuperscript{c}           &   $0.049\pm0.003$\,AU     \\
Eccentricity ($e$)\textsuperscript{c}           &   $0.011\pm0.006$    \\
Longitude of periastron ($\omega$)\textsuperscript{c}           &   $113.4\pm32.2^{\circ}$\\   
Epoch of periastron ($T_0$)\textsuperscript{c}           &   $2456400.94$\,$\pm$\,$0.30$\,(BJD)     \\
Inclination ($i$)\textsuperscript{f}         &    $43.5\pm0.7^{\circ}$   \\
Planet mass ($M_p$)\textsuperscript{f}        &     $6.24\pm0.23\,M_{\mathrm{Jup}}$    \\
\hline
\hline
\end{tabular}
\begin{tablenotes}
\small
\item{\textbf{References.}}
\textsuperscript{(a)}\cite{gray_physical_2001};
\textsuperscript{(b)}\cite{belle_directly_2009};
\textsuperscript{(c)}\cite{borsa_gaps_2015};
\textsuperscript{(d)}\cite{gaia_collaboration_gaia_2018};
\textsuperscript{(e)}\cite{brogi_signature_2012};
\textsuperscript{(f)}This work (Section~\ref{subsec:retrieval_analysis}).
\end{tablenotes}
\end{table}



\subsection{The \texorpdfstring{$\tau$}{}~Bo\"otis System}
The hot Jupiter \tauboo was one of the first exoplanets ever discovered~\citep{butler_three_1997} and remains one of the brightest and nearest exoplanet-bearing systems known to date.  
It was the first exoplanet to be successfully observed at high-resolution in thermal emission, with CO absorption detected on its dayside hemisphere~\citep{brogi_signature_2012, rodler_weighing_2012}.  
With evidence of water vapor also having been reported~\citep{lockwood_near-ir_2014}, \tauboo presents one of the best opportunities for atmospheric characterization and obtaining a precise measure of its C/O ratio.  The $\tau$~Boo planetary system is non-transiting~\citep{baliunas_properties_1997} and also hosts a widely separated M-dwarf companion~\citep{hale_orbital_1994}, with all three components orbiting in a coplanar configuration~\citep{justesen_constraining_2019}.
Systemic parameters of \tauboo and its host star are given in Table~\ref{tab:tauboo_params}.

\subsection{Observations} \label{subsec:Obs}

\begin{figure}[t!]
\begin{center}
\includegraphics[width=\linewidth]{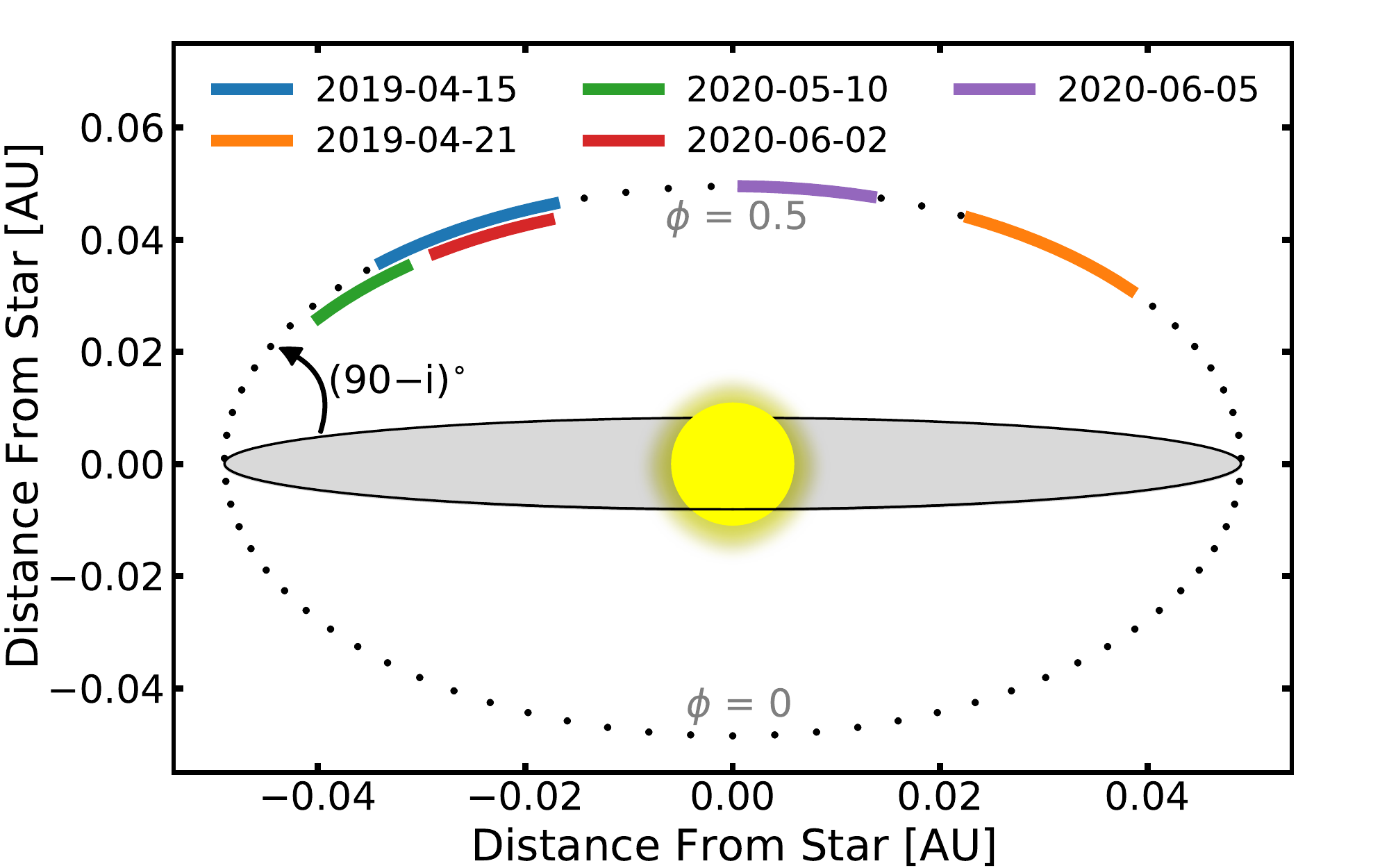}
\end{center}
\vspace{-2mm}
\caption{
Schematic illustration of the phases covered by our SPIRou observations of the non-transiting $\tau$~Boo planetary system.  
The trail of dots represents the planet's orbit, with each point tracing its displacement in approximately one hour.  An orbital phase of $\phi = 0$ corresponds to when \tauboo is closest to Earth while a phase of $\phi = 0.5$ is when its dayside is most visible.  From Earth, \tauboo is observed at an inclination $i$.  The observations span a total of 20 hours spread across five different nights, covering a large range of line-of-sight velocities and dayside viewing angles.   
\label{fig:ObsPlot}}
\end{figure}

We observed the extremely bright naked-eye $\tau$~Bo\"otis system 
with SPIRou on five different nights at orbital phases targeting the dayside of \tauboo (Figure~\ref{fig:ObsPlot}).  
We obtained continuous 5-hour observation sequences on the nights of 2019 April 15 and April 21 (Program 19AC29/19AC12, PI Pelletier/Herman), with additional 3.3-hour sequences obtained on 2020 May 10, June 2, and June 5 (Program 20AC32, PI Pelletier).  The epochs were chosen to sample a wide portion of \tauboos orbit while also avoiding regions where a larger fraction of the colder nightside is in view.  During the observations, the line-of-sight component of the planet's velocity shifts by approximately 6--8\,km\,s$^{-1}$ per hour, corresponding to roughly three full pixels on the SPIRou detector.  The observations were scheduled such that the radial velocity shift of \tauboos atmospheric trail does not overlap with the telluric rest frame, 
which could otherwise have resulted in leftover telluric residuals overlapping and interfering with the planetary signal of interest~\citep[e.g.,][]{brogi_exoplanet_2018, alonso-floriano_multiple_2019, sanchez-lopez_discriminating_2020}.

For each night we obtained 90 second exposures, averaging a signal-to-noise (S/N) per pixel of $\sim$265, with values typically ranging between 100 and 400 depending on the spectral order, with peak values being obtained in the $H$ and $K$ bands.  For these observations we opted not to use the polarimetric mode of SPIRou that measures the Stokes parameters.  While in theory this option provides polarimetric information for `free', it requires the position of the rhombs to be moved after each exposure.  To avoid having elements in the optical path move unnecessarily and to maximize stability, the rhombs were maintained at a fixed position for all observations.  
Sky conditions were dry and photometric throughout all the observations, with only the night of 2020 June 2 having elevated humidity levels. 

\subsection{Data Reduction} \label{subsec:data_reduction}

\begin{figure*}[t!]
\begin{center}
\includegraphics[width=\linewidth]{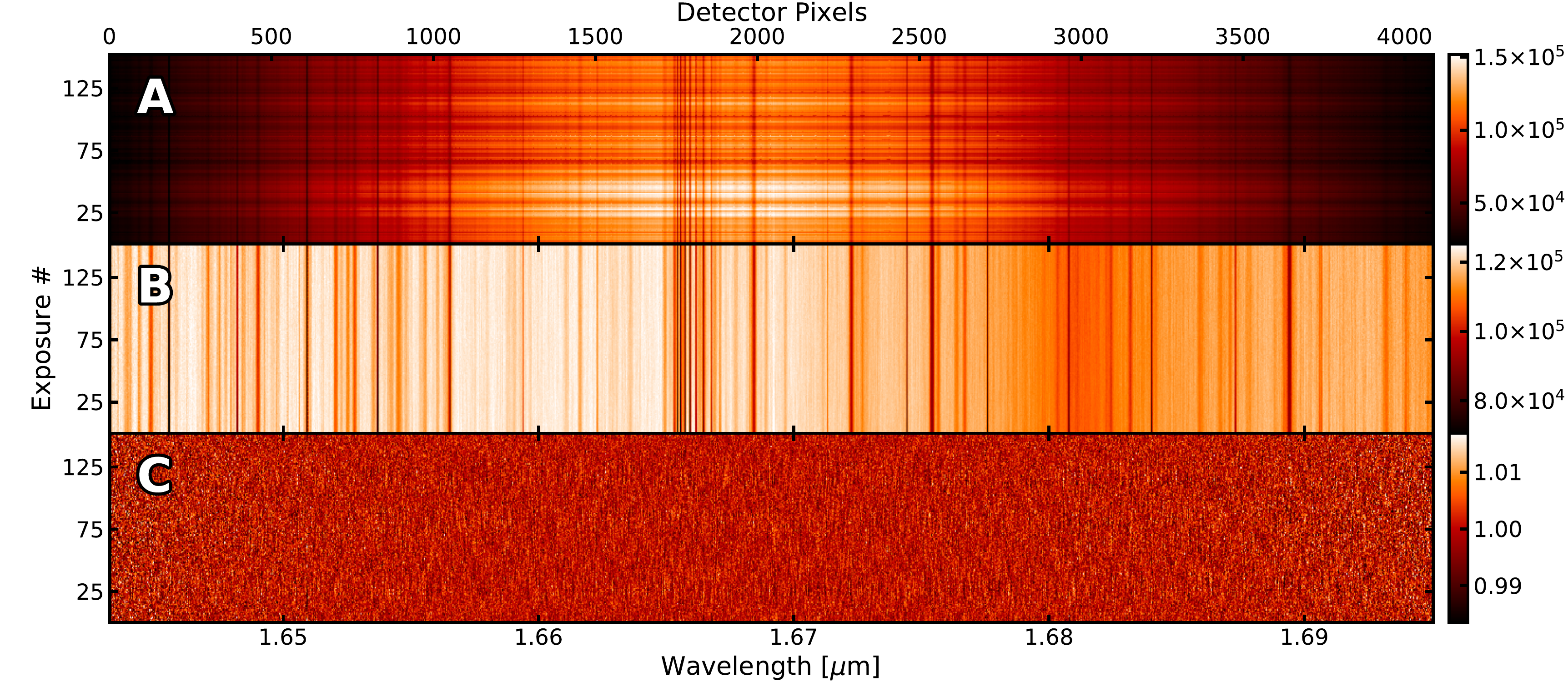}
\end{center}
\vspace{-2mm}
\caption{Detrending steps applied to the data to remove all contributions from the star, Earth's transmittance, and from systematics.  \textbf{A)} Example of a single SPIRou spectral order for the 5-hour observation sequence taken on 2019 April 15.  
Dark horizontal bands are exposures with lower signal-to-noise due to 
throughput variations while dark vertical lines are stellar and telluric absorption features.  \textbf{B)} After bad-pixel correction and continuum alignment.  \textbf{C)} Post removal of telluric and stellar contaminants via median spectrum and time polynomial fitting as well as PCA.  
The bottom panel contains the planetary signal buried in noise and serves as input for the cross-correlation and retrieval analyses. 
\label{fig:ReductionSteps}}
\end{figure*}

We reduced the raw SPIRou data using A PipelinE to Reduce Observations (\texttt{APERO}) v0.5~(\textcolor{citecolor}{Cook et al.}, in prep.), which outputs wavelength-calibrated and science-ready spectra for each order of each exposure.      
\texttt{APERO} also provides the data in an already telluric-corrected format, including the reconstructed transmission spectrum used to correct each exposure.  This reconstruction of Earth's transmittance is done via principal component analysis using a library of SPIRou observations of telluric standards collected at various airmasses and water vapor concentrations over many nights~(\citealt{artigau_telluric-line_2014}\textcolor{citecolor}{; Artigau et al.}, in prep.).  
For our analysis, we use the non-telluric-corrected version of the data 
and instead remove tellurics from the data itself 
as is commonly done in the literature~\citep[e.g.,][]{snellen_exoplanet_2010}.  This is done out of precaution, to avoid the automated removal of tellurics that may unexpectedly affect the underlying planetary signal in the data.  We note, however, that starting from the pre-telluric-corrected data product and performing the same detrending produces near-identical results and does not change the conclusions of this paper.  It may be beneficial to use this data product for SPIRou time series' with planetary trails that overlap in radial velocity shift with the telluric rest frame.
For our analysis we use the data in its extracted order-by-order flat-fielded format as provided by \texttt{APERO}.  An example time series for one order of the 2019 April 15 night of observations is shown in the top panel of Figure~\ref{fig:ReductionSteps}.

\subsubsection{Bad-pixel Correction}\label{subsubsec:bad_pixel_correction}

As a first step in the analysis, we identify and correct for outlier pixels remaining in the data.  
This is done by first dividing each exposure of each order by its median to bring all spectra to the same continuum level.  Then, each spectral channel is divided by its median in time.  In these data residuals, pixels deviating by more than 6$\sigma$ from the median of their channel are identified and flagged as bad pixels.  This procedure typically flags fewer than 0.05\% of pixels as being problematic.  As done in \cite{brogi_carbon_2014}, isolated single bad pixels are corrected by spline-interpolation while groups of 2--3 adjacent pixels are corrected via linear interpolation in the spectral direction.  Larger groups of 4 or more adjacent bad pixels are masked altogether. 
With all bad pixels either corrected or masked, the originally divided medians in time and wavelength are multiplied back to the now outlier-free residuals. 

\subsubsection{Blaze Removal and Continuum Alignment}\label{subsubsec:continuum_alignment}


As a next step, we divide out the \texttt{APERO}-provided blaze function and correct for continuum variations that can occur throughout an observation sequence.  To remove these small broadband discrepancies between different spectra, we first divide each spectral channel by its median in time. Then, for each exposure, a 21 pixel-wide box filter in the spectral direction smoothed by a Gaussian convolution with a standard deviation of 50 pixels is applied to each order individually. This filter is then divided out of the original exposures and the previously divided out medians in time are re-multiplied. This step brings all spectra to the same continuum level while also removing any broadband low-frequency trends without affecting the high-frequency planet signal of interest~\citep{gibson_revisiting_2019,gibson_detection_2020}.  The second panel of Figure~\ref{fig:ReductionSteps} shows the data after bad-pixel correction, blaze removal, and continuum alignment. 

\subsection{Removing Telluric \& Stellar Contaminants}\label{subsec:telluric_removal}
Before attempting to uncover the planetary signal, it is crucial to remove unwanted contributions from Earth's atmosphere and the host star. Many techniques exist to achieve such a `cleaning' of the data, all relying on the key concept that contrary to the Earth's and the star's atmospheres, which are stationary or quasi-stationary in wavelength, the exoplanet atmospheric features are rapidly being Doppler-shifted in time~\citep{birkby_spectroscopic_2018}. With an orbital period of just 3.31 days, $\tau$~Boo~b's orbital line-of-sight velocity accelerates by tens of km$\,$s$^{-1}$ over the course of the 3--5-hour long observations, moving across multiple pixels on the SPIRou detector. Therefore, the removal of all time varying effects that are constant in wavelength/pixel can get rid of nearly all traces of both the Earth's atmosphere and the star, while leaving the planetary signature mostly unaffected.  

\subsubsection{Median Spectrum \& Time Polynomial Fits}\label{subsubsec:mean_spec_time_fits}
To remove all flux contributions other than from $\tau$~Boo~b, we adapt and build on a combination of analysis methods available in the literature~\citep[e.g.,][]{brogi_retrieving_2019, damiano_principal_2019}. 
First, we divide each individual spectrum by a second-order polynomial fit of the median spectrum for that night.  
This first step acts as an approximate correction to both the stellar spectrum and Earth's transmittance.  Then, each spectral channel is divided by a second-order polynomial fit in time.  This is to account for remaining time-dependent residuals from poorly corrected stellar or telluric lines that still dominate the signal.  Stellar line residuals can remain because, while they are stationary in the star's rest frame, in the Earth frame of reference they shift by 0.4--0.6\,km\,s$^{-1}$ throughout the duration of the observations due to the barycentric motion of the Earth.  Meanwhile for telluric lines this is because of the fact that, while their position remains constant in wavelength, their depths and shapes change according to the varying temperature, wind profile, and molecular column densities in the line-of-sight.
At this stage, most smoothly varying effects of the star and Earth's atmosphere have been removed.  Any systematics that do not vary smoothly in time, however, can still remain in the data.  Indeed, clear localized non-white noise patterns that appear to be correlated with the movement of the telescope can be seen in some nights.  We discuss the potential source of these systematics 
in more detail in Appendix~\ref{sec:appendix_tel_angle}.  
The presence of such non-white noise in the detrended data can be problematic given that the signal of interest is at the $\sim$10$^{-4}$ contrast level.  However, while further cleaning of the data may help with removing any remaining unwanted features, it may also partially remove the planet's atmospheric signature.  The question then becomes: is further detrending actually beneficial for uncovering $\tau$~Boo~b's orbital trace?

\subsubsection{Principal Component Analysis}\label{subsubsec:pca}
In addition to the fitted median spectrum and polynomial in time, further cleaning of a spectral time series 
can be done by using principal component analysis (PCA).  
The basis of PCA is to decompose a data set into a set of orthogonal components ordered such that each captures as much of the variance in the data as possible.  A removal of the first few principal components should then eliminate unwanted effects that cause many different spectral channels to vary in a time-correlated way.  
Several forms of PCA have been used in the literature for analysing high-resolution data sets with great success (e.g., singular value decomposition~\citep{de_kok_detection_2013}, SYSREM~\citep{birkby_detection_2013}, eigenvalue decomposition~\citep{damiano_principal_2019}).  
Nevertheless, while PCA is optimal for removing stationary-in-wavelength spectral features and correlated noise, applying it too aggressively will at some point start to remove the signal of the planet~\citep{birkby_discovery_2017}.  This is due to the time-correlation caused by the finite amount of time it takes for spectral lines of the planet to be Doppler-shifted across the detector pixels.  In our case, the spectral lines of \tauboo shift by roughly 3--4 pixels per hour, resulting in unavoidable time-correlation on short timescales.  As a result, one must be careful in applying such a blind detrending algorithm.  

As opposed to the wavelength domain which would have more variables than observables (i.e., more pixels than spectra~\citep{damiano_principal_2019}), we perform our PCA analysis in the time domain, decomposing each data set into $N_{\mathrm{exposure}}$ components.  On top of the already removed fitted median spectrum and time polynomial, we remove a single principal component in the residual data sets that have strong correlated patterns present in them (2019 April 15, 2019 April 21, 2020 May 10). 
However, we do not remove any components from nights that have no clear systematics in their residuals (2020 June 2, 2020 June 5).  The hypothesized underlying cause of these observed non-white noise features in the time series observations of the first three nights of data, and how they were avoided in the subsequent nights is discussed in Appendix~\ref{sec:appendix_tel_angle}.  

To verify that correcting for the correlated noise seen in some observation sequences with the removal of a single principal component is justified and does not do more harm than good, we perform model injection/recovery tests~\citep[e.g][]{birkby_discovery_2017, nugroho_high-resolution_2017, hawker_evidence_2018}.  However, instead of using the known orbital solution of $\tau$~Boo~b, we use the negative Keplerian velocity $K_p$.  Performing this injection/recovery test at the known systemic velocity $V_{\mathrm{sys}}$ but at minus $K_p$ acts to avoid optimizing either the real signal or any underlying noise structures at the true location of the planet while also maintaining the same $\Delta$RV and position relative to the telluric rest frame as the real signal.  We regard this verification step as an essential cross-check for PCA-based HDCCS data analyses.  In practice, using the negative Keplerian velocity injects shifted planetary lines following an inverted trail that is redshifted instead of blueshifted (or vice versa).  We inject the model in the raw data to best represent a realistic signal and re-apply all detrending steps outlined so far. 
A cross-correlation analysis (as described in Section \ref{subsec:CrossCorrelation}) to estimate the signal-to-noise (S/N) at which the injected signal is recovered is then performed on the residual data with and without a principal component removed. 
This test indeed shows that removing a single component is helpful in better recovering an injected signal while removing additional components is not necessarily warranted.  On the nights that do not show discernible systematic noise patterns (2020 June 2 and June 5) it is not necessary to further detrend the data and remove any principal components, although doing so does not significantly affect the results.

\subsubsection{Masking}\label{subsubsec:masking}


Despite the procedures applied to correct for the effects of Earth's transmittance, regions of severe telluric contamination can still remain problematic for recovering a faint planetary signal.  
We opt to mask these regions for the subsequent analysis to ensure that there are no unwanted correlations with poorly corrected residual features from Earth's atmosphere.  To do this we use the reconstructed transmission spectrum of the Earth provided by \texttt{APERO} for each exposure and build a mask containing all pixels on which telluric absorption blocks more than 40\% of the continuum flux, or on which the sky emission (e.g., from OH emission lines) exceeds 110\%.  Such a mask typically removes 20--30\% of the data, depending on the humidity of the night, mostly in regions between the $Y,\, J,\, H$, and $K$ photometric bands.  
The choice of these percentage cutoffs is somewhat arbitrary and 
is a trade-off between having more data to work with and minimizing how much data is compromised by poorly-corrected tellurics.   Varying at what percentage these cutoffs are made does not significantly change our results.  Finally, noisy channels that have a standard deviation more than four times that of their spectral order are masked and not used for the remainder of the analysis.  This final mask typically contains less than 0.5\% of the data.  
We use this final detrended and masked data product (Figure~\ref{fig:ReductionSteps}, bottom panel) to search for and characterize the atmospheric signal of $\tau$~Boo~b.

\section{Modeling and High-Resolution Likelihood Prescription}
\label{sec:modeling}



With stellar and telluric features removed from the data, \tauboos atmospheric signature still remains buried in the noise and requires a cross-correlation analysis with a representative model to be uncovered.  
In this section we discuss how models of \tauboos atmosphere are generated, how these are processed to be representative of the signal in the detrended data, what likelihood prescription is used, and how the atmospheric retrieval is set up.

\subsection{Forward Model} \label{subsec:forward_model}

\begin{figure}[t]
\begin{center}
\includegraphics[width=\linewidth]{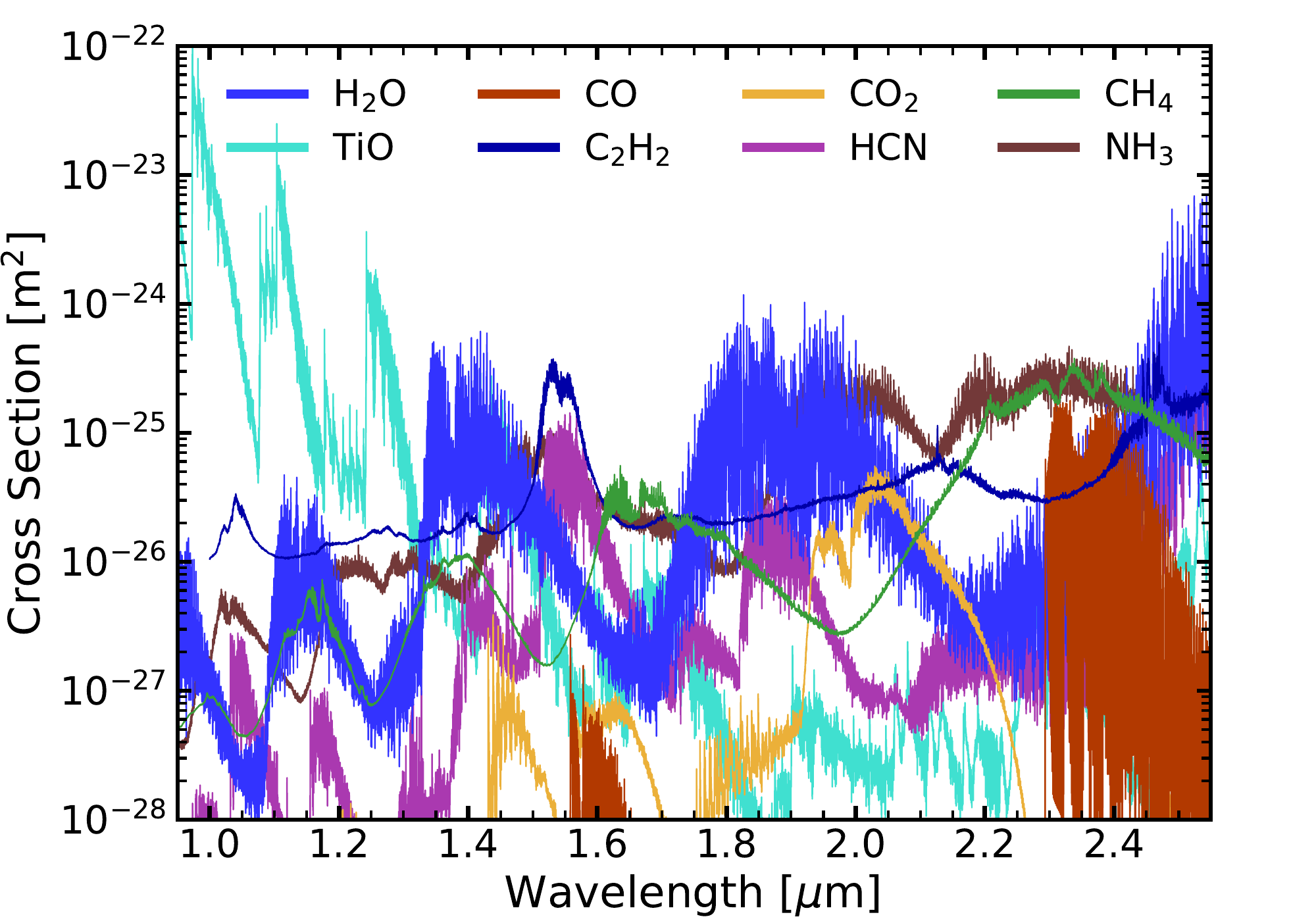}
\end{center}
\vspace{-2mm}
\caption{Molecular cross sections of species with prominent absorption bands in the 0.95--2.50\,$\mu$m wavelength range covered by SPIRou.  Values are computed at a temperature of 1500\,K and pressure of 0.1\,bar.  H$_2$O in particular has strong features throughout the near-infrared while CO has more localized absorption near 2.3\,$\mu$m.
\label{fig:mol_cross_sections}}
\end{figure}

We generate synthetic spectra of \tauboos atmosphere using the 
SCARLET 
framework~\citep{benneke_atmospheric_2012,benneke_how_2013,knutson_featureless_2014, kreidberg_clouds_2014,benneke_strict_2015,benneke_sub-neptune_2019,benneke_water_2019}.  SCARLET calculates high-resolution line-by-line thermal emission models for a given composition, thermal structure, and cloud properties.
Molecular opacities and associated line lists used in this work include H$_2$O~\citep{polyansky_exomol_2018,barber_high-accuracy_2006}, CO~\citep{rothman_hitemp_2010}, CH$_4$~\citep{yurchenko_exomol_2014}, CO$_2$~\citep{rothman_hitemp_2010}, HCN~\citep{barber_exomol_2014,harris_improved_2006}, TiO~\citep{mckemmish_exomol_2019}, C$_2$H$_2$~\citep{chubb_exomol_2020}, and NH$_3$~\citep{yurchenko_variationally_2011}.  Cross sections for these molecules are computed using the \texttt{HELIOS-K} opacity generator~\citep{grimm_helios-k_2015, grimm_helios-k_2021} and are accessible from the open-access DACE database\footnote{\texttt{https://dace.unige.ch/opacityDatabase}}.
All included molecules have prominent cross sections in the SPIRou bandpass and could be detectable if they are present in high enough volume mixing ratios (VMRs) in \tauboos atmosphere (Figure~\ref{fig:mol_cross_sections}).  
We do not add opacities from additional molecules or alkali metals as they are not expected to have noticeable effects in our data. 
H$_2$-H$_2$ and H$_2$-He collision-induced absorption is computed following \cite{borysow_collision-induced_2002}.  Models also include an optically thick gray cloud deck at a given cloud-top pressure $P_c$ in bars.  Assuming no other major carbon- or oxygen-bearing gaseous species are present in \tauboos atmosphere, the gas-phase C/O ratio can then be calculated as 
\begin{equation}\label{eq:CtoO}
     \rm (C/O)_{gas} = \frac{CO + CH_4 + CO_2 + HCN + 2C_2H_2}{H_2O + CO + 2CO_2 + TiO}.
\end{equation}


Given molecular abundances, a temperature-pressure (TP) profile, and a cloud-top pressure, all spectra are computed via line-by-line radiative transfer at a spectral resolution of $R = \lambda/\Delta \lambda = 250\,000$ and subsequently convolved with a Gaussian profile to match the $R=70\,000$ instrumental resolution of SPIRou.  The planetary thermal emission models $F_p(\lambda)$ are then scaled to the blackbody continuum level of the star and planet-to-star area ratio
\begin{equation}\label{eq:Fscaled}
    F_{\mathrm{scaled}}(\lambda) = \frac{F_p(\lambda)}{\pi B(\lambda,T_*)}\left(\frac{R_p}{R_*}\right)^2,
\end{equation}
where $B(\lambda,T_*)$ is the Planck function at the stellar effective temperature $T_*$,
while $R_p$ and $R_*$ are the planetary and stellar radii, respectively (Table~\ref{tab:tauboo_params}).  
As \tauboo is a non-transiting planet, its radius is not well constrained and thus we follow previous work and assume that $R_{p} = 1.15\,R_{\mathrm{Jup}}$, the average radius of all hot Jupiters in \tauboos mass regime measured from transiting planets~\citep{brogi_signature_2012, hoeijmakers_searching_2018}.  

\subsection{Accounting for Detrending Effects}\label{subsec:Tel_reconstruction}

While a generated forward model of \tauboos atmosphere may 
be a good description of the true astrophysical signal, it is not necessarily representative of the signal contained in the data after the removal of telluric and stellar contamination.  
This is because the detrending procedure that is applied to the data in Section~\ref{subsec:telluric_removal} to remove telluric and stellar features also stretches and scales the planetary signature (\citealt{brogi_retrieving_2019}, see also their \hbox{Figure 2}, \hbox{step 7} - noiseless).
This warping must be accounted for to produce an accurate atmospheric analysis and avoid introducing any biases when retrieving molecular abundances.  





To account for the distortions experienced by the underlying atmospheric signal during the telluric removal process, we replicate the same detrending procedure that was applied to the data onto each model. To achieve this, we first shift the scaled planetary model $F_{\mathrm{scaled}}(\lambda)$ in velocity space. For a given combination 
of $K_p$ and $V_{\mathrm{sys}}$, the radial velocity shift $V_p(t)$ that \tauboo would have at the observed epoch of each exposure is given by
\begin{equation}\label{eq:Vp}
    V_p(t) = K_p\big(\cos(f(t)+\omega) + e\cos\omega\big) + V_{\mathrm{sys}} + V_{\mathrm{bary}}(t),
\end{equation}
where $\omega$ is the longitude of periastron, $e$ is the eccentricity, $V_{\mathrm{bary}}(t)$ is the barycentric velocity of the Earth, and $f(t)$ is the true anomaly at the time of each observation.  This produces the expected planet-to-star contrast level $F_{\mathrm{scaled}}(\lambda, t)$ Doppler-shifted as a function of time for a given atmospheric model. 
We store the detrending steps outlined in Section~\ref{subsec:telluric_removal} 
that were divided out of the data to remove contributions from both the star $F_s(\lambda)$ and Earth's transmittance $T_{\mathrm{E}}(\lambda,t)$ and assume that the observed flux is of the form
\begin{equation}\label{eq:Fobs}
    F_{\mathrm{obs}}(\lambda,t) = F_s(\lambda)T_{\mathrm{E}}(\lambda,t)\big(1+ F_{\mathrm{scaled}}(\lambda, t) \big). 
\end{equation}
The reconstruction of the processing steps applied to the data, consisting of the fitted median spectrum, time polynomial, and principal components, then acts as an approximation of $F_s(\lambda)T_{\mathrm{E}}(\lambda,t)$.  
We can therefore inject each generated forward model into this reconstructed $F_s(\lambda)T_{\mathrm{E}}(\lambda,t)$, as per Equation~\ref{eq:Fobs}, and re-apply the same detrending steps as was done to the data in Section~\ref{subsec:telluric_removal}.  The result is then effectively $1+ F_{\mathrm{scaled}}(\lambda, t)$, but as the planet spectra would appear after having gone through the same detrending process as the real data.  Equation \ref{eq:Fobs}, but passed through the detrending procedure is then what serves as model input for the likelihood evaluation.

\subsection{Likelihood Prescription} \label{subsec:likelihood}

With the planetary model ($m$) now representative of what the atmospheric signal would be in the processed data ($d$), we calculate the likelihood ($L$) via the CCF-to-ln$L$ mapping 
\begin{equation}\label{eq:LogL}
    \mathrm{ln}(L) = - \frac{N}{2} \mathrm{ln} \big(s_d^2 - 2as_ds_m(v)\mathrm{CCF}(v) + a^2s_m^2(v)\big)
\end{equation}
derived in \cite{brogi_retrieving_2019}. Here $a$ is a scaling factor and the cross-correlation function is given by
\begin{equation}\label{eq:CCF}
    \mathrm{CCF}(v) = \frac{R(v) }{s_d s_m(v) },
\end{equation}
where
\begin{equation}\label{eq:R}
    R(v) = \frac{1}{N} \sum_n d(n) m(n-v),
\end{equation}
\begin{equation}\label{eq:s_f}
    s_d^2 = \frac{1}{N} \sum_n d^2(n),
\end{equation}
\begin{equation}\label{eq:s_g}
    s_m^2(v) = \frac{1}{N} \sum_n m^2(n-v),
\end{equation}
for $N$ number of spectral channels~\citep{zucker_cross-correlation_2003}. The CCF is then calculated at a given velocity shift ($v$) for each order of each exposure by summing over each wavelength bin ($n$).  Both $d$ and $m$ are mean normalized immediately before the CCF is computed.  
Keeping $a$ as a free parameter relaxes the presumption that the imposed planet-to-star line contrast is representative of the true signal. 
In particular for a non-transiting planet such as $\tau$~Boo~b, for which the precise planetary radius is unknown, fitting $a$ simultaneously is crucial to encompass modeling uncertainties.


\subsection{Retrieval Setup}\label{subsec:retrieval_setup}
With the CCF-to-ln$L$ mapping and an appropriate treatment of the generated models to replicate the detrending procedure added to the SCARLET framework, we can now perform atmospheric retrievals on high-resolution spectroscopy data sets. 
We opt for classical `free' retrievals that fit for molecular abundances directly to avoid making any assumptions on the chemistry on $\tau$~Boo~b.  Parameters explored in this analysis include molecular abundances, temperature structure points, cloud properties, orbital parameters, and the scaling factor $a$.
We adopt log-uniform priors on molecular volume mixing ratios between $10^{-10}$ and $1$, on the cloud top pressure $P_c$ between 0.1\,mbar and 100\,bar, as well as for the scaling parameter $a$ between 0.1 and 10.  Linear-uniform priors are used for 
$K_p$ and $V_{\mathrm{sys}}$ between $+50$ and $-50$\,km\,s$^{-1}$ from their respective known values.  


\subsection{Temperature Structure Parameterization}\label{subsec:TP_parameterization}
The precise shape of spectral lines probed in exoplanetary thermal emission spectra depends on the temperature profile of the planet.  For example, if a planet has a thermal inversion, with temperature increasing with altitude, spectral features can be observed in emission rather than absorption.  Similarly, the steepness of the temperature gradient dictates the narrowness of spectral lines.  Hot Jupiters in particular can have different temperature structures depending on the presence of ultraviolet and visible opacity sources driving at what altitude the stellar irradiation is absorbed~\citep{arcangeli_h_2018, merritt_non-detection_2020, nugroho_searching_2020}.  It is thus crucial that, when performing an atmospheric retrieval on high-resolution hot Jupiter thermal emission observations, an appropriate TP profile parameterization able to capture all possible spectral line shapes is adopted.

We parameterize the TP profile with $N$ free temperature points $\textbf{\textit{T}}$ uniformly distributed in log pressure between $p_{\mathrm{top}} =10^{-7}$ and $p_{\mathrm{bot}} = 10^{2}$\,bar, the respective pressures at the top and bottom of the atmosphere.  We impose a uniform prior between 0 and 5000\,K on each temperature point as well as a Gaussian prior of mean zero and variance $\sigma_s^2$ on the root mean square of the second derivative of the profile.  Dropping constant terms, this smoothing prior on the TP profile ($\pi$) is given by 
\begin{multline} \label{eq:TP_prior_1}
\ln  \pi(T(\log p)) = -\frac{1}{2 \sigma_s^2}  \left( \mathrm{RMS} \left( \frac{\mathrm{d}^2T}{\mathrm{d}(\log p)^2 } \right) \right)^2  \\
 = \frac{-1}{2 \sigma_s^2 \log (p_{\mathrm{bot}}/p_{\mathrm{top}}) } 
\int_{\log p_{\mathrm{top}}}^{\log p_{\mathrm{bot}}} 
 \left( \frac{\mathrm{d}^2T}{\mathrm{d}(\log p)^2 }\right)^2 \mathrm{d}(\log p), 
\end{multline}
where RMS is the root mean square of the continuous second derivative of the temperature-pressure profile $T(\log p)$.  
This smoothing prior effectively punishes TP profiles that have high second derivatives and prevents nonphysical zig-zaggy temperature oscillations in pressure regions that are less well constrained by the data.  The standard deviation $\sigma_s$ is physically motivated and has units of kelvin per pressure decade squared (K\,dex$^{-2}$), i.e., the unit of the second derivative. It controls how strongly the prior punishes the curvature in the temperature structure. Large values of $\sigma_s$ allow for more curvature and deliver profiles with high vertical resolution. Assuming grid points uniformly distributed in log pressure, Equation~\ref{eq:TP_prior_1} can be written in discretized form as
\begin{equation}\label{eq:TP_prior_2}
    \mathrm{ln} \, \pi(\textbf{\textit{T}}) = \frac{-1}{2 \sigma_s^2 \log (p_{\mathrm{bot}}/p_{\mathrm{top}})  }  \sum_{i=2}^{N-1}  \frac{(T_{i+1} - 2T_i + T_{i-1})^2}{(\Delta\log p)_i^3 },
\end{equation}
with the summation over each layer $i$ in the pressure interval between $p_{\mathrm{top}}$ and $p_{\mathrm{bot}}$.

A curvature-punishing smoothing prior was previously used for brown dwarf retrievals \citep{line_uniform_2015}; however, their parameterization using a $\gamma$ parameter was variant across different pressure grids and, moreover, we argue that $\sigma_s$ should not be simultaneously fit for most exoplanet applications. In particular, fitting $\sigma_s$ often leads to overly smooth profiles where extrapolation effects may falsely deliver tight constraints in pressure regions not probed by the data. This is particularly true for highly irradiated hot Jupiters whose dayside TP profiles are not as smooth as those of brown dwarfs. Instead, we argue that the $\sigma_s$ parameter is effectively a user choice that sets the trade-off between the vertical spatial resolution and the noisiness and stability of the temperature structure. For $\tau$~Boo~b, we explore different $\sigma_s$ values between 50\,K\,dex$^{-2}$ (strong smoothing) and 1000\,K\,dex$^{-2}$ (weak smoothing).  We find the smoothness prior to be too stringent for $\sigma_s \lesssim 80$\,K\,dex$^{-2}$ resulting in overly smooth temperature profiles that propagate constraints from the well-constrained mid-atmosphere into the deeper layers of the atmosphere. On the other extreme, we find nonphysical oscillations in pressure regions not readily probed by the data for $\sigma_s \gtrsim 500$\,K\,dex$^{-2}$. For our final analysis we opt for a value of $\sigma_s = 240$\,K\,dex$^{-2}$, which we find for this data set to be the best compromise between vertical resolution of the temperature structure and the noisiness and stability of the profile.  We note that the retrieved abundances do not vary significantly under different adopted smoothing factors values.

The advantages of our temperature retrieval prescription for exoplanets are that it is independent of the pressure grid and that it does not make any prior assumption on the overall shape of the temperature structure, which is generally a disadvantage of parametric forms for the temperature-pressure profile \citep[e.g.,][]{madhusudhan_temperature_2009, guillot_radiative_2010, benneke_atmospheric_2012, line_information_2012, line_systematic_2013, line_no_2016}.  We set the number of layers to $N=15$, enough to properly sample the TP profile for our \tauboo data set, while keeping the number of retrieval parameters small. 
We note that our physically motivated formulation of $\sigma_s$ in K\,dex$^{-2}$ is independent of the number of pressure levels and we obtain the same results by introducing more pressure levels. Too few temperature points, however, can result in an overly coarse TP profile being retrieved.

\subsection{Computation}

Performing full Bayesian atmospheric retrievals on HDS data sets is a computationally challenging task.  
Relative to low- or medium-resolution spectroscopic data sets from space-based instruments, the added computational cost of running HDS retrievals comes primarily from both the need to generate each forward model at very high spectral resolution ($R \geq 250\,000$) and the need to re-apply the detrending steps to each reconstructed data set.  
Calculated over the full 0.95--2.50\,$\mu$m wavelength range, a SCARLET 
model takes 2--3 seconds to compute when using only a single CPU core.  Processing such a model 
as per Section~\ref{subsec:Tel_reconstruction} for an observation sequence of 100--150 exposures further takes 2--5 seconds.  
The code is fully parallelizable to run simultaneously on many CPU cores on a cluster, but can still take up to multiple weeks to run to full convergence.
A typical SCARLET high-resolution retrieval has 27 parameters simultaneously being fitted, namely 15 temperature points, two orbital parameters, a cloud-top pressure, a scaling parameter, as well as the volume mixing ratio of eight molecules.

\section{Atmospheric Analysis} \label{sec:analysis}



At high spectral resolution, broad molecular bands can be resolved into forests of thousands of distinct spectral lines.  While individual lines may not have sufficient signal-to-noise to be detected, their individual contributions can be combined together via a cross-correlation function to produce a boosted signal~\citep[e.g.,][]{birkby_spectroscopic_2018}.  
As molecules each have unique spectral signatures, a CCF with different model templates can be used to identify which species are present in the observed exoplanet atmosphere.  In this section we first perform a blind cross-correlation search for molecules on $\tau$~Boo~b, finding a strong CO signal but a lack of absorption from H$_2$O or other molecules at the inferred orbital location.  We subsequently use the SCARLET retrieval framework adapted for high-resolution data sets to constrain the atmospheric and orbital properties of $\tau$~Boo~b.  Finally, we investigate the authenticity of the retrieved molecular abundances with injection/recovery tests.





\subsection{Initial Search: Cross-Correlation Analysis}\label{subsec:CrossCorrelation}

As \tauboo moves in time, the maximum of the CCF of each exposure should follow an orbital path as determined by the combination of the motion of the Earth around the Sun, of the $\tau$~Boo system relative to the Sun, and of \tauboo relative to its host star (Equation~\ref{eq:Vp}).  As such, a phase folding of the cross-correlation function for different combinations of $V_{\mathrm{sys}}$ and $K_p$ should reveal the strongest signal at the true values.  A comparison of the sum of the CCF in the known rest frame of the planet relative to erroneous velocity combinations can then give an estimate of the S/N of the atmospheric detection.

\begin{figure}[t]
\begin{center}
\includegraphics[width=\linewidth]{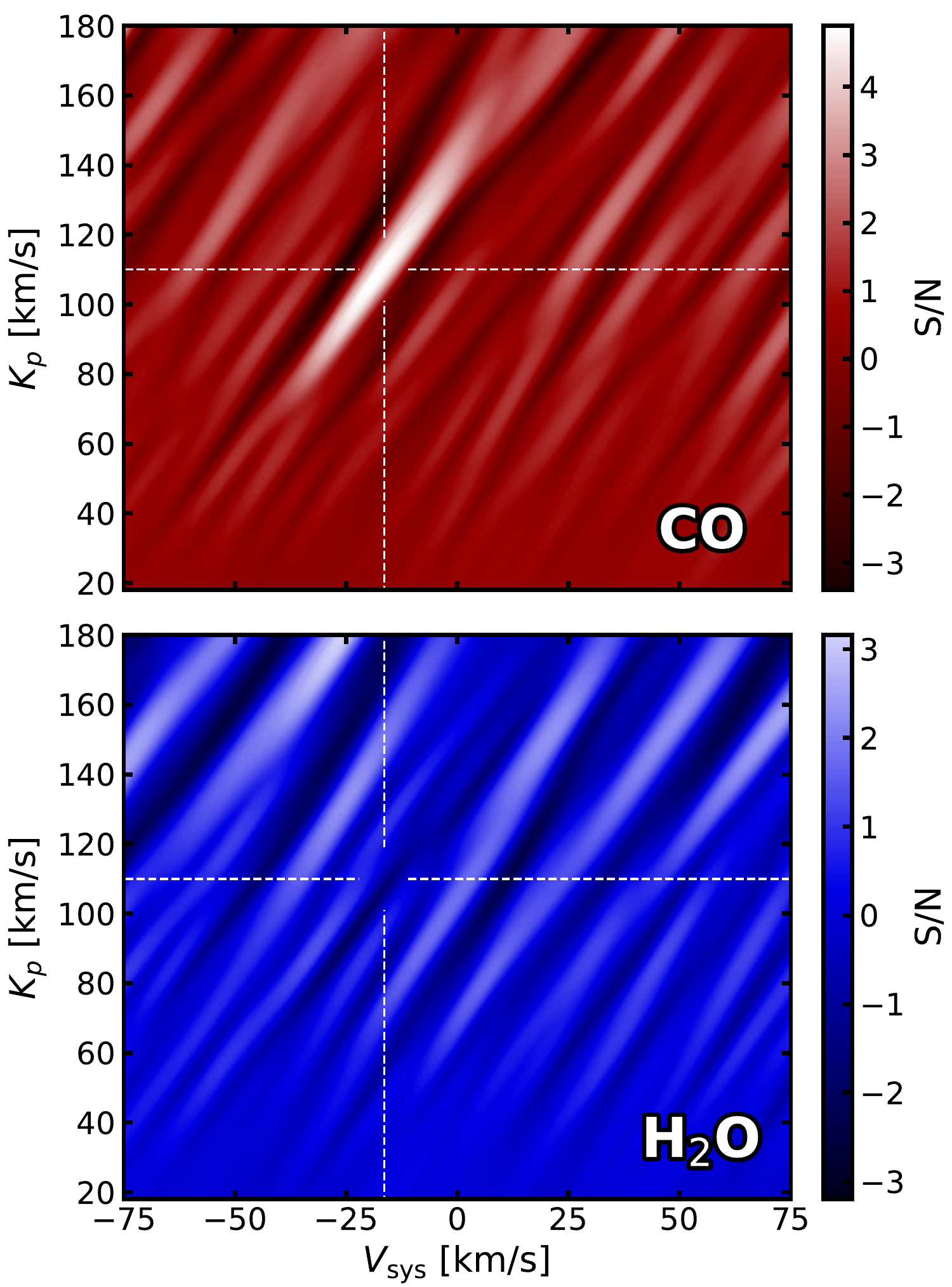}
\end{center}
\vspace{-2mm}
\caption{Cross-correlation signal-to-noise maps for CO and H$_2$O on $\tau$~Boo~b.  Shown here are phase-folded 2D velocity maps of the molecular signals of CO (top) and H$_2$O (bottom) relative to the known Keplerian and systemic velocities of the planet (white dashed lines) for the observations taken on 2019 April 15. A clear CO signal is present at the expected location of the planet but no evidence of water absorption can be seen.
\label{fig:KpVsys_CO_H2O}}
\end{figure}

We begin by performing a blind search for \tauboos atmospheric signature in velocity space.  The search is performed by setting up a velocity grid ranging from $-300$ to $+300$\,km\,s$^{-1}$ in steps of 2\,km\,s$^{-1}$ roughly corresponding to the velocity sampling of the SPIRou spectra on the detector pixel grid.  For each of these velocity shifts, a planetary model is Doppler-shifted and then interpolated to the same wavelength points as the data.  With both data and model on a common wavelength grid, the CCFs (Equation~\ref{eq:CCF}) of each spectral order of each exposure are calculated and combined. 
We convert summed CCFs into signal-to-noise estimates by dividing them by the standard deviation of all values excluding those near the expected location.

The resulting $K_p$\,$-$\,$V_{\mathrm{sys}}$ S/N maps derived from CO and H$_2$O templates for the night of 2019 April 15 are shown in Figure~\ref{fig:KpVsys_CO_H2O}.  
CO is clearly detected at $K_p$ and $V_{\mathrm{sys}}$ values consistent with the known radial velocity of the star and the $K_p$ inferred by previous CRIRES observations of $\tau$~Boo~b~\citep{brogi_signature_2012}.  On the contrary, even though the SPIRou data is more sensitive to H$_2$O than CO, no evidence for any water detection is seen.  Here we see SPIRou's capability for detecting atmospheric signatures in even a single high quality 5-hour observation sequence.

A similar search performed on the other four nights shows CO being added constructively at the expected location but, again, no sign of H$_2$O absorption is evident.  The strength of the CO signal varies from being strong to marginal depending on the night, although this is expected due to the varying observing conditions of each night.  The data quality obtained from each night and how sensitive each is to an underlying planetary signal due to the varying integration time, water column density in Earth's atmosphere, and presence of systematics is investigated in more detail in Appendix~\ref{sec:appendix_night_sensitivity}.

\begin{figure*}[t!]
\begin{center}
\includegraphics[width=\linewidth]{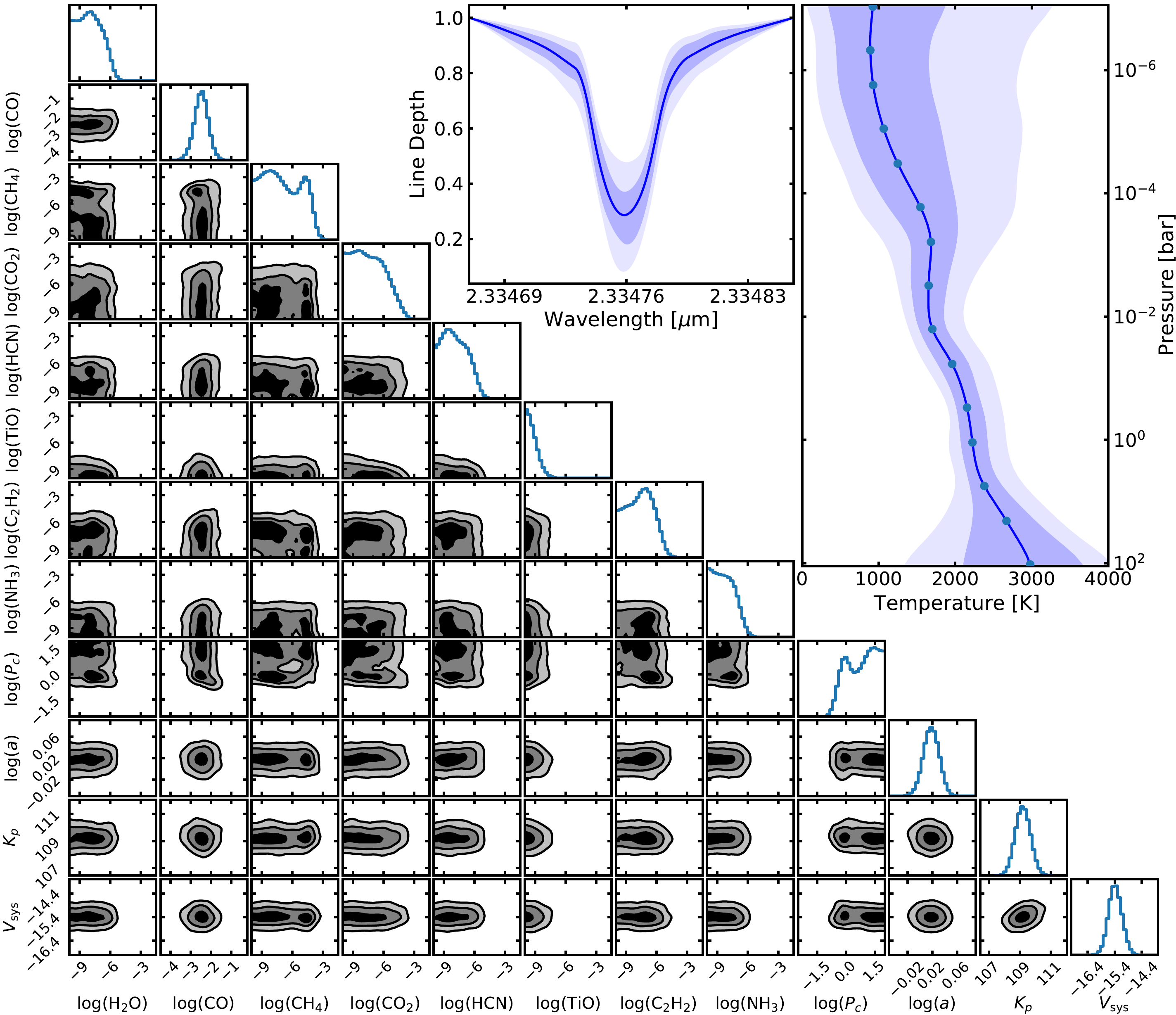}
\end{center}
\vspace{-2mm}
\caption{
Retrieved constraints on the atmospheric and orbital properties of \tauboo obtained from all five SPIRou nights of data combined.  The left corner plot shows the marginalized posterior distributions for the retrieved molecular abundances, the cloud-top pressure $P_c$ (in bars) and scaling parameter ($a$), as well as the orbital parameters $K_p$ and $V_{\mathrm{sys}}$. The black, dark-gray, and light-gray shaded regions respectively depict the 1$\sigma$ (39.3\%), 2$\sigma$ (86.5\%), and 3$\sigma$ (98.9\%) confidence intervals.
The top middle panel depicts the shape of a typical CO line calculated from drawing random model samples from the MCMC chain all normalized to the same line continuum.  
The top right panel shows the retrieved non-inverted temperature structure recovered from fitting 15 temperature points (dots) with a $\sigma_s = 240$\,K\,dex$^{-2}$ smoothing parameter (Equation~\ref{eq:TP_prior_2}).
The retrieved super-solar CO abundance combined with a lack of a significant concentration of H$_2$O or other species on \tauboo is consistent with an atmosphere with a slightly enriched Jupiter-like C/H ratio and a C/O ratio near unity. 
\label{fig:retrieval_results}}
\end{figure*}

Having prominent features in the wavelength range covered by SPIRou (Figure~\ref{fig:mol_cross_sections}), we also searched for CH$_4$, CO$_2$, HCN, TiO, C$_2$H$_2$, and NH$_3$ in all nights but found no statistically significant signal for any of these.  This is, however, not surprising as these molecules are not necessarily expected to be present in detectable levels in \tauboos atmosphere. 
We note that all conclusions regarding the detection or non-detection, as well as any abundance constraints, of molecular species is dependent on the accuracy of the line lists used (Section~\ref{subsec:forward_model}).

\subsection{Retrieval Analysis}\label{subsec:retrieval_analysis}

We calculate posterior distributions for the orbital and atmospheric parameters of $\tau$~Boo~b using the SCARLET retrieval framework adapted to analyze high-resolution spectroscopic data. 
The retrieved abundances, orbital velocity components, line shape, and temperature structure for all spectral orders of all five SPIRou nights combined are shown in Figure~\ref{fig:retrieval_results} and tabulated in Table~\ref{tab:retrieval_results}.  
We recover a slightly super-solar carbon monoxide abundance of log(CO) = $-2.46_{-0.29}^{+0.25}$ and find that models with very little water (3$\sigma$ upper limit of log(H$_2$O) = $-5.66$, corresponding to $0.0072\, \times$\,solar) best match the data. We find no evidence for a significant abundance of CH$_4$, CO$_2$, HCN, TiO, C$_2$H$_2$, or NH$_3$ in \tauboos atmosphere, with upper limits for each of these molecules reported in Table~\ref{tab:retrieval_results}.
The dayside atmosphere of \tauboo is consistent with being relatively cloud-free, with a cloud deck disfavored above the 250\,mbar -- 1\,bar pressure level.  The cloud-top pressure is partially degenerate with the CO abundance as the muting of spectral lines by a higher altitude cloud deck can be counteracted by an increase in line depth as a result of a higher CO abundance.  The recovered temperature structure is non-inverted, consistent with previous findings, and has a sub-adiabatic lapse rate of roughly d$T$/dlog$_{10}(P) \sim 300$\,K per pressure decade in the 10\,mbar -- 1\,bar regions most readily probed by the data.  A typical CO line drawn from a random sampling of the scaled model thermal spectra in the MCMC chain (all normalized to the same line continuum) shows a median depth of $\sim$70\% relative to the continuum and no evidence of partial emission in the line core~(Figure~\ref{fig:retrieval_results}, top-middle panel).

\begin{table}
\centering
\caption{\label{tab:retrieval_results} \tauboo Atmospheric Retrieval  Results}
\vspace{-3mm}
\renewcommand{\arraystretch}{1.1}
\begin{tabular}{lc}
\hline
\hline
Parameter               &  Value           \\
\hline
VMR log(CO)                              &  $-2.46_{-0.29}^{+0.25}$    \\
VMR log(H$_2$O)                          &  $<-$5.66 (3$\sigma$ upper limit)    \\
VMR log(CH$_4$)                          &  $<-$3.78 (3$\sigma$ upper limit)    \\
VMR log(CO$_2$)                          &  $<-$3.99 (3$\sigma$ upper limit)    \\
VMR log(HCN)                             &  $<-$5.37 (3$\sigma$ upper limit)    \\
VMR log(TiO)                             &  $<-$7.54 (3$\sigma$ upper limit)    \\
VMR log(C$_2$H$_2$)                      &  $<-$4.88 (3$\sigma$ upper limit)    \\
VMR log(NH$_3$)                          &  $<-$6.10 (3$\sigma$ upper limit)    \\
Cloud-Top Pressure ($P_c$)               &  $>0.26$\,bar (3$\sigma$ lower limit) \\
Scaling Parameter ($a$)                  &  $1.04\pm0.03$     \\
Keplerian Velocity ($K_p$)               &  $109.2\pm0.4$\,km\,s$^{-1}$    \\
Systemic Velocity ($V_{\mathrm{sys}}$)   &  $-15.4\pm0.2$\,km\,s$^{-1}$     \\
\hline
Derived parameters:\\
\hline
(C/H)$_{\mathrm{gas}}$  &   5.85$_{-2.82}^{+4.44}$\,$\times$\,solar   \\
(O/H)$_{\mathrm{gas}}$         &     3.21$_{-1.56}^{+2.43}$\,$\times$\,solar   \\
(C/O)$_{\mathrm{gas}}$         &     $1.00_{-0.00}^{+0.01}$   \\
(C/O)$_{\mathrm{all}}$         &     $>$0.60 (extreme limit)   \\
\hline
\hline
\end{tabular}
\end{table}

We recover a planet velocity semi-amplitude 
$K_p=109.2 \pm 0.4$\,km\,s$^{-1}$ 
that is consistent with previous results using CRIRES~\citep{brogi_signature_2012,cabot_robustness_2019, watson_doppler_2019}.  When combined with the stellar mass and stellar velocity semi-amplitude (Table~\ref{tab:tauboo_params}), the inferred $K_p$ of \tauboo corresponds to a mass of 
$6.24\pm0.23$\,$M_{\mathrm{Jup}}$ 
and an orbital inclination of $43.5\pm0.7^{\circ}$.  The velocity of the $\tau$~Boo system is constrained directly from the planetary signal to be 
$V_{\mathrm{sys}} = -15.4\pm0.2$\,km\,s$^{-1}$. 
This is slightly offset from previous values between $-16$ and $-16.4$\,km\,s$^{-1}$ derived from radial velocity measurements~\citep{nidever_radial_2002, donati_magnetic_2008, brogi_signature_2012, borsa_gaps_2015}, and from the predicted present-day systemic shift of roughly $-$17\,km\,s$^{-1}$ expected due to the motion of the wide-orbit M dwarf companion $\tau$~Boo~B~\citep[][]{justesen_constraining_2019}.  
The global scaling parameter does not show significant degeneracy with recovered molecular abundances and is retrieved to be $a = 1.04\pm0.03$, a value roughly consistent with unity perhaps indicative of a slight underestimation of \tauboos atmospheric scale height.  An insufficiently puffy modeled atmosphere may be caused by the assumed planet radius $R_{p} = 1.15\,R_{\mathrm{Jup}}$ (Section~\ref{subsec:forward_model}) being slightly too small.


\subsection{Signal Verification \& Water Sensitivity Test}\label{subsuc:Model_validation_h2o_sensitivity}
While a near-solar slightly enhanced abundance of CO on \tauboo is in line with expectations from both theory and previous results, the very low upper limit determined for the water abundance is somewhat surprising.  To better understand the retrieved results, we investigate how the best fit model fares when injected in the raw data before any processing has been applied.

\begin{figure}[t]
\begin{center}
\includegraphics[width=\linewidth]{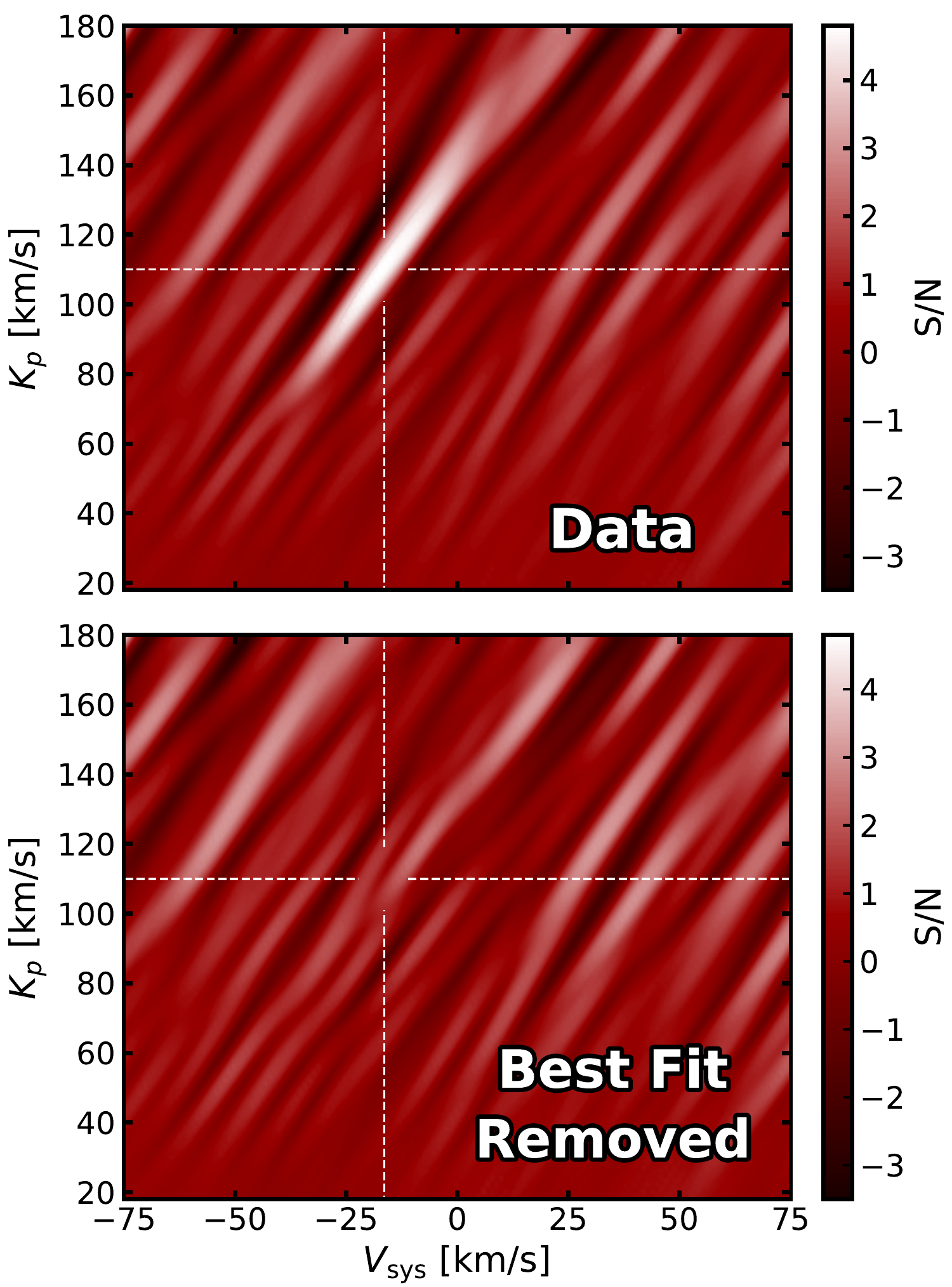}
\end{center}
\vspace{-2mm}
\caption{Canceling effect from the removal of the best fitting model from the 2019 April 15 night of observation.
Top: the CO signal of the data (same as Figure~\ref{fig:KpVsys_CO_H2O} top panel).  Bottom: same as above but with the best fitting model injected into the data at $\times-$1 nominal strength.  The model is injected prior to the detrending procedure.  The near perfect cancellation of the signal by the negatively injected best fit model shows that it well represents \tauboos CO signal.
\label{fig:CO_inj}}
\end{figure}

First we verify the planetary nature of the CO signal by injecting the negative of the retrieved best fit planetary model into the raw data at the inferred $K_p$ and $V_{\mathrm{sys}}$.  If the signal is indeed from the planet and the model is a good fit then the CO detection should be perfectly cancelled and disappear.  We find that this is indeed what happens.  After injecting the negative planet signal and reapplying the detrending procedure (Section~\ref{subsec:telluric_removal}) and repeating the cross-correlation analysis (Section~\ref{subsec:CrossCorrelation}), there is indeed no CO signature left in the data (Figure~\ref{fig:CO_inj}).  The near perfect cancellation of the signal indicates that the atmospheric model recovered is at a spectroscopic contrast that well matches \tauboos underlying signature.  We argue that this negative injection test is a verification procedure that should be widely applied in the literature to validate how well an atmospheric model truly represents the underlying planetary signal.



\begin{figure}[t]
\begin{center}
\includegraphics[width=\linewidth]{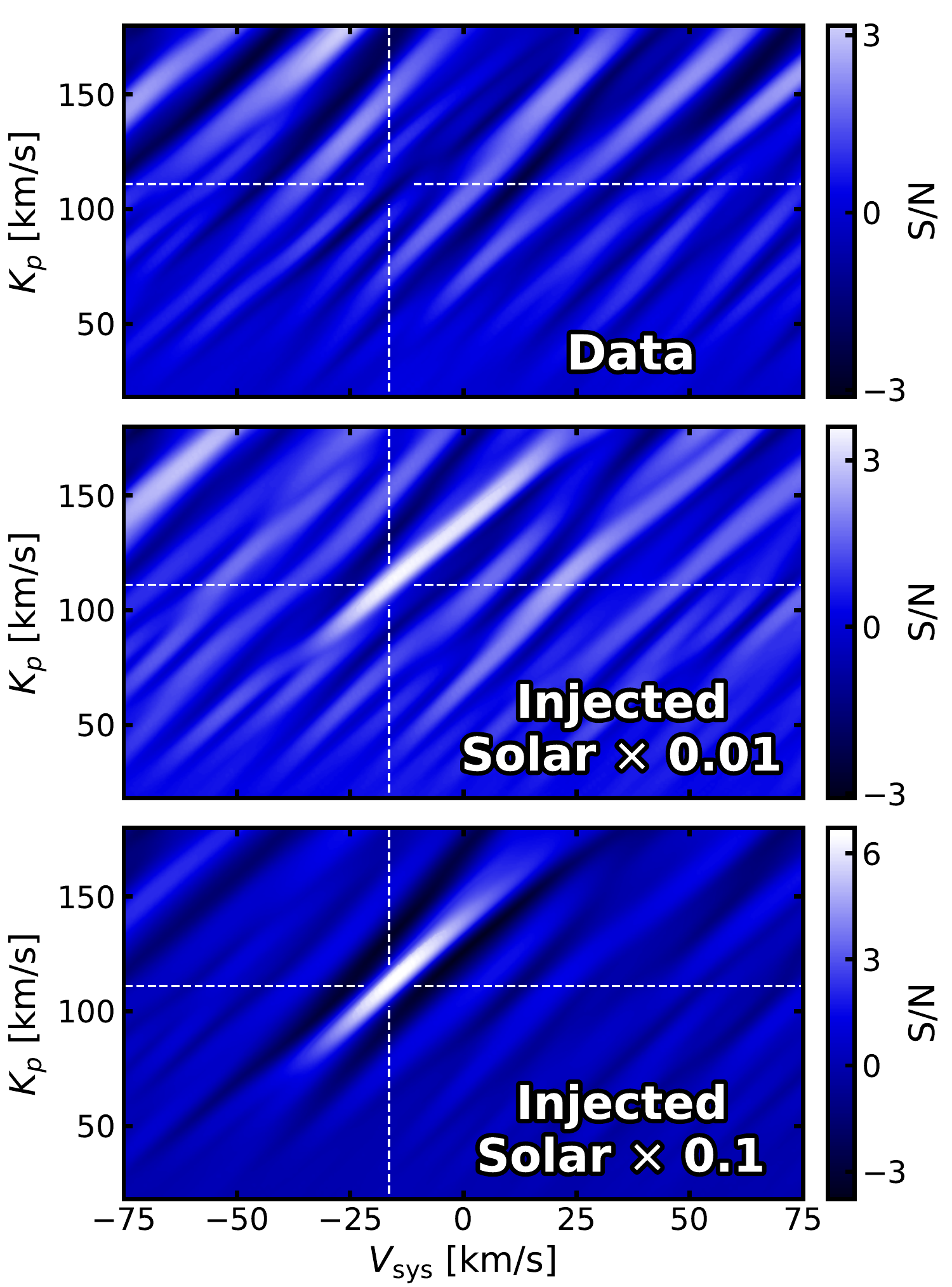}
\end{center}
\vspace{-2mm}
\caption{Sensitivity of the \tauboo data to injected water signals.  Top: H$_2$O map for the night of 2019 April 15 (same as Figure~\ref{fig:KpVsys_CO_H2O} bottom panel).  Subsequent panels: H$_2$O signal injected at increasing strengths demonstrating how sensitive even a single 5-hour SPIRou time series is to an underlying water signal.  Note the varying colorbar scale for each panel.  If present, even a sub-solar H$_2$O abundance should be easily detectable with SPIRou.
\label{fig:H2O_inj}}
\end{figure}

We also investigate the lack of a clear water signature seen in the data.  As H$_2$O has significant absorption throughout most of the SPIRou bandpass (Figure~\ref{fig:mol_cross_sections}), any underlying signal should be readily found in the data with even more sensitivity than CO.  To better understand the retrieved H$_2$O abundance upper limit, we explore how sensitive the data is to various injected water VMRs.  Using the best fit temperature structure, we inject simulated atmospheric models with different H$_2$O concentrations into the raw data and re-perform the full data reduction and cross-correlation analysis for each case.  The results of this water sensitivity test for the night of 2019 April 15 are shown in Figure~\ref{fig:H2O_inj}, where the observed water $K_p$ -- $V_{\mathrm{sys}}$ S/N map is compared to different injected sub-solar H$_2$O signals.  Here we conclude that the absence of a water detection in the data is not due to a lack of sensitivity as even an injected VMR$_{\mathrm{H}_2\mathrm{O}} = 0.01\times$\,solar can robustly be recovered from a single 5-hour SPIRou observation sequence.

\subsection{Elemental Abundances}\label{subsec:elemental_abundances}

\begin{figure}[ht!]
\begin{center}
\includegraphics[width=\linewidth]{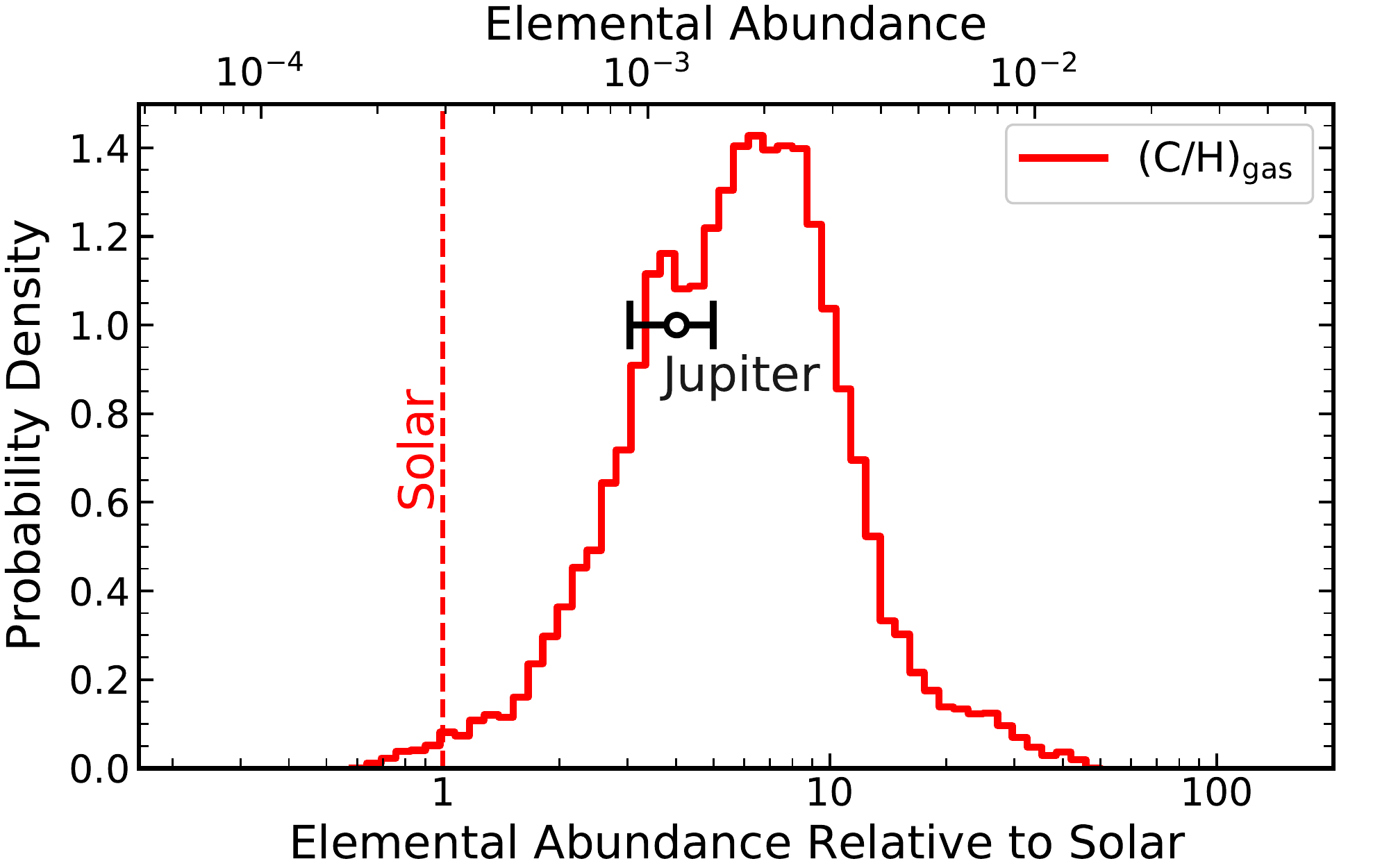}
\includegraphics[width=\linewidth]{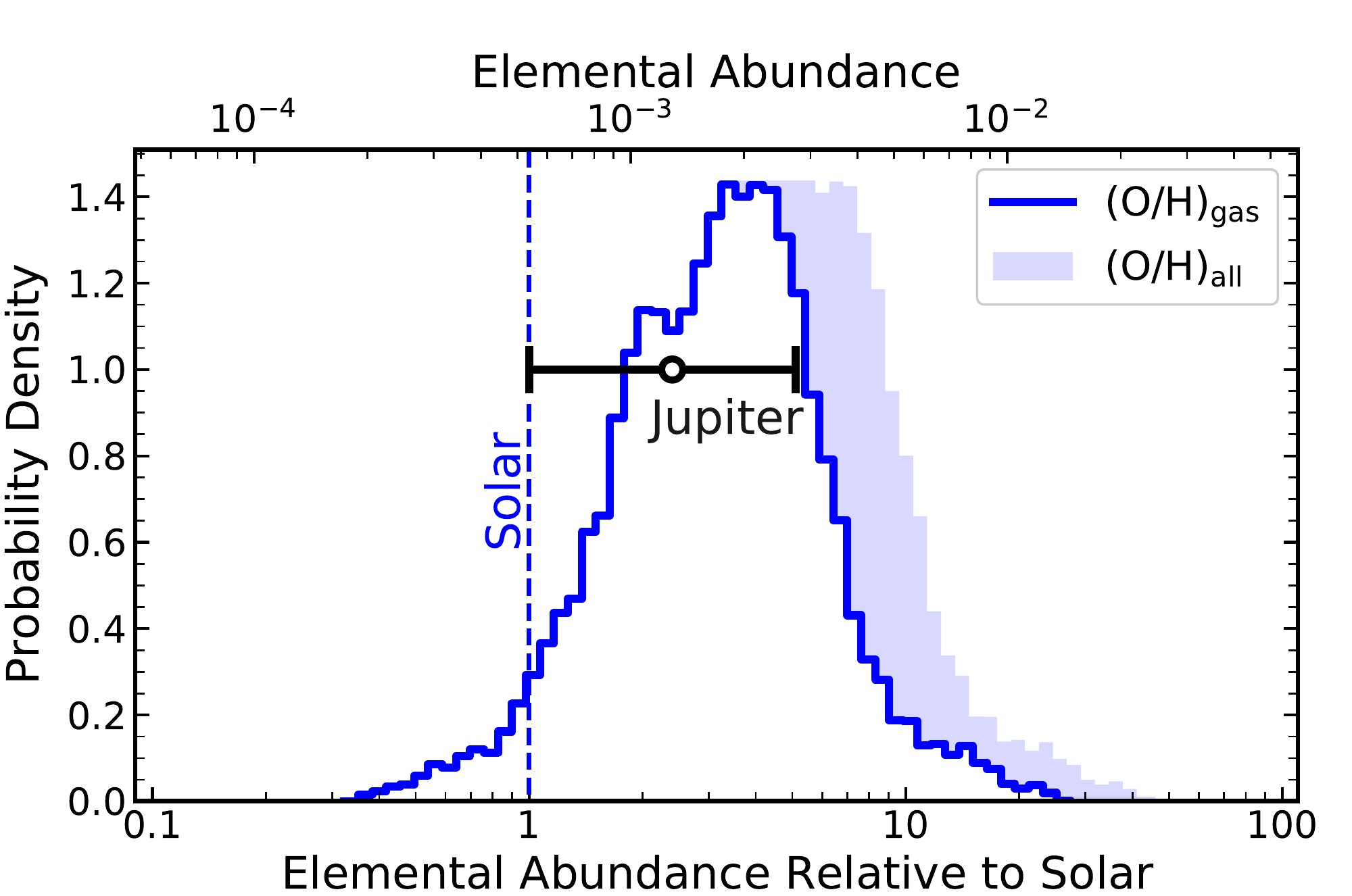}
\includegraphics[width=\linewidth]{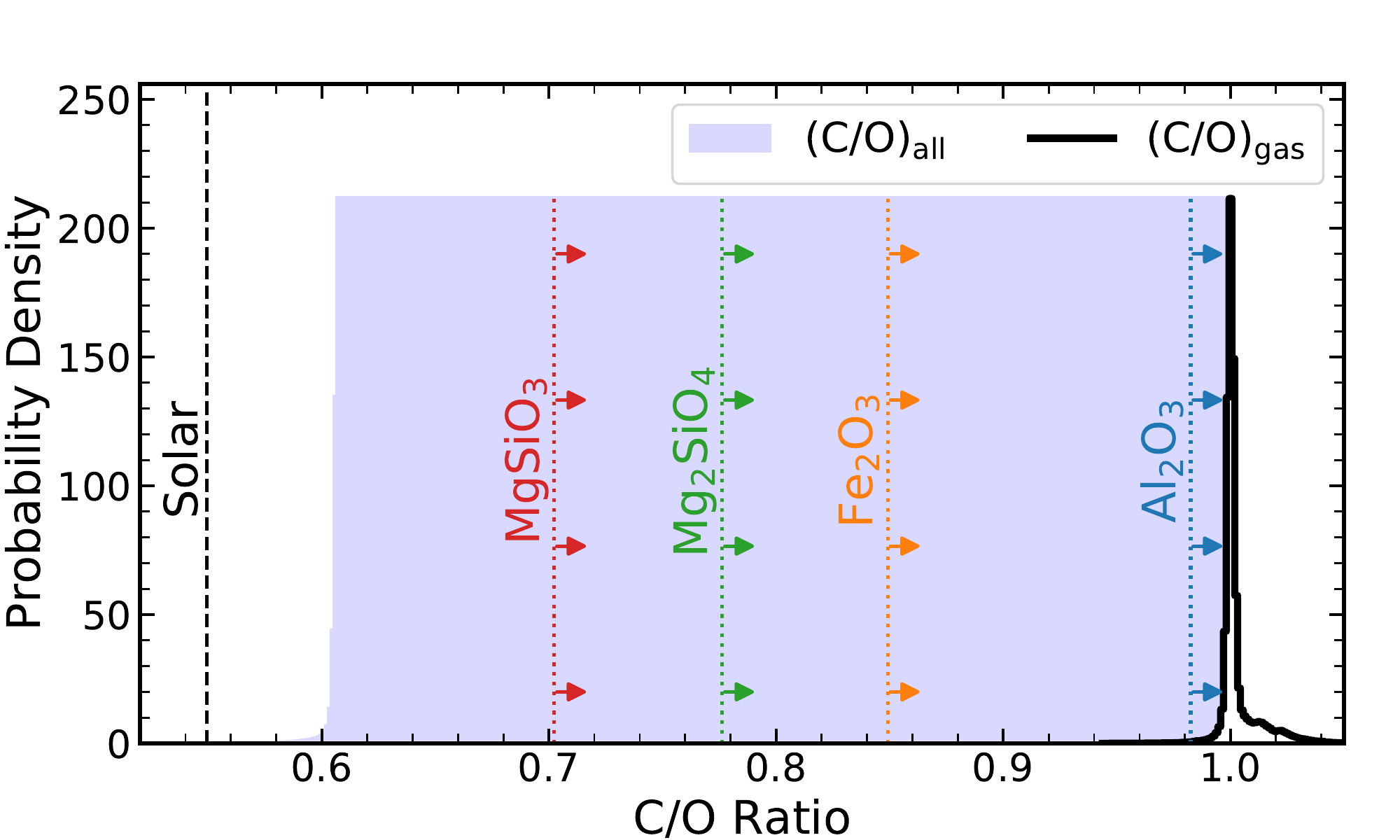}
\end{center}
\vspace{-3mm}
\caption{Probability distributions for inferred elemental abundance ratios on $\tau$~Boo~b.  
The top and middle panels respectively show the gas-phase C/H and O/H ratios on \tauboo relative to that of Jupiter~\citep{atreya_origin_2018, li_water_2020}.  The middle panel also shows the O/H ratio when taking into account oxygen that may be held up in condensates (shaded region).
The dashed lines depict the abundances expected from a protosolar-like composition~\citep{asplund_chemical_2009}.
With both the C/H and O/H principally constrained from the CO abundance, the C/O ratio (bottom panel) is tightly constrained to be $\sim$1.  
However, the C/O can be as low as $\sim$0.60 when accounting for the oxygen that may be condensed out of the gas phase in species such as Al$_2$O$_3$, Fe$_2$O$_3$, Mg$_2$SiO$_4$, and MgSiO$_3$ (dotted colored lines).
Even with 100\% efficient oxygen sequestration, the inferred C/O on \tauboo remains super-solar.
\label{fig:elemental_ratios}}
\end{figure}

With the retrieved volume mixing ratios of all major carbon- and oxygen-bearing gaseous species in \tauboos atmosphere (Section~\ref{subsec:retrieval_analysis}), we can 
calculate corresponding elemental abundance ratios.  
Relative to protosolar abundances (C/H$=2.95\times10^{-4}$, O/H$=5.37\times10^{-4}$,~\citealt[][]{asplund_chemical_2009}), we find a carbon abundance of 5.85$_{-2.82}^{+4.44}$\,$\times$\,solar, an oxygen abundance of 3.21$_{-1.56}^{+2.43}$\,$\times$\,solar, and a C/O ratio of $1.00_{-0.00}^{+0.01}$ (Figure~\ref{fig:elemental_ratios}). The inferred C/H and O/H ratios of \tauboo are super-solar and consistent with the corresponding values for Jupiter~\citep[][]{atreya_origin_2018, li_water_2020}.  With CH$_4$ as the next molecule after CO with the least strict upper limit abundance constraint, the distribution of inferred gas-phase C/O ratio values (Equation~\ref{eq:CtoO}) is slightly skewed towards carbon-rich (C/O $>$ 1) scenarios.  
However, as refractory species can condense out and preferentially remove oxygen from the gas phase, the directly measured O/H ratio is realistically only a lower limit. We must therefore account for this to infer the true atmospheric C/O ratio of \tauboos envelope.

\subsubsection{Oxygen Sequestration}\label{subsec:OxygenSequestration}
In exoplanet and brown dwarf atmospheres, certain refractory materials can condense out of the upper atmosphere gaseous phase and form clouds~\citep[e.g.,][]{fegley_chemical_1994, marley_clouds_2002, visscher_atmospheric_2010}.  
The exact species of clouds that can form depends on the refractory properties of available elements and the temperature-pressure profile.  For relatively hot atmospheres, expected prominent condensates include Al$_2$O$_3$, Fe$_2$O$_3$, Mg$_2$SiO$_4$, and MgSiO$_3$~\citep{lodders_atmospheric_2002, lodders_solar_2003,lodders_exoplanet_2010, wakeford_transmission_2015,  wakeford_high-temperature_2017}.  The formation of such oxides can then deplete the gas-phase O/H and cause the inferred C/O ratio to be overestimated.

To estimate by how much the O/H measured from gaseous species may be an underestimation of the bulk atmospheric abundance, we consider the quantity of available refractory material (e.g., Mg, Si, Fe, Al, Ca, Cr, Ti, V), assuming similar abundance ratios relative to carbon as in the solar system~(e.g., \citealt{lodders_exoplanet_2010}, see their Table 2).  From this we can calculate the total amount of oxygen that could theoretically be sequestered out of the upper atmosphere by forming oxides such as Al$_2$O$_3$, FeO, Fe$_2$O$_3$, CaAl$_{12}$O$_{19}$, CaTiO$_3$, Cr$_2$O$_3$, Mg$_2$SiO$_4$, and MgSiO$_3$ that each remove oxygen in corresponding stoichiometric proportions.  However, even assuming 100\% efficient oxygen removal from all available refractory elements, this process can only bind up to a maximum of $\sim$39\% of the total available oxygen.  Even under this most extreme scenario, the atmospheric C/O ratio of \tauboo would remain super-solar with a value of at least $\sim$0.60~(Figure~\ref{fig:elemental_ratios}).


Similarly, the measured gas-phase C/H ratio may be an underestimation of the bulk atmospheric amount.  However, carbon is generally not as efficiently removed as oxygen when forming condensates~\citep[][]{lodders_lanthanide_1993, lothringer_new_2020} and thus the measured gas-phase C/H is more likely to be well representative of the bulk atmosphere.

\section{Ruling Out Possible Causes for the low Water Measurement}\label{sec:lowwater}




As seen in Figure~\ref{fig:H2O_inj}, if there had been water on $\tau$~Boo~b this data set would have been highly sensitive to it, begging the question: does the lack of an observed water signal signify that \tauboo truly has an H$_2$O-poor atmosphere, or could it be that somehow the signal is not picked up?  
In particular, the inferred results under the presented analysis appear to be in contrast with previous work.  A multi-epoch analysis of five different nights of observations, each of approximately one hour (total of five hours), using the Keck/NIRSPEC spectrograph ($R\sim25\,000$) argued in favor of water absorption on \tauboo in the $L$-band~\citep{lockwood_near-ir_2014}.  Although this work does not directly constrain the H$_2$O abundance, it is difficult to reconcile its consistency with the SPIRou data.  A more direct comparison of these results is challenging due to the different combination of instrument, observing technique, wavelength range probed, atmospheric model, and analysis method.  Nevertheless, we note that differences between $K$- and $L$-band detections of water have previously been reported.  A CRIRES analysis of HD~189733b at 2.3\,$\mu$m by both \cite{de_kok_detection_2013} and \cite{brogi_retrieving_2019} shows no noticeable water signal (3$\sigma$ upper limit of $\sim$10$^{-4}$ in the latter case).  Meanwhile the 3.2\,$\mu$m data set shows a clear 4.8$\sigma$ H$_2$O detection~\citep{birkby_detection_2013}.  These HD~189733b results, however, are not necessarily inconsistent as the better contrast in the $L$ band may be sensitive to a modestly depleted H$_2$O abundance that the $K$-band data is not~\citep{brogi_retrieving_2019}.  In this section we explore different possible explanations for why a water signal may not have been detected with SPIRou.

\subsection{Water Line Lists}\label{subsec:water_line_lists}
One plausible explanation as to why a true underlying water signal is not seen in the SPIRou data would be if the analysis is performed using an erroneous line list.  While CO absorption produces a fewer number of deep, well defined spectral lines, the H$_2$O spectrum is much more convoluted, composed of millions of weaker lines blended together.  Variations across different line lists can cause significant biases in retrieved abundances~\citep{brogi_retrieving_2019}, or even cause strong detections 
to completely disappear 
such as in the case of HD~189733b~\citep{brogi_rotation_2016, flowers_high-resolution_2019}.  Most HDCCS water detections in the literature 
have utilized the HITEMP database based off of the BT2 line list~\citep{barber_high-accuracy_2006}.  Also having shown some success is the more recent ExoMol updated opacities based off of the POKAZATEL line list~\citep{polyansky_exomol_2018}.  
Tested on high-resolution data sets, these line lists have shown mostly consistent results in the wavelength ranges of $3.459$ -- $3.543$\,$\mu$m for HD~179949b~\citep{webb_weak_2020}, $0.92$ -- $2.45$\,$\mu$m for HD~209458b~\citep{giacobbe_five_2021}, and $2.27$ -- $2.35$\,$\mu$m as well as $3.18$ -- $3.27$\,$\mu$m for HD~179949b and HD~189733b respectively~\citep{gandhi_molecular_2020}.  We perform our analysis using both the POKAZATEL/ExoMol and BT2/HITEMP opacities, but find near identical results in both cases. 


\subsection{Effects of the Temperature Profile}\label{subsec:waterTP_profile}
Another possible explanation could be that the retrieved TP profile, here constrained from the shape and contrast of CO lines (Figure~\ref{fig:retrieval_results}), does not well-represent the water lines.  We investigate this by performing a free retrieval at the location of the planet and only including opacity contributions from water. 
Excluding other molecules allows the temperature structure to be fit entirely from H$_2$O lines, as opposed to be driven by the CO signal. 
Such a search, however, did not reveal any hint of appreciable water absorption or emission.  

Alternatively, as the core of CO and H$_2$O lines form for the most part at different pressure levels of the atmosphere, differences in low- and high-altitude winds may cause the H$_2$O signal to be blurred below a detectable level~\citep{brogi_retrieving_2019}.  Accounting for convoluted effects from contrasts in low- and high-altitudes winds would require the use of more sophisticated high-resolution 3D models~\citep[e.g.,][]{flowers_high-resolution_2019, beltz_significant_2021, harada_signatures_2021}. 
Recent work has shown that such 3D atmospheric models that take into account broadening effects from winds and rotation fare significantly better than 1D models at recovering molecular signatures in high-resolution dayside emission data sets~\citep{beltz_significant_2021}.
  


\begin{figure}[t]
\begin{center}
\includegraphics[width=\linewidth]{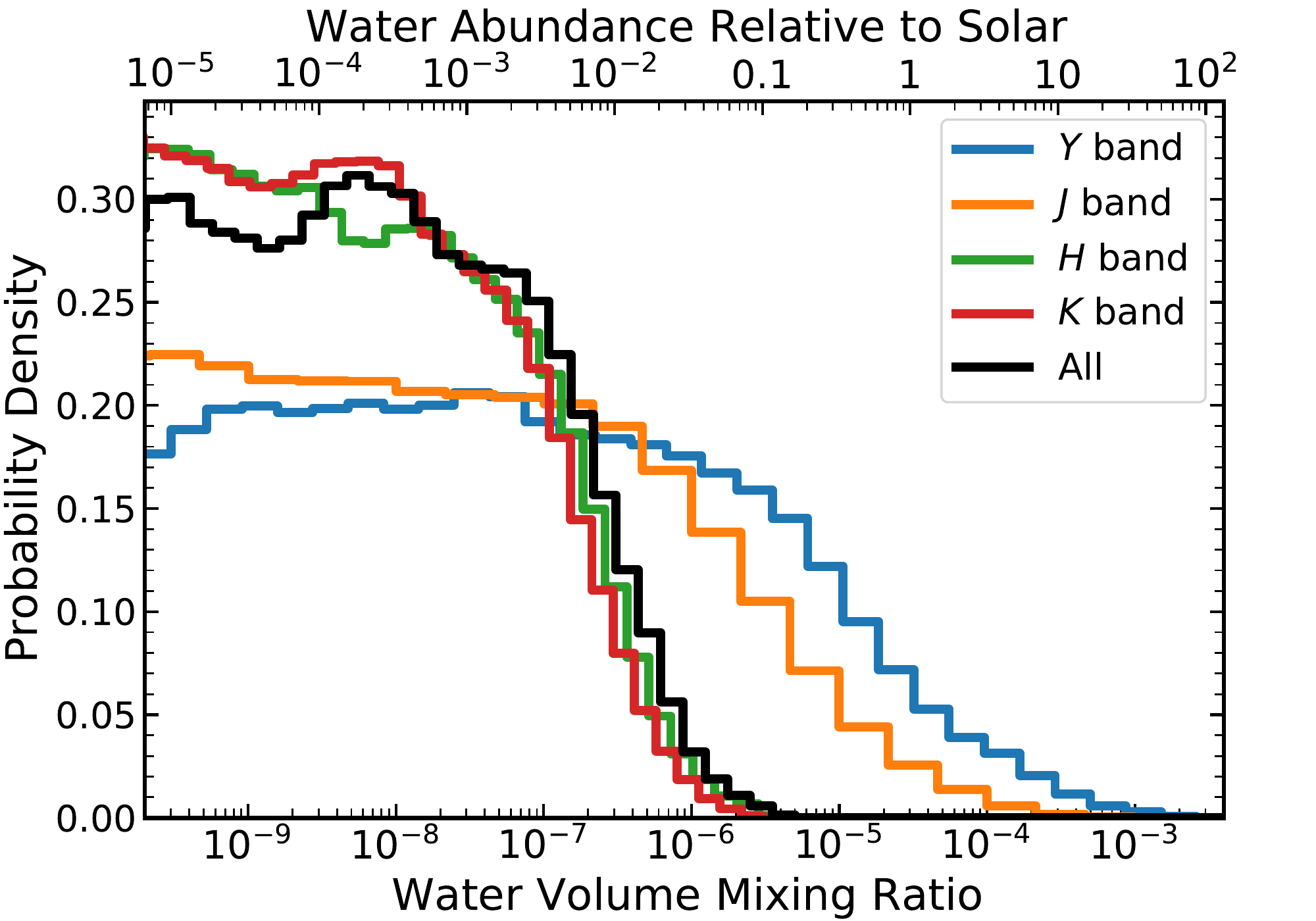}
\end{center}
\vspace{-2mm}
\caption{Retrieved posterior distribution for the water abundance on \tauboo for each photometric band.  A H$_2$O upper limit is inferred for each band individually, with the $H$ and $K$ bands providing the most stringent constraints.
\label{fig:YJHK}}
\end{figure}

\subsection{Noisy Spectral Orders}\label{subsec:water_noisy_orders}
In comparison to CO molecules that only have appreciable opacity near the 1.6 and 2.3\,$\mu$m spectral band heads, H$_2$O vapor has transition lines throughout the 
entire SPIRou wavelength range (Figure~\ref{fig:mol_cross_sections}).
However, the accuracy of the line list, the level of contamination from residual tellurics, and the presence of instrumental systematics can all vary greatly across different spectral orders.  It is therefore possible that positive correlations from water in some orders are drowned out 
by negative contributions from more problematic noisier orders.  
We investigate this by analyzing each of SPIRou's photometric bands ($Y,\, J,\, H,\, K$) individually.  For this we fix the temperature structure points at their median retrieved values and perform retrievals for each data subset at the location of the signal.  H$_2$O posterior distributions for each individual band of all five SPIRou nights combined are shown in Figure~\ref{fig:YJHK}.  No photometric band shows evidence of a water signal, with each providing an independent upper limit on the water abundance.  The $H$ and $K$ bands, where the SPIRou data has the highest signal-to-noise ratio and the planet-to-star flux contrast is more favorable, provides the most strict upper limits on the H$_2$O VMR.








\subsection{Data Reduction \& Detrending}\label{subsec:water_detrending}
It is also possible that the detrending algorithm used on the data fares well for uncovering CO lines but is not optimal for extracting a water signal. As some parts of the SPIRou bandpass are more plagued by tellurics than others, spectral orders falling in these heavily contaminated wavelength regions would potentially require a higher level of detrending or masking to uncover weaker absorption lines from H$_2$O on $\tau$~Boo~b. 
While it is common practice to optimize the number of PCA iterations or the telluric mask cutoff percentage of each order individually to maximize the retrieved significance of either the real signal, or of an injected signal, doing so can result in noise patterns being optimized into false positives~\citep{cabot_robustness_2019, zhang_platon_2020}.  This is particularly true given that SPIRou has many spectral orders, which would produce a high number of free parameters to be optimized. 
As the data already achieves a high degree of sensitivity to the water abundance under the general reduction of Section \ref{subsec:telluric_removal}, we prefer to avoid this practice and treat all orders equally to avoid introducing any biases.  

The potential partial inaccuracies of the line lists used, the neglect of 3D wind and rotation broadening effects, as well as the sub-optimal but unbiased generalized practice of treating all orders identically during the detrending process may all have some effect as to the exact value of the upper limit of the H$_2$O abundance.  However, even considering these caveats, it is clear from Figure~\ref{fig:H2O_inj} that even a modestly sub-solar abundance of water is hard to reconcile with the SPIRou data.



\subsection{Water Dissociation}\label{subsec:water_dissocition}
Finally, at high temperatures it is possible for H$_2$O molecules to thermally dissociate. This would act to reduce the strength of observed water spectral features.  Meanwhile, CO would remain mostly unaffected given its significantly stronger molecular bond~\citep{lodders_atmospheric_2002}.  
If the temperature is high enough on \tauboos dayside, it is therefore plausible that H$_2$O would be partially dissociated while CO molecules remain intact, thus naturally explaining the retrieved super-solar CO and significantly sub-solar water abundances from the SPIRou data.

Water dissociation is efficient in high temperature, low pressure environments such the upper atmospheres of ultra-hot Jupiters with strong thermal inversions.  Specifically, H$_2$O starts to break down above the $\sim100$\,mbar level at temperatures greater than about 2200\,K~\citep{parmentier_thermal_2018}.  However, \tauboo shows no sign of a strong temperature inversion and most likely does not reach high enough temperatures in its upper atmosphere for water dissociation to play a major role.  It is thus improbable that H$_2$O dissociation is prominent enough on \tauboo to explain the reported low water abundance.


\section{Implications For Planet Formation}\label{sec:implications}




Under the standard core accretion theory of formation~\citep[e.g.,][]{pollack_formation_1996} the present day envelope of a gaseous giant is the cumulation of the gas accreted during runaway growth and the metallic enrichment by planetesimals or material dredged up from the core.  A comparison of our inferred C/O on \tauboo to the expected C/O of the disk's gas and solid phases can thus provide insight as to how and where in the protoplanetary disk \tauboo may have accreted most of its atmosphere.  In addition, the measured atmospheric metallicity provides a measure of the importance of further metal enrichment either from planetesimals or from core erosion.  Then, with our measured super-solar C/H, O/H, and C/O ratios for $\tau$~Boo~b, what can we infer about how this planet formed?  Specifically, we would like to know where in the protoplanetary disk \tauboo accreted most of its envelope and whether its metallic enrichment is due to a gas- or solid-dominated accretion regime.  Obtaining such constraints would provide a valuable clue as to whether hot Jupiters commonly form in the distant disk and undergo migration, 
or rather form in situ~\citep[e.g.,][]{boley_situ_2016, batygin_situ_2016}.





\subsection{Disagreement with Formation in a Static Protoplanetary Disk}\label{subsec:StaticDiskDisagreement}

In a static protoplanetary disk, the different condensation temperatures of the main carbon- and oxygen-bearing ices introduce distinct gradients in the metallicity and C/O of the gas and solids~\citep{oberg_effects_2011}.
Primarily, the disk is comprised ($\sim$99\%) of molecular hydrogen and helium that remains mostly in a gaseous state.  The heavier elements (C, O, N, etc) making the remaining $\sim$1\%, however, can be either in gas or solid form, depending on the local temperature. Near the star it is warm enough that nearly all of the material is in vapor form part of the gaseous phase.  At larger distances from the star, where it is colder, different molecules condense out of the gas and add to the dust grains in the disk.  This causes the metallicity of the gas to drop with orbital separation as more and more metals freeze out. 
Since oxygen-rich molecules such as H$_2$O and CO$_2$ condense out at higher temperatures than CO, the C/O ratio decreases in solids and increases in the gas further out in the disk.  The result is then a protoplanetary disk with oxygen-rich, low C/O solids and metal-poor, super-stellar C/O gas beyond the water iceline~\citep[][]{madhusudhan_toward_2014}.  



Within this simple picture of a static disk model, different regimes of gas- or planetesimal-dominated accretion occurring at distinct orbital separations can be inferred from the present-day atmospheric compositions of giant planets.
In a gas-dominated accretion regime (no substantial planetesimal accretion), the planet's present-day envelope may reflect the gas phase C/O of the disk at the location where the runaway growth occurred~\citep{oberg_effects_2011}.  Such a scenario would most likely occur if the planet migrated to its current orbit 
without accreting a significant amount of solids (disk-free migration).  
Assuming an isolated core, gas-dominated accretion beyond the water ice line preferably enables the formation of low metallicity, high C/O planets.  

Alternatively, in a regime dominated by planetesimal accretion, the results will be an atmosphere that is heavily polluted by metals from captured solid.  Planetesimal accretion can occur during the envelope contraction phase of runaway growth, during the clearing of the feeding zone, and while the planet migrates to becoming a hot Jupiter while interacting with the still-present protoplanetary disk (Type II migration).  Solid-dominated accretion typically results in planets that are metal-rich but have low C/O ratios~\citep{madhusudhan_toward_2014, madhusudhan_atmospheric_2017, cridland_composition_2016, cridland_connecting_2019, cridland_connecting_2020, mordasini_imprint_2016, espinoza_metal_2017, ali-dib_disentangling_2017, nowak_peering_2020}.
Core erosion can also contribute to increase the metallic content of the envelope with stellar or sub-stellar C/O material, depending on where in the disk the core formed~\citep[e.g.,][]{oberg_jupitertextquotesingles_2019}.  However, by building a planet from metal-poor, super-stellar C/O gas with metal enrichment from sub-stellar to stellar C/O solids either from the core of from planetesimals, there is no outcome that allows for the formation of high metallicity, high C/O planets like \tauboo regardless of formation location and gas/solid accretion fractions~\citep{madhusudhan_toward_2014, madhusudhan_atmospheric_2017}. 


\subsection{Support for Pebble Drift and Gas-Dominated Accretion}\label{subsec:PebbleDrift}

One possible mechanism that naturally explains a carbon-rich, high C/O gas giant envelope is by enrichment of the gas in the disk due to pebble drift across condensation fronts~\citep{oberg_excess_2016, booth_chemical_2017}.  Indeed observations show that protoplanetary disks contain large reservoirs of mm- to cm-sized pebbles~\citep[e.g.,][]{rodmann_large_2006}.  Such grains are large enough to decouple from the gas and feel a headwind that causes them to drift radially inwards.  The migration of icy grains to the warmer inner regions of the disk can then cause the transported frozen volatiles to evaporate back to the gas phase and enhance the local chemical abundance near condensation front~\citep{ali-dib_carbon-rich_2014}.  For example, pebbles drifting inward from the outer disk 
will begin to sublimate once they cross the CO snowline which will cause a local metal enrichment of the gas phase.  The deposited excess CO molecules can then diffuse and advect throughout the disk, increasing the metallicity of the gas phase inward of the CO snowline while maintaining a C/O $\sim$ 1~\citep{oberg_excess_2016}. 
Evidence of pebble drift in the form of excess C/H within the CO snowline was recently observed in the HD\,163296 disk~\citep{zhang_excess_2020}.
Gas-dominated accretion in a pebble drift induced CO-enriched environment provides a natural mechanism to form metal-rich, high C/O planets of the likes of $\tau$~Boo~b, without the need for substantial enrichment from planetesimal accretion or core dredging. In the scenario where the C/O ratio is overestimated (but still super-solar) due to significant oxygen condensed out of the gas phase, pebble drift past the CO or CO$_2$ snowlines would still be favored, although a more significant enrichment from low C/O, high O/H planetesimals may have occurred.











\begin{figure}[t]
\begin{center}
\includegraphics[width=\linewidth]{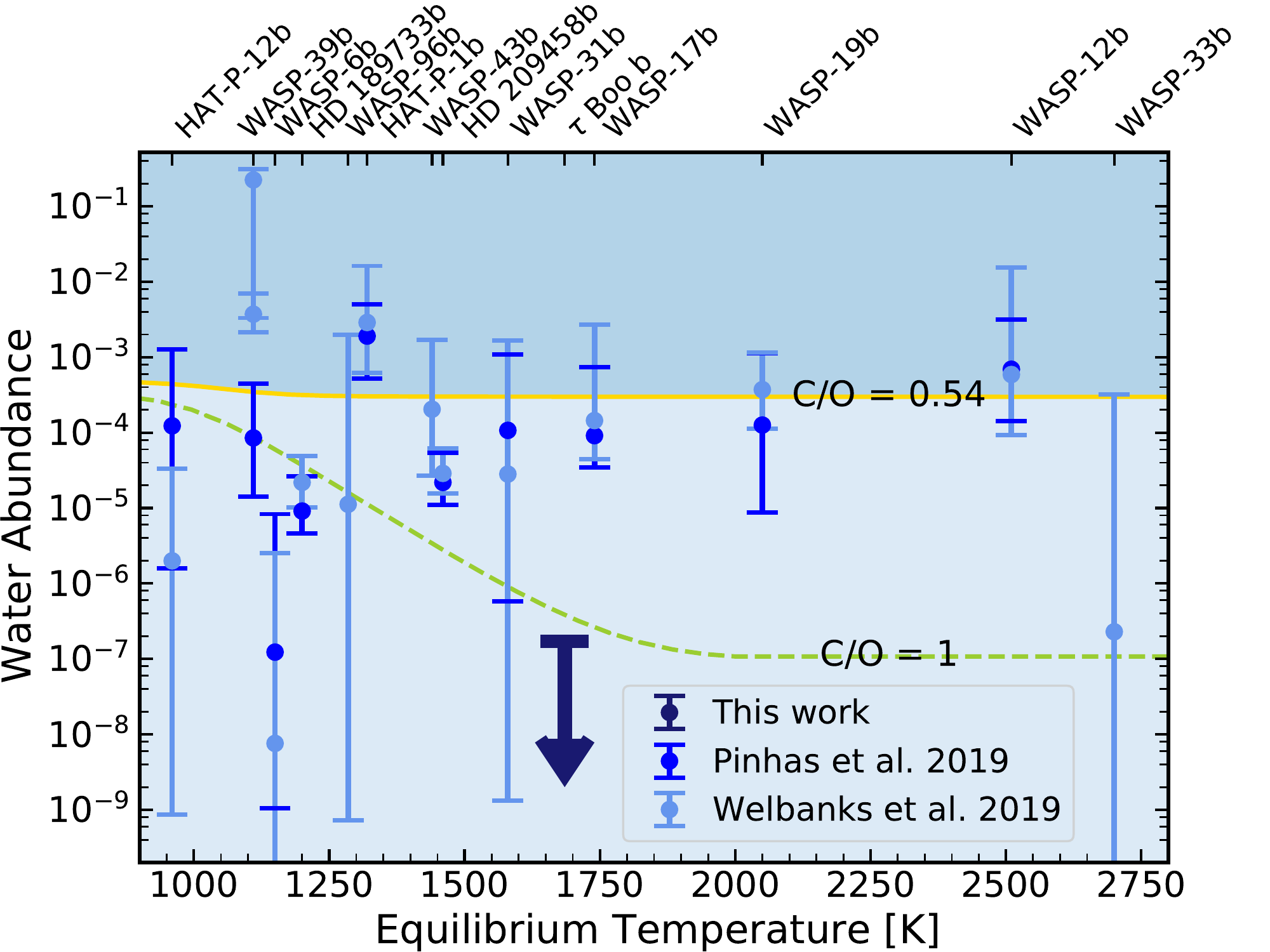}
\end{center}
\vspace{-2mm}
\caption{
Comparison of \tauboos retrieved H$_2$O (1$\sigma$) upper limit to water abundances of a sample of hot Jupiters determined from transit spectroscopy with \textit{HST}.
The solid gold and dashed green lines shows the H$_2$O abundances that would be expected under chemical equilibrium assuming a solar C/H ratio and a C/O ratio of either 0.54 or 1, respectively.  Light and dark shaded areas correspond to sub- and super-solar water volume mixing ratios for a solar C/O ratio.  Most reported abundances other than HAT-P-1b and some cases of WASP-39b are consistent with a sub-solar composition to within 1$\sigma$.  The low H$_2$O abundance on \tauboo is likely evidence for a super-solar C/O ratio rather than a low O/H ratio.
\label{fig:HJ_water_abun}}
\end{figure}

We note that planet formation is likely to be much more complex than the simple picture depicted here and a number of processes can act to alter our conclusions.  For one, protoplanetary disks can undergo significant evolution, changing the chemical composition and position of snowlines over time~\citep{eistrup_setting_2016, eistrup_molecular_2018, cridland_composition_2016, cridland_connecting_2019, cridland_connecting_2020}.  It is also possible that the contribution of accreted planetesimals to the net envelope metallic enrichment is overestimated if captured planetesimals reach the core instead of being dissolved and mixed into the atmosphere~\citep{pinhas_efficiency_2016}.  Similarly, the degree to which the core interacts and exchanges material with the envelope is uncertain.  The assumed oxygen-rich, low C/O composition of the solids past the water ice line may also be erroneous if more carbon than expected is held in refractory organic material~\citep[e.g.,][]{lodders_jupiter_2004}.  All interpretations also rest on the fundamental assumption that the probed atmospheric abundances in hot Jupiters are representative of the bulk composition of the planet, which may not necessarily be the case~\citep{leconte_new_2012, vazan_jupiters_2018}. Finally, \tauboo may have formed in a truncated disk due to the system's highly eccentric wide-orbit companion, which would likely further complicate this simple picture~\citep[][]{justesen_constraining_2019}.

Another possible pathway to form giant planets is via gravitational instability~\citep[e.g.,][]{boss_giant_1997, boss_gas_2001}.  This is a relatively short timescale process that results in an envelope initially representative of the local bulk composition of the disk.  With subsequent enrichment from solids, planets formed via gravitational instability are expected to have stellar or super-stellar metallicities and C/O ratios that are stellar or sub-stellar~\citep{madhusudhan_toward_2014}.  It is unclear how a gravitational instability formation scenario could produce a $\tau$~Boo~b-like high C/H, high C/O planet.



\subsection{\texorpdfstring{$\tau$}{}~Boo~b in Context}\label{subsec:context}

\begin{figure}[t]
\begin{center}
\includegraphics[width=\linewidth]{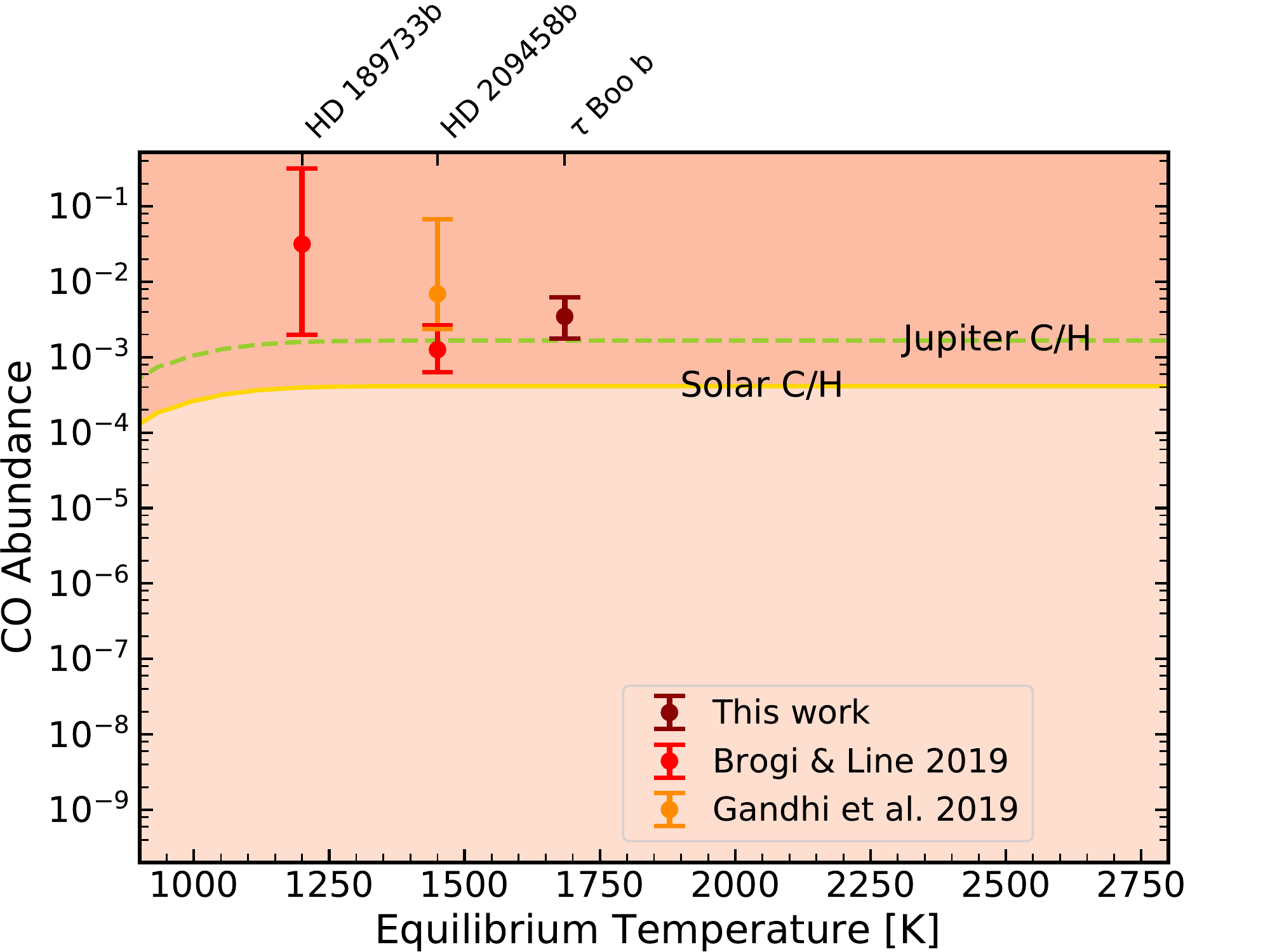}
\end{center}
\vspace{-2mm}
\caption{
Similar to Figure~\ref{fig:HJ_water_abun}, but for CO abundances reported on hot Jupiters from high-resolution spectroscopy.  Although measurements are sparse, carbon monoxide volume mixing ratios favor slightly super-solar C/H values, similarly to Jupiter which has a C/H ratio of roughly 4$\times$ protosolar~\citep{atreya_origin_2018}.  In combination with the overall tendency for low H$_2$O abundances, this trend points towards elevated C/O ratios being common in hot Jupiter atmospheres.
\label{fig:HJ_co_abun}}
\end{figure}

So how does \tauboo and its Jupiter-like C/H and elevated C/O fit in relative to other hot Jupiters?  While the lack of water observed in the SPIRou data is unexpected, sub-solar H$_2$O abundances on exoplanets are not uncommon.  
Systematic analyses of a sample of hot Jupiters observed in transit spectroscopy have shown a favorability for water volume mixing ratios that are depleted relative to expectations from solar~\citep{barstow_consistent_2017, pinhas_h2o_2019, welbanks_massmetallicity_2019}.  A comparison of the retrieved H$_2$O VMR 
upper limit on \tauboo to water measurements on hot Jupiters obtained from transmission spectroscopy with \textit{HST}/WFC3  
is shown in Figure~\ref{fig:HJ_water_abun}.  While most reported H$_2$O abundances are consistent with being sub-solar, the stringent upper limit derived from the SPIRou data is lower than any of these.  However, the low water abundance is not necessarily evidence of a low O/H if instead of in H$_2$O, the oxygen budget of \tauboo is held in other molecules such as CO.  This is why constraining the abundance of multiple species is crucial for understanding the chemical budget of exoplanetary atmospheres.  

With a measure of its CO abundance, \tauboo joins HD\,\,189733b and HD\,\,209458b in having precise constraints on both H$_2$O and CO obtained from high-resolution spectroscopy observations~\citep{brogi_retrieving_2019, gandhi_hydra-h_2019}.  With only these very few robust and precise measurements that presently exist, current trends of CO abundances on hot Jupiters as determined from high-resolution spectroscopy favor slightly enhanced Jupiter-like C/H ratios (Figure~\ref{fig:HJ_co_abun}).  As the respective main carriers of O and C under equilibrium chemistry, this allows for the C/O ratios of these planets to be estimated.  In all three cases, the CO abundances is significantly higher than the H$_2$O and the CO/(H$_2$O+CO) ratio, a proxy for the C/O in the absence of condensates or other carbon- and oxygen-bearing species, can constrained to be $\sim$1.  In the case of $\tau$~Boo~b, gas-phase C/O ratios significantly deviating from unity are disfavored due to the upper limit constraints derived on other potential carbon- or oxygen-bearing molecules CH$_4$, CO$_2$, HCN, C$_2$H$_2$, or TiO.

\section{Conclusion}\label{sec:conclusion}

We analyzed a large SPIRou data set totalling 20 hours targeting the dayside emission of $\tau$~Boo~b across five different nights.  
We find a near-solar but slightly enhanced CO abundance of log(CO) = $-2.62_{-0.29}^{+0.25}$, consistent with a C/H ratio similar to that of Jupiter, as well as a strong water depletion (3$\sigma$ upper limit of $0.0072\, \times$\,solar).
Our Bayesian atmospheric retrieval analysis on the full data set also favors a non-inverted temperature structure in an atmosphere with no optically thick cloud deck above the 250\,mbar -- 1\,bar pressure level.  
No statistically significant evidence for the presence of  CH$_4$, CO$_2$, HCN, TiO, C$_2$H$_2$, or NH$_3$ in \tauboos atmosphere was found, with upper limits on the abundance of each of these derived.  The retrieved Keplerian velocity of $K_p=109.2\pm0.4$\,km\,s$^{-1}$ is consistent with previous work and corresponds to a planetary mass of $6.24\pm0.23\,M_{\mathrm{Jup}}$.  
We recover \tauboos scaling parameter $a = 1.04\pm0.03$ at a precision that is encouraging for a non-transiting planet. 
The ability to precisely constrain molecular abundances from dayside emission observations without an \textit{a priori} radius measurement further opens up the possibility for comparative exo-planetology of bright non-transiting systems to more well known transiting hot Jupiters. 

The depleted water content on \tauboo is more severe, but fits in a trend of sub-solar H$_2$O abundances reported on a sample of hot Jupiters observed in transmission.  When combined with the super-solar CO abundance, the lack of water can be interpreted as the results of an elevated C/O ratio as opposed to a low O/H ratio.  The super-solar C/H, O/H, and C/O ratios found on \tauboo are difficult to explain under the standard core accretion model of planet formation in a static disk.  Instead, a natural formation scenario for a metal-rich, high C/O planet is via gas-dominated accretion of pebble drift induced enriched gas, followed by limited enrichment from oxygen-rich material.  Such a formation history is most readily explained if \tauboo accreted most of its envelope in the outer disk past the H$_2$O snowline and subsequently underwent a disk-free migration process involving limited capturing of planetesimals and a mostly isolated core.  




\acknowledgments
Based on observations obtained at the Canada-France-Hawaii Telescope (CFHT) which is operated from the summit of Maunakea by the National Research Council of Canada, the Institut National des Sciences de l'Univers of the Centre National de la Recherche Scientifique of France, and the University of Hawaii. The observations at the Canada-France-Hawaii Telescope were performed with care and respect from the summit of Maunakea which is a significant cultural and historic site.  S.P., A.D.-B. and C.P. acknowledge funding from the Technologies for Exo-Planetary Science (TEPS) CREATE program.  C.P. and B.B. acknowledges funding from the Fonds de Recherche Qu\'eb\'ecois-Nature et Technologie (FRQNT; Qu\'ebec). B.B. also recognizes financial support from the Natural Sciences and Engineering Research Council (NSERC) of Canada. R.A. acknowledges support from the Trottier Family Foundation.  X.D. acknowledges funding from the French National Research Agency (ANR) under contract number ANR-18-CE31-0019 (SPlaSH) and also in the framework of the Investissements d'Avenir program (ANR-15-IDEX-02), through the funding of the ``Origin of Life'' project of the Univ. Grenoble-Alpes.  Based on observations obtained with SPIRou, an international project led by Institut de Recherche en Astrophysique et Plan\'etologie, Toulouse, France.  The SPIRou project is funded by the IDEX initiative at UFTMP, UPS, the DIM-ACAV program in R\'egion \^{I}le-de-France, the MIDEX initiative at AMU, the Labex@OSUG2020 program, UGA, INSU/CNRS, CFI, CFHT, LNA, CAUP, and DIAS. We are also grateful for generous amounts of in-kind human effort allocated to SPIRou by OMP/IRAP, OHP/LAM, IPAG, CFHT, NRC-H, UdeM, UL, OG, LNA, and ASIAA.  We wish to extend our gratitude to the CFHT QSO team for executing the observations.  We thank Greg Barrick and Tom Vermeulen at CFHT for their help both observing with the Atmospheric Dispersion Corrector held fixed and investigating the source of observed systematics in the SPIRou data. We are also thankful to Matteo Brogi for thoughtful and constructive feedback that helped us improve the quality of this manuscript.  

%

\vspace{5mm}
\facility{CFHT (SPIRou) }


\software{\texttt{Astropy}~\citep{robitaille_astropy_2013, price-whelan_astropy_2018}, \texttt{NumPy}~\citep{harris_array_2020}, \texttt{SciPy}~\citep{virtanen_scipy_2020}, \texttt{Matplotlib}~\citep{hunter_matplotlib_2007}, \texttt{emcee}~\citep{foreman-mackey_emcee_2013}, \texttt{corner}~\citep{foreman-mackey_cornerpy_2016}}






\appendix

\section{Telescope Angle Systematics}\label{sec:appendix_tel_angle}
Strong systematics correlated with telescope position are seen in some of the time series observations taken with the SPIRou spectrograph.  These can introduce percent-level variations visible in the raw data that compromise the quality of the data obtained and should be taken into consideration when planning observations. 

\begin{figure}[t!]
\begin{center}
\includegraphics[width=\linewidth]{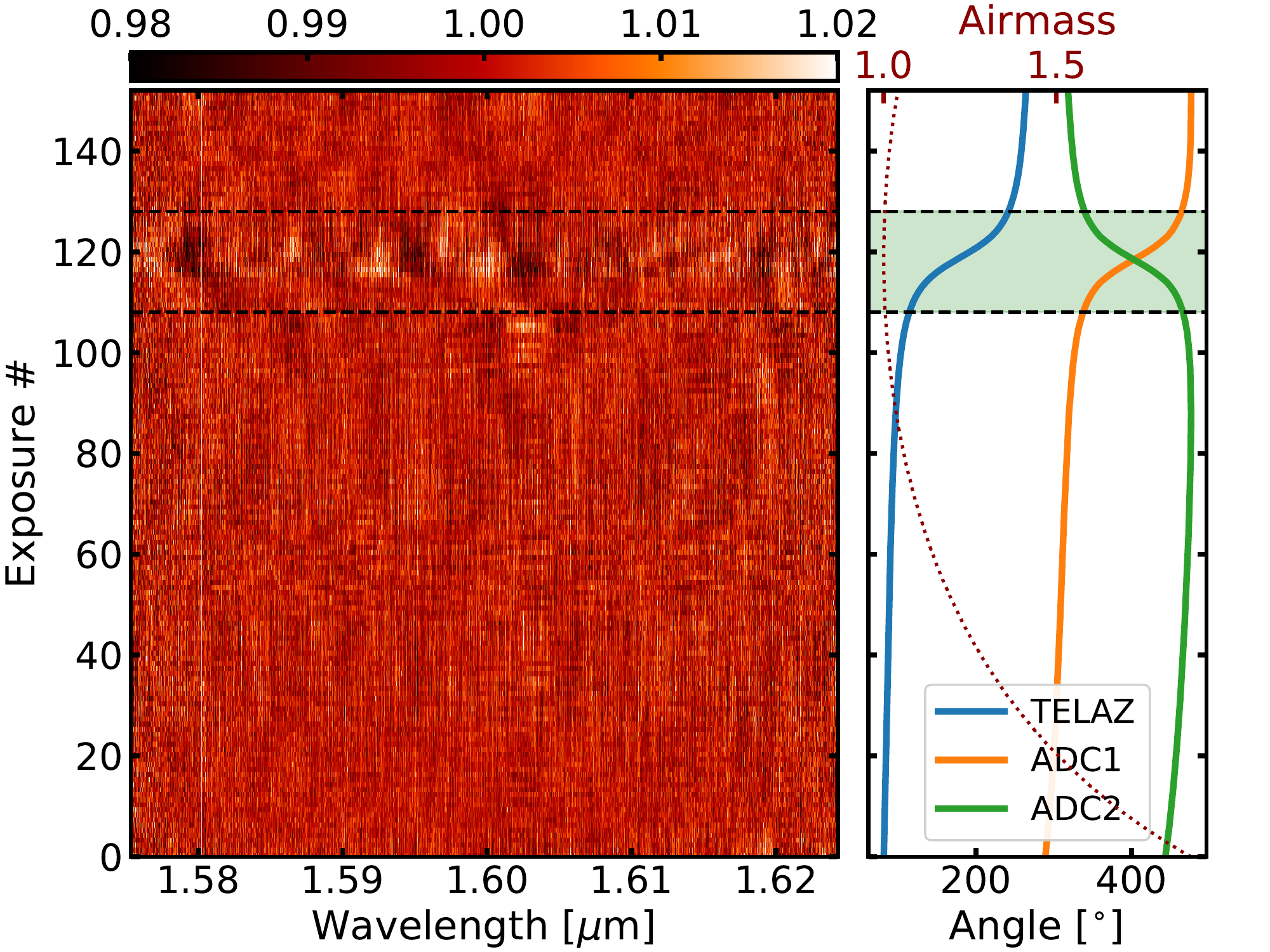}
\end{center}
\vspace{-2mm}
\caption{Residuals of the observation sequence taken on 2019 April 21.  The left panel shows one SPIRou spectral order after detrending.  The telescope azimuth and atmospheric dispersion corrector angles as well as the airmass are shown in the right panel.  Strong systematics can clearly be seen in the data near exposure 120, coinciding to where the target crosses the zenith (between the dashed lines, at TELAZ = $180\pm65^{\circ}$ and airmass $<$ 1.005) and where the telescope and ADC angles change most rapidly.  The mean nightly seeing is 0.57.
\label{fig:tel_angles_apr21}}
\end{figure}

\begin{figure}[ht]
\begin{center}
\includegraphics[width=\linewidth]{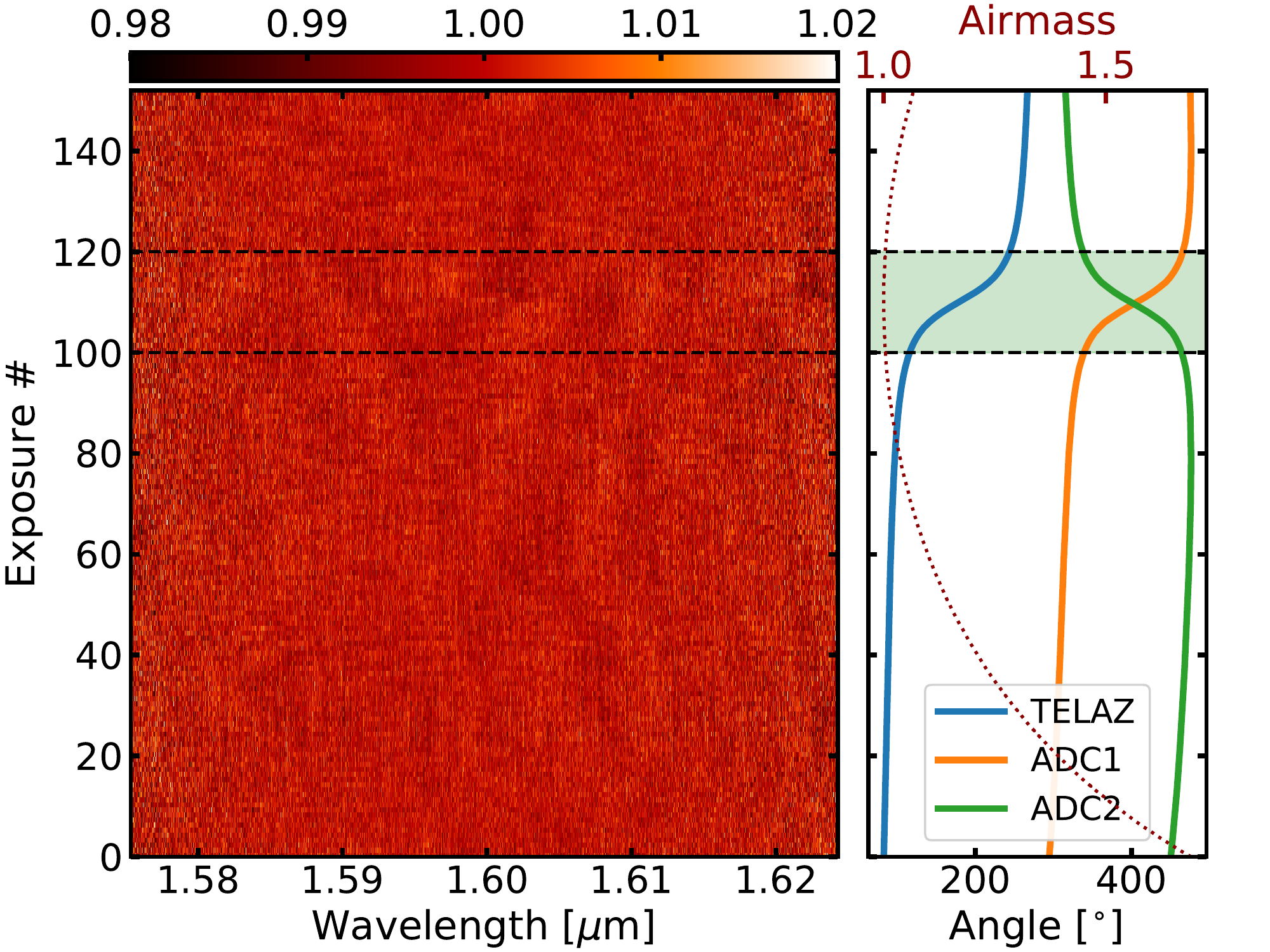}
\end{center}
\vspace{-2mm}
\caption{Same as Figure~\ref{fig:tel_angles_apr21} but for the night of 2019 April 15.  Non-white noise patterns are still present when the target is near the zenith (shaded region), but much less pronounced than in the night of 2019 April 21.  The mean nightly seeing is 1.02.  
\label{fig:tel_angles_apr15}}
\end{figure}

From the data analysis and injection/recovery tests, 
it is clear that not every observation taken of \tauboo contributes proportionally to the overall signal.  Moreover, the data quality of different observations cannot be explained only by differences in sky conditions (Appendix~\ref{sec:appendix_night_sensitivity}).  A closer look at the data after telluric removal (as described in Sections~\ref{subsec:data_reduction} \& \ref{subsec:telluric_removal}, but using a wider filter for the continuum alignment to better visualize systematics) shows that the final output of the reduction for some of the observation sequences have clear non-white noise components (e.g., Figure~\ref{fig:tel_angles_apr21}).  While the exact underlying cause for this effect is unknown, the red noise tends to be localized to exposures taken when the target is near the zenith (airmass $\sim$\,1).  This corresponds to when the telescope azimuth (TELAZ) angle as well as the Atmospheric Dispersion Corrector (ADC) angles change the most rapidly.  Both the 2019 April 15 and 21 nights show non-white noise patterns at regions of greatest angle change rates, although the effect is much more pronounced in the latter case (Figures~\ref{fig:tel_angles_apr15} and \ref{fig:tel_angles_apr21} respectively).

\begin{figure}[ht]
\begin{center}
\includegraphics[width=\linewidth]{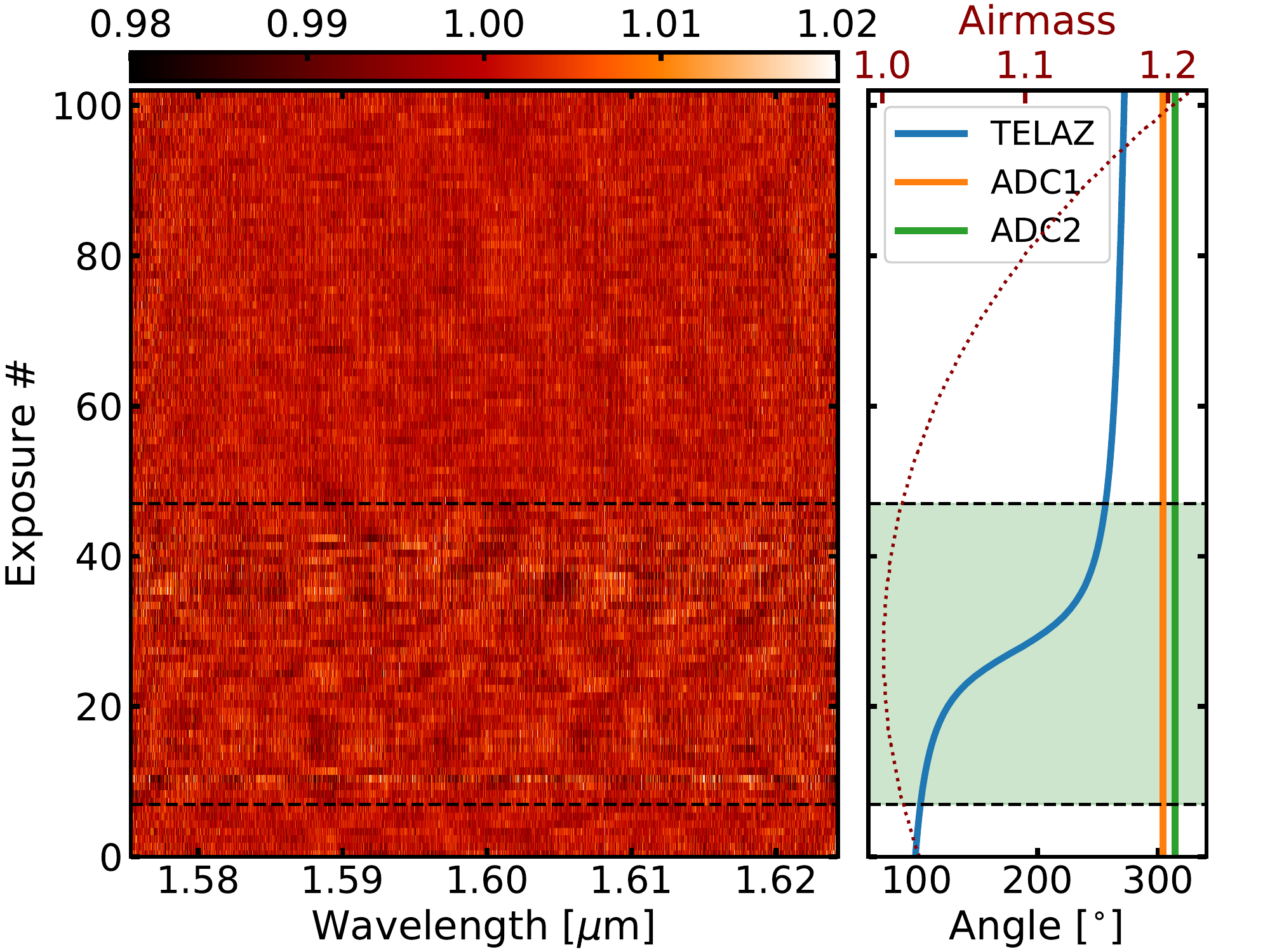}
\end{center}
\vspace{-2mm}
\caption{Same as Figures~\ref{fig:tel_angles_apr21} \& \ref{fig:tel_angles_apr15} but for the night of 2020 May 10.  This observation sequence was taken with the ADC angles held fixed throughout the time series.  Despite this, increased noise can still be seen in the region of greatest telescope azimuth angle change. 
It is thus unlikely that the ADC is the source of the problem.  The mean nightly seeing is 0.65.    
\label{fig:tel_angles_may10}}
\end{figure}

Investigating the possibility that these systematics are introduced by the rotating ADC, we scheduled the observation sequence of 2020 May 10 to be taken while keeping these fixed at their zenith positions.  However, even with the ADC angles held constant, the reduced time series still shows compromised data quality in the region where the azimuth angle changes most rapidly (Figure~\ref{fig:tel_angles_may10}).

It is unclear why the observations on April 21 and May 10 are affected more drastically by this effect than on April 15, although it is possible that this is correlated to the different point spread function (PSF) widths.  The more affected nights of April 21 and May 10 have `better' mean seeings (0.57 and 0.65 respectively) while the less affected `worse' night has an average seeing of 1.02.  It is possible that the wider (more blurred) PSF in the latter case acts to dampen this effect.  In contrast, a narrower PSF may be more susceptible to slight changes in how the light enters or travels through the fiber.  This effect is observed in both the combined as well as the individual science channels.

\begin{table*}[t]
\centering
\caption{\label{tab:night_info} Properties \& Sensitivity of the SPIRou Nights}
\vspace{-3mm}
\renewcommand{\arraystretch}{1.1}
\begin{tabular}{c|cccccccc}
\hline
\hline
  Night   &       Time (h)\textsuperscript{(a)}   &   S/N\textsuperscript{(b)}  & Mean Seeing (") &  Humidity\textsuperscript{(c)} & Zenith\textsuperscript{(d)} & Systemics\textsuperscript{(e)} & ADC\textsuperscript{(f)} & Sensitivity\textsuperscript{(g)}  \\
\hline
2019 Apr 15  &  5    &   239  &  1.02  & 0.60 &  Yes  &  Mild    &  Moving  &  9.1\\
2019 Apr 21  &  5    &   278  &  0.57 &  0.93 &  Yes  &  Severe  &  Moving  &  7.0\\
2020 May 10  &  3.3  &   306  &  0.65 &  0.82  & Yes  &  Severe  &  Fixed   &  4.1\\
2020 Jun 02  &  3.3  &   271  &  0.72 &  4.74  &  No   &  None    &  Moving  &  5.7\\
2020 Jun 05  &  3.3  &   238  &  1.25 &  0.84  &  No   &  None    &  Moving  &  7.5\\
\hline
\hline
\end{tabular}
\begin{tablenotes}
\small
\item{\textsuperscript{(a)} Total integration time.  \textsuperscript{(b)} Average signal-to-noise per detecctor pixel.
\textsuperscript{(c)} Mean water column density in the line-of-sight measured relative to an average Maunakea night as calculated by \texttt{APERO}.
\textsuperscript{(d)} Whether the observing sequence crosses the zenith.
\textsuperscript{(e)} Whether the observing sequence has any systematics in the data and if so, how severe.
\textsuperscript{(f)} Status of the ADC during the time series.
\textsuperscript{(g)} Average S/N at which an injected planetary signal is recovered from each data set.}
\end{tablenotes}
\end{table*}

While applying a high-pass filter and principal component analysis can help in dealing with the induced systematics, the effect is not fully removed and the data quality remains sub-par.  To avoid this issue entirely, the nights of June 2 and June 5 were scheduled at sub-optimal elevations so that \tauboo was not observed near the zenith when it would go through an airmass close to one and the telescope azimuth angle would change most rapidly.  No systematics were observed in either of those observations in which the TELAZ angle varies only very slowly.  We note that this same type of systematics has been observed in other SPIRou time series and the localized effects of the introduced noise patterns always coincides with the region of lowest airmass when the most rapid change in telescope angle occurs.   




\section{Night Sensitivity}\label{sec:appendix_night_sensitivity}

How sensitive a given observation sequence is to an underlying planetary signal depends on the time series' total integration time and S/N, the airmass and sky conditions under which the observations were taken, as well as whether there are any instrumental systematics compromising the quality of the data set.  As such, it is not necessarily expected that all obtained observations, when combined, will contribute proportionally to the total signal.   

To assess the `data quality' of each night, we perform injection-recovery tests, injecting a planet signal at 100 orbital velocities away from the known value and recording the significance at which the signal can be recovered.  This acts as an effective measure of how sensitive a given observation sequence is expected to be to an underlying real signal.   
The average recovered S/N (sensitivity) of these injected signals as well as the properties of each night is shown in Table~\ref{tab:night_info}.  Even for identical total integration times, large variations in how sensitive different nights are exist.
Most of these differences can be readily explained.  
The observation sequences of 2019 April 15 and 2020 May 10 are both plagued by severe systematics that severely degrade the data quality (Figures \ref{fig:tel_angles_apr21} and \ref{fig:tel_angles_may10}, respectively) and their sensitivity is accordingly lower (Table~\ref{tab:night_info}).  
These systematics, while still present, only appear mildly in the night of 2019 April 15 and the data set is significantly more sensitive to an injected signal as a result.   
The observations taken on 2020 June 2 are less sensitive than those on 2020 June 5 mainly due to the significantly higher mean water column density. 
Given its higher sensitivity, the night of 2019 April 15 is expected to provide the strongest detection, meanwhile little to no signal is expected to be found in the observations of 2020 May 10.  
The reduction and analysis procedure for how these S/N sensitivity tests are performed is outlined in Sections \ref{subsec:data_reduction}, \ref{subsec:telluric_removal} and \ref{subsec:CrossCorrelation} in the main text.

\bibliography{references}{}

\begin{thebibliography}{}
\providecommand\natexlab[1]{#1}
\providecommand\JournalTitle[1]{#1}

\bibitem[{Ali-Dib(2017)}]{ali-dib_disentangling_2017}
Ali-Dib, M. 2017,
  \href{http://dx.doi.org/10.1093/mnras/stx260}{\JournalTitle{Monthly Notices
  of the Royal Astronomical Society}, 467, 2845}

\bibitem[{Ali-Dib {et~al.}(2014)Ali-Dib, Mousis, Petit, \&
  Lunine}]{ali-dib_carbon-rich_2014}
Ali-Dib, M., Mousis, O., Petit, J.-M., \& Lunine, J.~I. 2014,
  \href{http://dx.doi.org/10.1088/0004-637X/785/2/125}{\JournalTitle{The
  Astrophysical Journal}, 785, 125}

\bibitem[{Alonso-Floriano {et~al.}(2019)Alonso-Floriano, Sánchez-López,
  Snellen, López-Puertas, Nagel, Amado, Bauer, Caballero, Czesla, Nortmann,
  Pallé, Salz, Reiners, Ribas, Quirrenbach, Aceituno, Anglada-Escudé, Béjar,
  Guenther, Henning, Kaminski, Kürster, Lampón, Lara, Montes, Morales,
  Tal-Or, Schmitt, Osorio, \& Zechmeister}]{alonso-floriano_multiple_2019}
Alonso-Floriano, F.~J., Sánchez-López, A., Snellen, I. a.~G., {et~al.} 2019,
  \href{http://dx.doi.org/10.1051/0004-6361/201834339}{\JournalTitle{Astronomy
  \& Astrophysics}, 621, A74}

\bibitem[{Arcangeli {et~al.}(2018)Arcangeli, Désert, Line, Bean, Parmentier,
  Stevenson, Kreidberg, Fortney, Mansfield, \& Showman}]{arcangeli_h_2018}
Arcangeli, J., Désert, J.-M., Line, M.~R., {et~al.} 2018,
  \href{http://dx.doi.org/10.3847/2041-8213/aab272}{\JournalTitle{The
  Astrophysical Journal Letters}, 855, L30}

\bibitem[{Artigau {et~al.}(2014{\natexlab{a}})Artigau, Kouach, Donati, Doyon,
  Delfosse, Baratchart, Lacombe, Moutou, Rabou, Parès, Micheau, Thibault,
  Reshetov, Dubois, Hernandez, Vallée, Wang, Dolon, Pepe, Bouchy, Striebig,
  Hénault, Loop, Saddlemyer, Barrick, Vermeulen, Dupieux, Hébrard, Boisse,
  Martioli, Alencar, Nascimento, \& Figueira}]{artigau_spirou_2014}
Artigau, E., Kouach, D., Donati, J.-F., {et~al.} 2014{\natexlab{a}},
  \href{http://dx.doi.org/10.1117/12.2055663}{in Ground-based and {Airborne}
  {Instrumentation} for {Astronomy} {V}, Vol. 9147} (International Society for
  Optics and Photonics), 914715

\bibitem[{Artigau {et~al.}(2014{\natexlab{b}})Artigau, Astudillo-Defru,
  Delfosse, Bouchy, Bonfils, Lovis, Pepe, Moutou, Donati, Doyon, \&
  Malo}]{artigau_telluric-line_2014}
Artigau, E., Astudillo-Defru, N., Delfosse, X., {et~al.} 2014{\natexlab{b}},
  \href{http://dx.doi.org/10.1117/12.2056385}{in Observatory {Operations}:
  {Strategies}, {Processes}, and {Systems} {V}, Vol. 9149} (International
  Society for Optics and Photonics), 914905

\bibitem[{Asplund {et~al.}(2009)Asplund, Grevesse, Sauval, \&
  Scott}]{asplund_chemical_2009}
Asplund, M., Grevesse, N., Sauval, A.~J., \& Scott, P. 2009,
  \href{http://dx.doi.org/10.1146/annurev.astro.46.060407.145222}{\JournalTitle{Annual
  Review of Astronomy and Astrophysics}, 47, 481}, publisher: Annual Reviews

\bibitem[{Atreya {et~al.}(2018)Atreya, Crida, Guillot, Lunine, Madhusudhan, \&
  Mousis}]{atreya_origin_2018}
Atreya, S.~K., Crida, A., Guillot, T., {et~al.} 2018,
  \href{http://dx.doi.org/10.1017/9781316227220.002}{in Saturn in the 21st
  {Century}, Cambridge {Planetary} {Science}} (Cambridge: Cambridge University
  Press), 5

\bibitem[{Atreya \& Wong(2005)}]{atreya_coupled_2005}
Atreya, S.~K. \& Wong, A.-S. 2005,
  \href{http://dx.doi.org/10.1007/s11214-005-1951-5}{\JournalTitle{Space
  Science Reviews}, 116, 121}

\bibitem[{Baliunas {et~al.}(1997)Baliunas, Henry, Donahue, Fekel, \&
  Soon}]{baliunas_properties_1997}
Baliunas, S.~L., Henry, G.~W., Donahue, R.~A., Fekel, F.~C., \& Soon, W.~H.
  1997, \href{http://dx.doi.org/10.1086/310442}{\JournalTitle{The Astrophysical
  Journal Letters}, 474, L119}, publisher: IOP Publishing

\bibitem[{Barber {et~al.}(2014)Barber, Strange, Hill, Polyansky, Mellau,
  Yurchenko, \& Tennyson}]{barber_exomol_2014}
Barber, R.~J., Strange, J.~K., Hill, C., {et~al.} 2014,
  \href{http://dx.doi.org/10.1093/mnras/stt2011}{\JournalTitle{Monthly Notices
  of the Royal Astronomical Society}, 437, 1828}

\bibitem[{Barber {et~al.}(2006)Barber, Tennyson, Harris, \&
  Tolchenov}]{barber_high-accuracy_2006}
Barber, R.~J., Tennyson, J., Harris, G.~J., \& Tolchenov, R.~N. 2006,
  \href{http://dx.doi.org/10.1111/j.1365-2966.2006.10184.x}{\JournalTitle{Monthly
  Notices of the Royal Astronomical Society}, 368, 1087}, publisher: Oxford
  Academic

\bibitem[{Barman {et~al.}(2015)Barman, Konopacky, Macintosh, \&
  Marois}]{barman_simultaneous_2015}
Barman, T.~S., Konopacky, Q.~M., Macintosh, B., \& Marois, C. 2015,
  \href{http://dx.doi.org/10.1088/0004-637X/804/1/61}{\JournalTitle{The
  Astrophysical Journal}, 804, 61}, publisher: IOP Publishing

\bibitem[{Barstow {et~al.}(2017)Barstow, Aigrain, Irwin, \&
  Sing}]{barstow_consistent_2017}
Barstow, J.~K., Aigrain, S., Irwin, P. G.~J., \& Sing, D.~K. 2017,
  \href{http://dx.doi.org/10.3847/1538-4357/834/1/50}{\JournalTitle{The
  Astrophysical Journal}, 834, 50}, publisher: American Astronomical Society

\bibitem[{Batygin {et~al.}(2016)Batygin, Bodenheimer, \&
  Laughlin}]{batygin_situ_2016}
Batygin, K., Bodenheimer, P.~H., \& Laughlin, G.~P. 2016,
  \href{http://dx.doi.org/10.3847/0004-637X/829/2/114}{\JournalTitle{The
  Astrophysical Journal}, 829, 114}, publisher: American Astronomical Society

\bibitem[{Belle \& Braun(2009)}]{belle_directly_2009}
Belle, G. T.~v. \& Braun, K.~v. 2009,
  \href{http://dx.doi.org/10.1088/0004-637X/694/2/1085}{\JournalTitle{The
  Astrophysical Journal}, 694, 1085}, publisher: American Astronomical Society

\bibitem[{Beltz {et~al.}(2021)Beltz, Rauscher, Brogi, \&
  Kempton}]{beltz_significant_2021}
Beltz, H., Rauscher, E., Brogi, M., \& Kempton, E. M.-R. 2021,
  \href{http://dx.doi.org/10.3847/1538-3881/abb67b}{\JournalTitle{The
  Astronomical Journal}, 161, 1}, publisher: American Astronomical Society

\bibitem[{Benneke(2015)}]{benneke_strict_2015}
Benneke, B. 2015,
  \href{http://arxiv.org/abs/1504.07655}{\JournalTitle{arXiv:1504.07655
  [astro-ph]}}, arXiv: 1504.07655

\bibitem[{Benneke \& Seager(2012)}]{benneke_atmospheric_2012}
Benneke, B. \& Seager, S. 2012,
  \href{http://dx.doi.org/10.1088/0004-637X/753/2/100}{\JournalTitle{The
  Astrophysical Journal}, 753, 100}

\bibitem[{Benneke \& Seager(2013)}]{benneke_how_2013}
Benneke, B. \& Seager, S. 2013,
  \href{http://dx.doi.org/10.1088/0004-637X/778/2/153}{\JournalTitle{The
  Astrophysical Journal}, 778, 153}

\bibitem[{Benneke {et~al.}(2019{\natexlab{a}})Benneke, Knutson, Lothringer,
  Crossfield, Moses, Morley, Kreidberg, Fulton, Dragomir, Howard, Wong,
  Désert, McCullough, Kempton, Fortney, Gilliland, Deming, \&
  Kammer}]{benneke_sub-neptune_2019}
Benneke, B., Knutson, H.~A., Lothringer, J., {et~al.} 2019{\natexlab{a}},
  \href{http://dx.doi.org/10.1038/s41550-019-0800-5}{\JournalTitle{Nature
  Astronomy}, 3, 813}

\bibitem[{Benneke {et~al.}(2019{\natexlab{b}})Benneke, Wong, Piaulet, Knutson,
  Lothringer, Morley, Crossfield, Gao, Greene, Dressing, Dragomir, Howard,
  McCullough, Kempton, Fortney, \& Fraine}]{benneke_water_2019}
Benneke, B., Wong, I., Piaulet, C., {et~al.} 2019{\natexlab{b}},
  \href{http://dx.doi.org/10.3847/2041-8213/ab59dc}{\JournalTitle{The
  Astrophysical Journal}, 887, L14}

\bibitem[{Birkby(2018)}]{birkby_spectroscopic_2018}
Birkby, J.~L. 2018, \href{http://dx.doi.org/10.1007/978-3-319-55333-7_16}{in
  Handbook of {Exoplanets}, ed. H.~J. Deeg \& J.~A. Belmonte} (Cham: Springer
  International Publishing), 1485

\bibitem[{Birkby {et~al.}(2017)Birkby, Kok, Brogi, Schwarz, \&
  Snellen}]{birkby_discovery_2017}
Birkby, J.~L., Kok, R. J.~d., Brogi, M., Schwarz, H., \& Snellen, I. A.~G.
  2017, \href{http://dx.doi.org/10.3847/1538-3881/aa5c87}{\JournalTitle{The
  Astronomical Journal}, 153, 138}

\bibitem[{Birkby {et~al.}(2013)Birkby, Kok, J, Brogi, Mooij, W, Schwarz,
  Albrecht, \& Snellen}]{birkby_detection_2013}
Birkby, J.~L., Kok, D., J, R., {et~al.} 2013,
  \href{http://dx.doi.org/10.1093/mnrasl/slt107}{\JournalTitle{Monthly Notices
  of the Royal Astronomical Society: Letters}, 436, L35}

\bibitem[{Boley {et~al.}(2016)Boley, Contreras, \& Gladman}]{boley_situ_2016}
Boley, A.~C., Contreras, A. P.~G., \& Gladman, B. 2016,
  \href{http://dx.doi.org/10.3847/2041-8205/817/2/L17}{\JournalTitle{The
  Astrophysical Journal}, 817, L17}, publisher: American Astronomical Society

\bibitem[{Booth {et~al.}(2017)Booth, Clarke, Madhusudhan, \&
  Ilee}]{booth_chemical_2017}
Booth, R.~A., Clarke, C.~J., Madhusudhan, N., \& Ilee, J.~D. 2017,
  \href{http://dx.doi.org/10.1093/mnras/stx1103}{\JournalTitle{Monthly Notices
  of the Royal Astronomical Society}, 469, 3994}, publisher: Oxford Academic

\bibitem[{Borsa {et~al.}(2015)Borsa, Scandariato, Rainer, Bignamini, Maggio,
  Poretti, Lanza, Mauro, Benatti, Biazzo, Bonomo, Damasso, Esposito, Gratton,
  Affer, Barbieri, Boccato, Claudi, Cosentino, Covino, Desidera, Fiorenzano,
  Gandolfi, Harutyunyan, Maldonado, Micela, Molaro, Molinari, Pagano,
  Pillitteri, Piotto, Shkolnik, Silvotti, Smareglia, Southworth, Sozzetti, \&
  Stelzer}]{borsa_gaps_2015}
Borsa, F., Scandariato, G., Rainer, M., {et~al.} 2015,
  \href{http://dx.doi.org/10.1051/0004-6361/201525741}{\JournalTitle{Astronomy
  \& Astrophysics}, 578, A64}, publisher: EDP Sciences

\bibitem[{Borysow(2002)}]{borysow_collision-induced_2002}
Borysow, A. 2002,
  \href{http://dx.doi.org/10.1051/0004-6361:20020555}{\JournalTitle{Astronomy
  \& Astrophysics}, 390, 779}

\bibitem[{Boss(1997)}]{boss_giant_1997}
Boss, A.~P. 1997,
  \href{http://dx.doi.org/10.1126/science.276.5320.1836}{\JournalTitle{Science},
  276, 1836}

\bibitem[{Boss(2001)}]{boss_gas_2001}
Boss, A.~P. 2001, \href{http://dx.doi.org/10.1086/323694}{\JournalTitle{The
  Astrophysical Journal}, 563, 367}, publisher: IOP Publishing

\bibitem[{Brewer {et~al.}(2017)Brewer, Fischer, \&
  Madhusudhan}]{brewer_co_2017}
Brewer, J.~M., Fischer, D.~A., \& Madhusudhan, N. 2017,
  \href{http://dx.doi.org/10.3847/1538-3881/153/2/83}{\JournalTitle{The
  Astronomical Journal}, 153, 83}, publisher: American Astronomical Society

\bibitem[{Brogi {et~al.}(2018)Brogi, Giacobbe, Guilluy, Kok, Sozzetti, Mancini,
  \& Bonomo}]{brogi_exoplanet_2018}
Brogi, M., Giacobbe, P., Guilluy, G., {et~al.} 2018,
  \href{http://dx.doi.org/10.1051/0004-6361/201732189}{\JournalTitle{Astronomy
  \& Astrophysics}, 615, A16}, publisher: EDP Sciences

\bibitem[{Brogi {et~al.}(2016)Brogi, Kok, Albrecht, Snellen, Birkby, \&
  Schwarz}]{brogi_rotation_2016}
Brogi, M., Kok, R. J.~d., Albrecht, S., {et~al.} 2016,
  \href{http://dx.doi.org/10.3847/0004-637X/817/2/106}{\JournalTitle{The
  Astrophysical Journal}, 817, 106}

\bibitem[{Brogi {et~al.}(2014)Brogi, Kok, Birkby, Schwarz, \&
  Snellen}]{brogi_carbon_2014}
Brogi, M., Kok, R. J.~d., Birkby, J.~L., Schwarz, H., \& Snellen, I. a.~G.
  2014,
  \href{http://dx.doi.org/10.1051/0004-6361/201423537}{\JournalTitle{Astronomy
  \& Astrophysics}, 565, A124}

\bibitem[{Brogi {et~al.}(2017)Brogi, Line, Bean, Désert, \&
  Schwarz}]{brogi_framework_2017}
Brogi, M., Line, M., Bean, J., Désert, J.-M., \& Schwarz, H. 2017,
  \href{http://dx.doi.org/10.3847/2041-8213/aa6933}{\JournalTitle{The
  Astrophysical Journal Letters}, 839, L2}

\bibitem[{Brogi \& Line(2019)}]{brogi_retrieving_2019}
Brogi, M. \& Line, M.~R. 2019,
  \href{http://dx.doi.org/10.3847/1538-3881/aaffd3}{\JournalTitle{The
  Astronomical Journal}, 157, 114}

\bibitem[{Brogi {et~al.}(2012)Brogi, Snellen, Kok, Albrecht, Birkby, \&
  Mooij}]{brogi_signature_2012}
Brogi, M., Snellen, I. A.~G., Kok, R. J.~d., {et~al.} 2012,
  \href{http://dx.doi.org/10.1038/nature11161}{\JournalTitle{Nature}, 486, 502}

\bibitem[{Brogi {et~al.}(2013)Brogi, Snellen, Kok, Albrecht, Birkby, \&
  Mooij}]{brogi_detection_2013}
Brogi, M., Snellen, I. A.~G., Kok, R. J.~d., {et~al.} 2013,
  \href{http://dx.doi.org/10.1088/0004-637X/767/1/27}{\JournalTitle{The
  Astrophysical Journal}, 767, 27}

\bibitem[{Burrows \& Sharp(1999)}]{burrows_chemical_1999}
Burrows, A. \& Sharp, C.~M. 1999,
  \href{http://dx.doi.org/10.1086/306811}{\JournalTitle{The Astrophysical
  Journal}, 512, 843}, publisher: IOP Publishing

\bibitem[{Butler {et~al.}(1997)Butler, Marcy, Williams, Hauser, \&
  Shirts}]{butler_three_1997}
Butler, R.~P., Marcy, G.~W., Williams, E., Hauser, H., \& Shirts, P. 1997,
  \href{http://dx.doi.org/10.1086/310444}{\JournalTitle{The Astrophysical
  Journal Letters}, 474, L115}, publisher: IOP Publishing

\bibitem[{Buzard {et~al.}(2020)Buzard, Finnerty, Piskorz, Pelletier, Benneke,
  Bender, Lockwood, Wallack, Wilkins, \& Blake}]{buzard_simulating_2020}
Buzard, C., Finnerty, L., Piskorz, D., {et~al.} 2020,
  \href{http://dx.doi.org/10.3847/1538-3881/ab8f9c}{\JournalTitle{The
  Astronomical Journal}, 160, 1}, publisher: American Astronomical Society

\bibitem[{Cabot {et~al.}(2019)Cabot, Madhusudhan, Hawker, \&
  Gandhi}]{cabot_robustness_2019}
Cabot, S. H.~C., Madhusudhan, N., Hawker, G.~A., \& Gandhi, S. 2019,
  \href{http://dx.doi.org/10.1093/mnras/sty2994}{\JournalTitle{Monthly Notices
  of the Royal Astronomical Society}, 482, 4422}

\bibitem[{Chubb {et~al.}(2020)Chubb, Tennyson, \&
  Yurchenko}]{chubb_exomol_2020}
Chubb, K.~L., Tennyson, J., \& Yurchenko, S.~N. 2020,
  \href{http://dx.doi.org/10.1093/mnras/staa229}{\JournalTitle{Monthly Notices
  of the Royal Astronomical Society}, 493, 1531}

\bibitem[{Cridland {et~al.}(2019)Cridland, Dishoeck, Alessi, \&
  Pudritz}]{cridland_connecting_2019}
Cridland, A.~J., Dishoeck, E. F.~v., Alessi, M., \& Pudritz, R.~E. 2019,
  \href{http://dx.doi.org/10.1051/0004-6361/201936105}{\JournalTitle{Astronomy
  \& Astrophysics}, 632, A63}

\bibitem[{Cridland {et~al.}(2020)Cridland, Dishoeck, Alessi, \&
  Pudritz}]{cridland_connecting_2020}
Cridland, A.~J., Dishoeck, E. F.~v., Alessi, M., \& Pudritz, R.~E. 2020,
  \href{http://dx.doi.org/10.1051/0004-6361/202038767}{\JournalTitle{Astronomy
  \& Astrophysics}, 642, A229}, publisher: EDP Sciences

\bibitem[{Cridland {et~al.}(2016)Cridland, Pudritz, \&
  Alessi}]{cridland_composition_2016}
Cridland, A.~J., Pudritz, R.~E., \& Alessi, M. 2016,
  \href{http://dx.doi.org/10.1093/mnras/stw1511}{\JournalTitle{Monthly Notices
  of the Royal Astronomical Society}, 461, 3274}, publisher: Oxford Academic

\bibitem[{Crossfield {et~al.}(2012)Crossfield, Barman, Hansen, Tanaka, \&
  Kodama}]{crossfield_re-evaluating_2012}
Crossfield, I. J.~M., Barman, T., Hansen, B. M.~S., Tanaka, I., \& Kodama, T.
  2012, \href{http://dx.doi.org/10.1088/0004-637X/760/2/140}{\JournalTitle{The
  Astrophysical Journal}, 760, 140}, publisher: IOP Publishing

\bibitem[{Damiano {et~al.}(2019)Damiano, Micela, \&
  Tinetti}]{damiano_principal_2019}
Damiano, M., Micela, G., \& Tinetti, G. 2019,
  \href{http://dx.doi.org/10.3847/1538-4357/ab22b2}{\JournalTitle{The
  Astrophysical Journal}, 878, 153}

\bibitem[{de~Kok {et~al.}(2013)de~Kok, Brogi, Snellen, Birkby, Albrecht, \&
  Mooij}]{de_kok_detection_2013}
de~Kok, R.~J., Brogi, M., Snellen, I. a.~G., {et~al.} 2013,
  \href{http://dx.doi.org/10.1051/0004-6361/201321381}{\JournalTitle{Astronomy
  \& Astrophysics}, 554, A82}

\bibitem[{Donati {et~al.}(2008)Donati, Moutou, Farès, Bohlender, Catala,
  Deleuil, Shkolnik, Cameron, Jardine, \& Walker}]{donati_magnetic_2008}
Donati, J.-F., Moutou, C., Farès, R., {et~al.} 2008,
  \href{http://dx.doi.org/10.1111/j.1365-2966.2008.12946.x}{\JournalTitle{Monthly
  Notices of the Royal Astronomical Society}, 385, 1179}, publisher: Oxford
  Academic

\bibitem[{Donati {et~al.}(2018)Donati, Kouach, Lacombe, Baratchart, Doyon,
  Delfosse, Artigau, Moutou, Hébrard, Bouchy, Bouvier, Alencar, Saddlemyer,
  Parès, Rabou, Micheau, Dolon, Barrick, Hernandez, Wang, Reshetov, Striebig,
  Challita, Carmona, Tibault, Martioli, Figueira, Boisse, \&
  Pepe}]{donati_spirou_2018}
Donati, J.-F., Kouach, D., Lacombe, M., {et~al.} 2018,
  \href{http://dx.doi.org/10.1007/978-3-319-30648-3_107-1}{in Handbook of
  {Exoplanets}, ed. H.~J. Deeg \& J.~A. Belmonte} (Cham: Springer International
  Publishing), 1

\bibitem[{Donati {et~al.}(2020)Donati, Kouach, Moutou, Doyon, Delfosse,
  Artigau, Baratchart, Lacombe, Barrick, Hébrard, Bouchy, Saddlemyer, Parès,
  Rabou, Micheau, Dolon, Reshetov, Challita, Carmona, Striebig, Thibault,
  Martioli, Cook, Fouqué, Vermeulen, Wang, Arnold, Pepe, Boisse, Figueira,
  Bouvier, Ray, Feugeade, Morin, Alencar, Hobson, Castilho, Udry, Santos,
  Hernandez, Benedict, Vallée, Gallou, Dupieux, Larrieu, Perruchot, Sottile,
  Moreau, Usher, Baril, Wildi, Chazelas, Malo, Bonfils, Loop, Kerley, Wevers,
  Dunn, Pazder, Macdonald, Dubois, Carrié, Valentin, Henault, Yan, \&
  Steinmetz}]{donati_spirou_2020}
Donati, J.-F., Kouach, D., Moutou, C., {et~al.} 2020,
  \href{http://dx.doi.org/10.1093/mnras/staa2569}{\JournalTitle{Monthly Notices
  of the Royal Astronomical Society}, 498, 5684}, publisher: Oxford Academic

\bibitem[{Eistrup {et~al.}(2016)Eistrup, Walsh, \&
  Dishoeck}]{eistrup_setting_2016}
Eistrup, C., Walsh, C., \& Dishoeck, E. F.~v. 2016,
  \href{http://dx.doi.org/10.1051/0004-6361/201628509}{\JournalTitle{Astronomy
  \& Astrophysics}, 595, A83}, publisher: EDP Sciences

\bibitem[{Eistrup {et~al.}(2018)Eistrup, Walsh, \&
  Dishoeck}]{eistrup_molecular_2018}
Eistrup, C., Walsh, C., \& Dishoeck, E. F.~v. 2018,
  \href{http://dx.doi.org/10.1051/0004-6361/201731302}{\JournalTitle{Astronomy
  \& Astrophysics}, 613, A14}

\bibitem[{Espinoza {et~al.}(2017)Espinoza, Fortney, Miguel, Thorngren, \&
  Murray-Clay}]{espinoza_metal_2017}
Espinoza, N., Fortney, J.~J., Miguel, Y., Thorngren, D., \& Murray-Clay, R.
  2017, \href{http://dx.doi.org/10.3847/2041-8213/aa65ca}{\JournalTitle{The
  Astrophysical Journal}, 838, L9}, publisher: American Astronomical Society

\bibitem[{Fegley \& Lodders(1994)}]{fegley_chemical_1994}
Fegley, B. \& Lodders, K. 1994,
  \href{http://dx.doi.org/10.1006/icar.1994.1111}{\JournalTitle{Icarus}, 110,
  117}

\bibitem[{Flagg {et~al.}(2019)Flagg, Johns-Krull, Nofi, Llama, Prato, Sullivan,
  Jaffe, \& Mace}]{flagg_co_2019}
Flagg, L., Johns-Krull, C.~M., Nofi, L., {et~al.} 2019,
  \href{http://dx.doi.org/10.3847/2041-8213/ab276d}{\JournalTitle{The
  Astrophysical Journal}, 878, L37}, publisher: American Astronomical Society

\bibitem[{Flowers {et~al.}(2019)Flowers, Brogi, Rauscher, Kempton, \&
  Chiavassa}]{flowers_high-resolution_2019}
Flowers, E., Brogi, M., Rauscher, E., Kempton, E. M.-R., \& Chiavassa, A. 2019,
  \href{http://dx.doi.org/10.3847/1538-3881/ab164c}{\JournalTitle{The
  Astronomical Journal}, 157, 209}, publisher: American Astronomical Society

\bibitem[{Foreman-Mackey(2016)}]{foreman-mackey_cornerpy_2016}
Foreman-Mackey, D. 2016,
  \href{http://dx.doi.org/10.21105/joss.00024}{\JournalTitle{Journal of Open
  Source Software}, 1, 24}

\bibitem[{Foreman-Mackey {et~al.}(2013)Foreman-Mackey, Hogg, Lang, \&
  Goodman}]{foreman-mackey_emcee_2013}
Foreman-Mackey, D., Hogg, D.~W., Lang, D., \& Goodman, J. 2013,
  \href{http://dx.doi.org/10.1086/670067}{\JournalTitle{Publications of the
  Astronomical Society of the Pacific}, 125, 306}, publisher: IOP Publishing

\bibitem[{Föhring {et~al.}(2013)Föhring, Dhillon, Madhusudhan, Marsh,
  Copperwheat, Littlefair, \& Wilson}]{fohring_ultracam_2013}
Föhring, D., Dhillon, V.~S., Madhusudhan, N., {et~al.} 2013,
  \href{http://dx.doi.org/10.1093/mnras/stt1443}{\JournalTitle{Monthly Notices
  of the Royal Astronomical Society}, 435, 2268}, publisher: Oxford Academic

\bibitem[{Gaia~Collaboration {et~al.}(2018)Gaia~Collaboration, Brown,
  Vallenari, Prusti, Bruijne, Babusiaux, Bailer-Jones, Biermann, Evans, Eyer,
  Jansen, Jordi, Klioner, Lammers, Lindegren, Luri, Mignard, Panem, Pourbaix,
  Randich, Sartoretti, Siddiqui, Soubiran, Leeuwen, Walton, Arenou, Bastian,
  Cropper, Drimmel, Katz, Lattanzi, Bakker, Cacciari, Castañeda, Chaoul,
  Cheek, Angeli, Fabricius, Guerra, Holl, Masana, Messineo, Mowlavi,
  Nienartowicz, Panuzzo, Portell, Riello, Seabroke, Tanga, Thévenin,
  Gracia-Abril, Comoretto, Garcia-Reinaldos, Teyssier, Altmann, Andrae, Audard,
  Bellas-Velidis, Benson, Berthier, Blomme, Burgess, Busso, Carry, Cellino,
  Clementini, Clotet, Creevey, Davidson, Ridder, Delchambre, Dell’Oro,
  Ducourant, Fernández-Hernández, Fouesneau, Frémat, Galluccio,
  García-Torres, González-Núñez, González-Vidal, Gosset, Guy, Halbwachs,
  Hambly, Harrison, Hernández, Hestroffer, Hodgkin, Hutton, Jasniewicz,
  Jean-Antoine-Piccolo, Jordan, Korn, Krone-Martins, Lanzafame, Lebzelter,
  Löffler, Manteiga, Marrese, Martín-Fleitas, Moitinho, Mora, Muinonen,
  Osinde, Pancino, Pauwels, Petit, Recio-Blanco, Richards, Rimoldini, Robin,
  Sarro, Siopis, Smith, Sozzetti, Süveges, Torra, Reeven, Abbas, Aramburu,
  Accart, Aerts, Altavilla, Álvarez, Alvarez, Alves, Anderson, Andrei, Varela,
  Antiche, Antoja, Arcay, Astraatmadja, Bach, Baker, Balaguer-Núñez, Balm,
  Barache, Barata, Barbato, Barblan, Barklem, Barrado, Barros, Barstow, Muñoz,
  Bassilana, Becciani, Bellazzini, Berihuete, Bertone, Bianchi, Bienaymé,
  Blanco-Cuaresma, Boch, Boeche, Bombrun, Borrachero, Bossini, Bouquillon,
  Bourda, Bragaglia, Bramante, Breddels, Bressan, Brouillet, Brüsemeister,
  Brugaletta, Bucciarelli, Burlacu, Busonero, Butkevich, Buzzi, Caffau,
  Cancelliere, Cannizzaro, Cantat-Gaudin, Carballo, Carlucci, Carrasco,
  Casamiquela, Castellani, Castro-Ginard, Charlot, Chemin, Chiavassa, Cocozza,
  Costigan, Cowell, Crifo, Crosta, Crowley, Cuypers†, Dafonte, Damerdji,
  Dapergolas, David, David, Laverny, Luise, March, Martino, Souza, Torres,
  Debosscher, Pozo, Delbo, Delgado, Delgado, Matteo, Diakite, Diener,
  Distefano, Dolding, Drazinos, Durán, Edvardsson, Enke, Eriksson, Esquej,
  Bontemps, Fabre, Fabrizio, Faigler, Falcão, Casas, Federici, Fedorets,
  Fernique, Figueras, Filippi, Findeisen, Fonti, Fraile, Fraser, Frézouls,
  Gai, Galleti, Garabato, García-Sedano, Garofalo, Garralda, Gavel, Gavras,
  Gerssen, Geyer, Giacobbe, Gilmore, Girona, Giuffrida, Glass, Gomes, Granvik,
  Gueguen, Guerrier, Guiraud, Gutiérrez-Sánchez, Haigron, Hatzidimitriou,
  Hauser, Haywood, Heiter, Helmi, Heu, Hilger, Hobbs, Hofmann, Holland, Huckle,
  Hypki, Icardi, Janßen, Fombelle, Jonker, Juhász, Julbe, Karampelas, Kewley,
  Klar, Kochoska, Kohley, Kolenberg, Kontizas, Kontizas, Koposov, Kordopatis,
  Kostrzewa-Rutkowska, Koubsky, Lambert, Lanza, Lasne, Lavigne, Fustec,
  Poncin-Lafitte, Lebreton, Leccia, Leclerc, Lecoeur-Taibi, Lenhardt, Leroux,
  Liao, Licata, Lindstrøm, Lister, Livanou, Lobel, López, Managau, Mann,
  Mantelet, Marchal, Marchant, Marconi, Marinoni, Marschalkó, Marshall,
  Martino, Marton, Mary, Massari, Matijevič, Mazeh, McMillan, Messina,
  Michalik, Millar, Molina, Molinaro, Molnár, Montegriffo, Mor, Morbidelli,
  Morel, Morris, Mulone, Muraveva, Musella, Nelemans, Nicastro, Noval,
  O’Mullane, Ordénovic, Ordóñez-Blanco, Osborne, Pagani, Pagano, Pailler,
  Palacin, Palaversa, Panahi, Pawlak, Piersimoni, Pineau, Plachy, Plum, Poggio,
  Poujoulet, Prša, Pulone, Racero, Ragaini, Rambaux, Ramos-Lerate, Regibo,
  Reylé, Riclet, Ripepi, Riva, Rivard, Rixon, Roegiers, Roelens,
  Romero-Gómez, Rowell, Royer, Ruiz-Dern, Sadowski, Sellés, Sahlmann,
  Salgado, Salguero, Sanna, Santana-Ros, Sarasso, Savietto, Schultheis,
  Sciacca, Segol, Segovia, Ségransan, Shih, Siltala, Silva, Smart, Smith,
  Solano, Solitro, Sordo, Nieto, Souchay, Spagna, Spoto, Stampa, Steele,
  Steidelmüller, Stephenson, Stoev, Suess, Surdej, Szabados, Szegedi-Elek,
  Tapiador, Taris, Tauran, Taylor, Teixeira, Terrett, Teyssandier, Thuillot,
  Titarenko, Clotet, Turon, Ulla, Utrilla, Uzzi, Vaillant, Valentini, Valette,
  Elteren, Hemelryck, Leeuwen, Vaschetto, Vecchiato, Veljanoski, Viala,
  Vicente, Vogt, Essen, Voss, Votruba, Voutsinas, Walmsley, Weiler, Wertz,
  Wevers, Wyrzykowski, Yoldas, Žerjal, Ziaeepour, Zorec, Zschocke, Zucker,
  Zurbach, \& Zwitter}]{gaia_collaboration_gaia_2018}
Gaia~Collaboration, T., Brown, A. G.~A., Vallenari, A., {et~al.} 2018,
  \href{http://dx.doi.org/10.1051/0004-6361/201833051}{\JournalTitle{Astronomy
  \& Astrophysics}, 616, A1}, publisher: EDP Sciences

\bibitem[{Gandhi {et~al.}(2019)Gandhi, Madhusudhan, Hawker, \&
  Piette}]{gandhi_hydra-h_2019}
Gandhi, S., Madhusudhan, N., Hawker, G., \& Piette, A. 2019,
  \href{http://dx.doi.org/10.3847/1538-3881/ab4efc}{\JournalTitle{The
  Astronomical Journal}, 158, 228}, publisher: American Astronomical Society

\bibitem[{Gandhi {et~al.}(2020)Gandhi, Brogi, Yurchenko, Tennyson, Coles, Webb,
  Birkby, Guilluy, Hawker, Madhusudhan, Bonomo, \&
  Sozzetti}]{gandhi_molecular_2020}
Gandhi, S., Brogi, M., Yurchenko, S.~N., {et~al.} 2020,
  \href{http://dx.doi.org/10.1093/mnras/staa981}{\JournalTitle{Monthly Notices
  of the Royal Astronomical Society}, 495, 224}, publisher: Oxford Academic

\bibitem[{Giacobbe {et~al.}(2021)Giacobbe, Brogi, Gandhi, Cubillos, Bonomo,
  Sozzetti, Fossati, Guilluy, Carleo, Rainer, Harutyunyan, Borsa, Pino,
  Nascimbeni, Benatti, Biazzo, Bignamini, Chubb, Claudi, Cosentino, Covino,
  Damasso, Desidera, Fiorenzano, Ghedina, Lanza, Leto, Maggio, Malavolta,
  Maldonado, Micela, Molinari, Pagano, Pedani, Piotto, Poretti, Scandariato,
  Yurchenko, Fantinel, Galli, Lodi, Sanna, \& Tozzi}]{giacobbe_five_2021}
Giacobbe, P., Brogi, M., Gandhi, S., {et~al.} 2021,
  \href{http://dx.doi.org/10.1038/s41586-021-03381-x}{\JournalTitle{Nature},
  592, 205}, number: 7853 Publisher: Nature Publishing Group

\bibitem[{Gibson {et~al.}(2019)Gibson, de~Mooij, Evans, Merritt, Nikolov, Sing,
  \& Watson}]{gibson_revisiting_2019}
Gibson, N.~P., de~Mooij, E. J.~W., Evans, T.~M., {et~al.} 2019,
  \href{http://dx.doi.org/10.1093/mnras/sty2722}{\JournalTitle{Monthly Notices
  of the Royal Astronomical Society}, 482, 606}, publisher: Oxford Academic

\bibitem[{Gibson {et~al.}(2020)Gibson, Merritt, Nugroho, Cubillos, de~Mooij,
  Mikal-Evans, Fossati, Lothringer, Nikolov, Sing, Spake, Watson, \&
  Wilson}]{gibson_detection_2020}
Gibson, N.~P., Merritt, S., Nugroho, S.~K., {et~al.} 2020,
  \href{http://dx.doi.org/10.1093/mnras/staa228}{\JournalTitle{Monthly Notices
  of the Royal Astronomical Society}, 493, 2215}, publisher: Oxford Academic

\bibitem[{Gray {et~al.}(2001)Gray, Napier, \& Winkler}]{gray_physical_2001}
Gray, R.~O., Napier, M.~G., \& Winkler, L.~I. 2001,
  \href{http://dx.doi.org/10.1086/319956}{\JournalTitle{The Astronomical
  Journal}, 121, 2148}, publisher: IOP Publishing

\bibitem[{Grimm \& Heng(2015)}]{grimm_helios-k_2015}
Grimm, S.~L. \& Heng, K. 2015,
  \href{http://dx.doi.org/10.1088/0004-637X/808/2/182}{\JournalTitle{The
  Astrophysical Journal}, 808, 182}

\bibitem[{Grimm {et~al.}(2021)Grimm, Malik, Kitzmann, Guzmán-Mesa,
  Hoeijmakers, Fisher, Mendonça, Yurchenko, Tennyson, Alesina, Buchschacher,
  Burnier, Segransan, Kurucz, \& Heng}]{grimm_helios-k_2021}
Grimm, S.~L., Malik, M., Kitzmann, D., {et~al.} 2021,
  \href{http://dx.doi.org/10.3847/1538-4365/abd773}{\JournalTitle{The
  Astrophysical Journal Supplement Series}, 253, 30}, publisher: American
  Astronomical Society

\bibitem[{Guillot(2010)}]{guillot_radiative_2010}
Guillot, T. 2010,
  \href{http://dx.doi.org/10.1051/0004-6361/200913396}{\JournalTitle{Astronomy
  \& Astrophysics}, 520, A27}, publisher: EDP Sciences

\bibitem[{Guilluy {et~al.}(2019)Guilluy, Sozzetti, Brogi, Bonomo, Giacobbe,
  Claudi, \& Benatti}]{guilluy_exoplanet_2019}
Guilluy, G., Sozzetti, A., Brogi, M., {et~al.} 2019,
  \href{http://dx.doi.org/10.1051/0004-6361/201834615}{\JournalTitle{Astronomy
  \& Astrophysics}, 625, A107}, publisher: EDP Sciences

\bibitem[{Hale(1994)}]{hale_orbital_1994}
Hale, A. 1994, \href{http://dx.doi.org/10.1086/116855}{\JournalTitle{The
  Astronomical Journal}, 107, 306}

\bibitem[{Harada {et~al.}(2021)Harada, Kempton, Rauscher, Roman, Malsky,
  Brinjikji, \& DiTomasso}]{harada_signatures_2021}
Harada, C.~K., Kempton, E. M.-R., Rauscher, E., {et~al.} 2021,
  \href{http://dx.doi.org/10.3847/1538-4357/abdc22}{\JournalTitle{The
  Astrophysical Journal}, 909, 85}, publisher: American Astronomical Society

\bibitem[{Harris {et~al.}(2020)Harris, Millman, van~der Walt, Gommers,
  Virtanen, Cournapeau, Wieser, Taylor, Berg, Smith, Kern, Picus, Hoyer, van
  Kerkwijk, Brett, Haldane, del Río, Wiebe, Peterson, Gérard-Marchant,
  Sheppard, Reddy, Weckesser, Abbasi, Gohlke, \& Oliphant}]{harris_array_2020}
Harris, C.~R., Millman, K.~J., van~der Walt, S.~J., {et~al.} 2020,
  \href{http://dx.doi.org/10.1038/s41586-020-2649-2}{\JournalTitle{Nature},
  585, 357}, number: 7825 Publisher: Nature Publishing Group

\bibitem[{Harris {et~al.}(2006)Harris, Tennyson, Kaminsky, Pavlenko, \&
  Jones}]{harris_improved_2006}
Harris, G.~J., Tennyson, J., Kaminsky, B.~M., Pavlenko, Y.~V., \& Jones, H.
  R.~A. 2006,
  \href{http://dx.doi.org/10.1111/j.1365-2966.2005.09960.x}{\JournalTitle{Monthly
  Notices of the Royal Astronomical Society}, 367, 400}

\bibitem[{Hawker {et~al.}(2018)Hawker, Madhusudhan, Cabot, \&
  Gandhi}]{hawker_evidence_2018}
Hawker, G.~A., Madhusudhan, N., Cabot, S. H.~C., \& Gandhi, S. 2018,
  \href{http://dx.doi.org/10.3847/2041-8213/aac49d}{\JournalTitle{The
  Astrophysical Journal Letters}, 863, L11}, publisher: IOP Publishing

\bibitem[{Hoeijmakers {et~al.}(2018)Hoeijmakers, Snellen, \&
  Terwisga}]{hoeijmakers_searching_2018}
Hoeijmakers, H.~J., Snellen, I. a.~G., \& Terwisga, S. E.~v. 2018,
  \href{http://dx.doi.org/10.1051/0004-6361/201731192}{\JournalTitle{Astronomy
  \& Astrophysics}, 610, A47}

\bibitem[{Hunter(2007)}]{hunter_matplotlib_2007}
Hunter, J.~D. 2007,
  \href{http://dx.doi.org/10.1109/MCSE.2007.55}{\JournalTitle{Computing in
  Science Engineering}, 9, 90}, conference Name: Computing in Science
  Engineering

\bibitem[{Justesen \& Albrecht(2019)}]{justesen_constraining_2019}
Justesen, A.~B. \& Albrecht, S. 2019,
  \href{http://dx.doi.org/10.1051/0004-6361/201834368}{\JournalTitle{Astronomy
  \& Astrophysics}, 625, A59}, publisher: EDP Sciences

\bibitem[{Knutson {et~al.}(2014)Knutson, Benneke, Deming, \&
  Homeier}]{knutson_featureless_2014}
Knutson, H.~A., Benneke, B., Deming, D., \& Homeier, D. 2014,
  \href{http://dx.doi.org/10.1038/nature12887}{\JournalTitle{Nature}, 505, 66},
  number: 7481 Publisher: Nature Publishing Group

\bibitem[{Konopacky {et~al.}(2013)Konopacky, Barman, Macintosh, \&
  Marois}]{konopacky_detection_2013}
Konopacky, Q.~M., Barman, T.~S., Macintosh, B.~A., \& Marois, C. 2013,
  \href{http://dx.doi.org/10.1126/science.1232003}{\JournalTitle{Science (New
  York, N.Y.)}, 339, 1398}

\bibitem[{Kreidberg {et~al.}(2014)Kreidberg, Bean, Désert, Benneke, Deming,
  Stevenson, Seager, Berta-Thompson, Seifahrt, \&
  Homeier}]{kreidberg_clouds_2014}
Kreidberg, L., Bean, J.~L., Désert, J.-M., {et~al.} 2014,
  \href{http://dx.doi.org/10.1038/nature12888}{\JournalTitle{Nature}, 505, 69}

\bibitem[{Kreidberg {et~al.}(2015)Kreidberg, Line, Bean, Stevenson, Désert,
  Madhusudhan, Fortney, Barstow, Henry, Williamson, \&
  Showman}]{kreidberg_detection_2015}
Kreidberg, L., Line, M.~R., Bean, J.~L., {et~al.} 2015,
  \href{http://dx.doi.org/10.1088/0004-637X/814/1/66}{\JournalTitle{The
  Astrophysical Journal}, 814, 66}, publisher: IOP Publishing

\bibitem[{Lavie {et~al.}(2017)Lavie, Mendonça, Mordasini, Malik, Bonnefoy,
  Demory, Oreshenko, Grimm, Ehrenreich, \&
  Heng}]{lavie_heliosretrievalopen-source_2017}
Lavie, B., Mendonça, J.~M., Mordasini, C., {et~al.} 2017,
  \href{http://dx.doi.org/10.3847/1538-3881/aa7ed8}{\JournalTitle{The
  Astronomical Journal}, 154, 91}, publisher: American Astronomical Society

\bibitem[{Leconte \& Chabrier(2012)}]{leconte_new_2012}
Leconte, J. \& Chabrier, G. 2012,
  \href{http://dx.doi.org/10.1051/0004-6361/201117595}{\JournalTitle{Astronomy
  \& Astrophysics}, 540, A20}, publisher: EDP Sciences

\bibitem[{Lee {et~al.}(2013)Lee, Heng, \& Irwin}]{lee_atmospheric_2013}
Lee, J.-M., Heng, K., \& Irwin, P. G.~J. 2013,
  \href{http://dx.doi.org/10.1088/0004-637X/778/2/97}{\JournalTitle{The
  Astrophysical Journal}, 778, 97}, publisher: American Astronomical Society

\bibitem[{Li {et~al.}(2020)Li, Ingersoll, Bolton, Levin, Janssen, Atreya,
  Lunine, Steffes, Brown, Guillot, Allison, Arballo, Bellotti, Adumitroaie,
  Gulkis, Hodges, Li, Misra, Orton, Oyafuso, Santos-Costa, Waite, \&
  Zhang}]{li_water_2020}
Li, C., Ingersoll, A., Bolton, S., {et~al.} 2020,
  \href{http://dx.doi.org/10.1038/s41550-020-1009-3}{\JournalTitle{Nature
  Astronomy}, 1}, publisher: Nature Publishing Group

\bibitem[{Line {et~al.}(2014)Line, Knutson, Wolf, \&
  Yung}]{line_systematic_2014}
Line, M.~R., Knutson, H., Wolf, A.~S., \& Yung, Y.~L. 2014,
  \href{http://dx.doi.org/10.1088/0004-637X/783/2/70}{\JournalTitle{The
  Astrophysical Journal}, 783, 70}, publisher: IOP Publishing

\bibitem[{Line {et~al.}(2015)Line, Teske, Burningham, Fortney, \&
  Marley}]{line_uniform_2015}
Line, M.~R., Teske, J., Burningham, B., Fortney, J.~J., \& Marley, M.~S. 2015,
  \href{http://dx.doi.org/10.1088/0004-637X/807/2/183}{\JournalTitle{The
  Astrophysical Journal}, 807, 183}, publisher: IOP Publishing

\bibitem[{Line {et~al.}(2012)Line, Zhang, Vasisht, Natraj, Chen, \&
  Yung}]{line_information_2012}
Line, M.~R., Zhang, X., Vasisht, G., {et~al.} 2012,
  \href{http://dx.doi.org/10.1088/0004-637X/749/1/93}{\JournalTitle{The
  Astrophysical Journal}, 749, 93}, publisher: American Astronomical Society

\bibitem[{Line {et~al.}(2013)Line, Wolf, Zhang, Knutson, Kammer, Ellison,
  Deroo, Crisp, \& Yung}]{line_systematic_2013}
Line, M.~R., Wolf, A.~S., Zhang, X., {et~al.} 2013,
  \href{http://dx.doi.org/10.1088/0004-637X/775/2/137}{\JournalTitle{The
  Astrophysical Journal}, 775, 137}, publisher: American Astronomical Society

\bibitem[{Line {et~al.}(2016)Line, Stevenson, Bean, Desert, Fortney, {Laura
  Kreidberg}, Madhusudhan, Showman, \& Diamond-Lowe}]{line_no_2016}
Line, M.~R., Stevenson, K.~B., Bean, J., {et~al.} 2016,
  \href{http://dx.doi.org/10.3847/0004-6256/152/6/203}{\JournalTitle{The
  Astronomical Journal}, 152, 203}

\bibitem[{Lockwood {et~al.}(2014)Lockwood, Johnson, Bender, Carr, Barman,
  Richert, \& Blake}]{lockwood_near-ir_2014}
Lockwood, A.~C., Johnson, J.~A., Bender, C.~F., {et~al.} 2014,
  \href{http://dx.doi.org/10.1088/2041-8205/783/2/L29}{\JournalTitle{The
  Astrophysical Journal Letters}, 783, L29}

\bibitem[{Lodders(2003)}]{lodders_solar_2003}
Lodders, K. 2003, \href{http://dx.doi.org/10.1086/375492}{\JournalTitle{The
  Astrophysical Journal}, 591, 1220}, publisher: IOP Publishing

\bibitem[{Lodders(2004)}]{lodders_jupiter_2004}
Lodders, K. 2004, \href{http://dx.doi.org/10.1086/421970}{\JournalTitle{The
  Astrophysical Journal}, 611, 587}, publisher: IOP Publishing

\bibitem[{Lodders(2010)}]{lodders_exoplanet_2010}
Lodders, K. 2010,
  \href{https://onlinelibrary.wiley.com/doi/abs/10.1002/9783527629763.ch8}{in
  Formation and {Evolution} of {Exoplanets}} (John Wiley \& Sons, Ltd), 157,
  section: 8

\bibitem[{Lodders \& Fegley(1993)}]{lodders_lanthanide_1993}
Lodders, K. \& Fegley, B. 1993,
  \href{http://dx.doi.org/10.1016/0012-821X(93)90122-P}{\JournalTitle{Earth and
  Planetary Science Letters}, 117, 125}

\bibitem[{Lodders \& Fegley(2002)}]{lodders_atmospheric_2002}
Lodders, K. \& Fegley, B. 2002,
  \href{http://dx.doi.org/10.1006/icar.2001.6740}{\JournalTitle{Icarus}, 155,
  393}

\bibitem[{Lothringer {et~al.}(2020)Lothringer, Rustamkulov, Sing, Gibson,
  Wilson, \& Schlaufman}]{lothringer_new_2020}
Lothringer, J.~D., Rustamkulov, Z., Sing, D.~K., {et~al.} 2020,
  \href{http://arxiv.org/abs/2011.10626}{\JournalTitle{arXiv:2011.10626
  [astro-ph]}}, arXiv: 2011.10626

\bibitem[{Madhusudhan(2012)}]{madhusudhan_co_2012}
Madhusudhan, N. 2012,
  \href{http://dx.doi.org/10.1088/0004-637X/758/1/36}{\JournalTitle{The
  Astrophysical Journal}, 758, 36}, publisher: IOP Publishing

\bibitem[{Madhusudhan {et~al.}(2014{\natexlab{a}})Madhusudhan, Amin, \&
  Kennedy}]{madhusudhan_toward_2014}
Madhusudhan, N., Amin, M.~A., \& Kennedy, G.~M. 2014{\natexlab{a}},
  \href{http://dx.doi.org/10.1088/2041-8205/794/1/L12}{\JournalTitle{The
  Astrophysical Journal}, 794, L12}, publisher: IOP Publishing

\bibitem[{Madhusudhan {et~al.}(2017)Madhusudhan, Bitsch, Johansen, \&
  Eriksson}]{madhusudhan_atmospheric_2017}
Madhusudhan, N., Bitsch, B., Johansen, A., \& Eriksson, L. 2017,
  \href{http://dx.doi.org/10.1093/mnras/stx1139}{\JournalTitle{Monthly Notices
  of the Royal Astronomical Society}, 469, 4102}

\bibitem[{Madhusudhan {et~al.}(2014{\natexlab{b}})Madhusudhan, Crouzet,
  McCullough, Deming, \& Hedges}]{madhusudhan_h2o_2014}
Madhusudhan, N., Crouzet, N., McCullough, P.~R., Deming, D., \& Hedges, C.
  2014{\natexlab{b}},
  \href{http://dx.doi.org/10.1088/2041-8205/791/1/L9}{\JournalTitle{The
  Astrophysical Journal}, 791, L9}

\bibitem[{Madhusudhan {et~al.}(2011)Madhusudhan, Mousis, Johnson, \&
  Lunine}]{madhusudhan_carbon-rich_2011}
Madhusudhan, N., Mousis, O., Johnson, T.~V., \& Lunine, J.~I. 2011,
  \href{http://dx.doi.org/10.1088/0004-637X/743/2/191}{\JournalTitle{The
  Astrophysical Journal}, 743, 191}, publisher: IOP Publishing

\bibitem[{Madhusudhan \& Seager(2009)}]{madhusudhan_temperature_2009}
Madhusudhan, N. \& Seager, S. 2009,
  \href{http://dx.doi.org/10.1088/0004-637X/707/1/24}{\JournalTitle{The
  Astrophysical Journal}, 707, 24}, publisher: American Astronomical Society

\bibitem[{Marley {et~al.}(2002)Marley, Seager, Saumon, Lodders, Ackerman,
  Freedman, \& Fan}]{marley_clouds_2002}
Marley, M.~S., Seager, S., Saumon, D., {et~al.} 2002,
  \href{http://dx.doi.org/10.1086/338800}{\JournalTitle{The Astrophysical
  Journal}, 568, 335}, publisher: IOP Publishing

\bibitem[{McKemmish {et~al.}(2019)McKemmish, Masseron, Hoeijmakers,
  Pérez-Mesa, Grimm, Yurchenko, \& Tennyson}]{mckemmish_exomol_2019}
McKemmish, L.~K., Masseron, T., Hoeijmakers, H.~J., {et~al.} 2019,
  \href{http://dx.doi.org/10.1093/mnras/stz1818}{\JournalTitle{Monthly Notices
  of the Royal Astronomical Society}, 488, 2836}, publisher: Oxford Academic

\bibitem[{Merritt {et~al.}(2020)Merritt, Gibson, Nugroho, Mooij, Hooton,
  Matthews, McKemmish, Mikal-Evans, Nikolov, Sing, Spake, \&
  Watson}]{merritt_non-detection_2020}
Merritt, S.~R., Gibson, N.~P., Nugroho, S.~K., {et~al.} 2020,
  \href{http://dx.doi.org/10.1051/0004-6361/201937409}{\JournalTitle{Astronomy
  \& Astrophysics}, 636, A117}, publisher: EDP Sciences

\bibitem[{Mollière {et~al.}(2020)Mollière, Stolker, Lacour, Otten, Shangguan,
  Charnay, Molyarova, Nowak, Henning, Semenov, Dishoeck, Eisenhauer, Garcia,
  Greenbaum, Kervella, Kreidberg, Maire, \&
  Nasedkin}]{molliere_retrieving_2020}
Mollière, P., Stolker, T., Lacour, S., {et~al.} 2020,
  \href{http://dx.doi.org/10.1051/0004-6361/202038325}{\JournalTitle{Astronomy
  \& Astrophysics}}, publisher: EDP Sciences

\bibitem[{Mordasini {et~al.}(2016)Mordasini, Boekel, Mollière, Henning, \&
  Benneke}]{mordasini_imprint_2016}
Mordasini, C., Boekel, R.~v., Mollière, P., Henning, T., \& Benneke, B. 2016,
  \href{http://dx.doi.org/10.3847/0004-637X/832/1/41}{\JournalTitle{The
  Astrophysical Journal}, 832, 41}

\bibitem[{Nidever {et~al.}(2002)Nidever, Marcy, Butler, Fischer, \&
  Vogt}]{nidever_radial_2002}
Nidever, D.~L., Marcy, G.~W., Butler, R.~P., Fischer, D.~A., \& Vogt, S.~S.
  2002, \href{http://dx.doi.org/10.1086/340570}{\JournalTitle{The Astrophysical
  Journal Supplement Series}, 141, 503}, publisher: IOP Publishing

\bibitem[{Niemann {et~al.}(1998)Niemann, Atreya, Carignan, Donahue, Haberman,
  Harpold, Hartle, Hunten, Kasprzak, Mahaffy, Owen, \&
  Way}]{niemann_composition_1998}
Niemann, H.~B., Atreya, S.~K., Carignan, G.~R., {et~al.} 1998,
  \href{http://dx.doi.org/10.1029/98JE01050}{\JournalTitle{Journal of
  Geophysical Research: Planets}, 103, 22831}

\bibitem[{Nowak {et~al.}(2020)Nowak, Lacour, Mollière, Wang, Charnay,
  Dishoeck, Abuter, Amorim, Berger, Beust, Bonnefoy, Bonnet, Brandner, Buron,
  Cantalloube, Collin, Chapron, Clénet, Foresto, Zeeuw, Dembet, Dexter,
  Duvert, Eckart, Eisenhauer, Schreiber, Fédou, Lopez, Gao, Gendron, Genzel,
  Gillessen, Haußmann, Henning, Hippler, Hubert, Jocou, Kervella, Lagrange,
  Lapeyrère, Bouquin, Léna, Maire, Ott, Paumard, Paladini, Perraut, Perrin,
  Pueyo, Pfuhl, Rabien, Rau, Rodríguez-Coira, Rousset, Scheithauer, Shangguan,
  Straub, Straubmeier, Sturm, Tacconi, Vincent, Widmann, Wieprecht, Wiezorrek,
  Woillez, Yazici, \& Ziegler}]{nowak_peering_2020}
Nowak, M., Lacour, S., Mollière, P., {et~al.} 2020,
  \href{http://dx.doi.org/10.1051/0004-6361/201936898}{\JournalTitle{Astronomy
  \& Astrophysics}, 633, A110}, publisher: EDP Sciences

\bibitem[{Nugroho {et~al.}(2020)Nugroho, Gibson, de Mooij, Watson, Kawahara,
  \& Merritt}]{nugroho_searching_2020}
Nugroho, S.~K., Gibson, N.~P., de Mooij, E. J.~W., {et~al.} 2020,
  \href{http://dx.doi.org/10.1093/mnras/staa1459}{\JournalTitle{Monthly Notices
  of the Royal Astronomical Society}, 496, 504}

\bibitem[{Nugroho {et~al.}(2017)Nugroho, Kawahara, Masuda, Hirano, Kotani, \&
  Tajitsu}]{nugroho_high-resolution_2017}
Nugroho, S.~K., Kawahara, H., Masuda, K., {et~al.} 2017,
  \href{http://dx.doi.org/10.3847/1538-3881/aa9433}{\JournalTitle{The
  Astronomical Journal}, 154, 221}

\bibitem[{Nugroho {et~al.}(2021)Nugroho, Kawahara, Gibson, Mooij, Hirano,
  Kotani, Kawashima, Masuda, Brogi, Birkby, Watson, Tamura, Zwintz, Harakawa,
  Kudo, Kuzuhara, Hodapp, Ishizuka, Jacobson, Konishi, Kurokawa, Nishikawa,
  Omiya, Serizawa, Ueda, \& Vievard}]{nugroho_first_2021}
Nugroho, S.~K., Kawahara, H., Gibson, N.~P., {et~al.} 2021,
  \href{http://dx.doi.org/10.3847/2041-8213/abec71}{\JournalTitle{The
  Astrophysical Journal Letters}, 910, L9}, publisher: American Astronomical
  Society

\bibitem[{Orton {et~al.}(1998)Orton, Fisher, Baines, Stewart, Friedson, Ortiz,
  Marinova, Ressler, Dayal, Hoffmann, Hora, Hinkley, Krishnan, Masanovic,
  Tesic, Tziolas, \& Parija}]{orton_characteristics_1998}
Orton, G.~S., Fisher, B.~M., Baines, K.~H., {et~al.} 1998,
  \href{http://dx.doi.org/10.1029/98JE02380}{\JournalTitle{Journal of
  Geophysical Research: Planets}, 103, 22791}

\bibitem[{Owen {et~al.}(1999)Owen, Mahaffy, Niemann, Atreya, Donahue, Bar-Nun,
  \& de~Pater}]{owen_low-temperature_1999}
Owen, T., Mahaffy, P., Niemann, H.~B., {et~al.} 1999,
  \href{http://dx.doi.org/10.1038/46232}{\JournalTitle{Nature}, 402, 269}

\bibitem[{Parmentier {et~al.}(2018)Parmentier, Line, Bean, Mansfield,
  Kreidberg, Lupu, Visscher, Désert, Fortney, Deleuil, Arcangeli, Showman, \&
  Marley}]{parmentier_thermal_2018}
Parmentier, V., Line, M.~R., Bean, J.~L., {et~al.} 2018,
  \href{http://dx.doi.org/10.1051/0004-6361/201833059}{\JournalTitle{Astronomy
  \& Astrophysics}, 617, A110}

\bibitem[{Pinhas {et~al.}(2016)Pinhas, Madhusudhan, \&
  Clarke}]{pinhas_efficiency_2016}
Pinhas, A., Madhusudhan, N., \& Clarke, C. 2016,
  \href{http://dx.doi.org/10.1093/mnras/stw2239}{\JournalTitle{Monthly Notices
  of the Royal Astronomical Society}, 463, 4516}, publisher: Oxford Academic

\bibitem[{Pinhas {et~al.}(2019)Pinhas, Madhusudhan, Gandhi, \&
  MacDonald}]{pinhas_h2o_2019}
Pinhas, A., Madhusudhan, N., Gandhi, S., \& MacDonald, R. 2019,
  \href{http://dx.doi.org/10.1093/mnras/sty2544}{\JournalTitle{Monthly Notices
  of the Royal Astronomical Society}, 482, 1485}

\bibitem[{Piskorz {et~al.}(2016)Piskorz, Benneke, Crockett, Lockwood, Blake,
  Barman, Bender, Bryan, Carr, Fischer, Howard, {Howard Isaacson}, \&
  Johnson}]{piskorz_evidence_2016}
Piskorz, D., Benneke, B., Crockett, N.~R., {et~al.} 2016,
  \href{http://dx.doi.org/10.3847/0004-637X/832/2/131}{\JournalTitle{The
  Astrophysical Journal}, 832, 131}

\bibitem[{Piskorz {et~al.}(2017)Piskorz, Benneke, Crockett, Lockwood, Blake,
  Barman, Bender, Carr, \& Johnson}]{piskorz_detection_2017}
Piskorz, D., Benneke, B., Crockett, N.~R., {et~al.} 2017,
  \href{http://dx.doi.org/10.3847/1538-3881/aa7dd8}{\JournalTitle{The
  Astronomical Journal}, 154, 78}

\bibitem[{Piskorz {et~al.}(2018)Piskorz, Buzard, Line, Knutson, Benneke,
  Crockett, Lockwood, Blake, Barman, Bender, Deming, \&
  Johnson}]{piskorz_ground-_2018}
Piskorz, D., Buzard, C., Line, M.~R., {et~al.} 2018,
  \href{http://dx.doi.org/10.3847/1538-3881/aad781}{\JournalTitle{The
  Astronomical Journal}, 156, 133}

\bibitem[{Pollack {et~al.}(1996)Pollack, Hubickyj, Bodenheimer, Lissauer,
  Podolak, \& Greenzweig}]{pollack_formation_1996}
Pollack, J.~B., Hubickyj, O., Bodenheimer, P., {et~al.} 1996,
  \href{http://dx.doi.org/10.1006/icar.1996.0190}{\JournalTitle{Icarus}, 124,
  62}

\bibitem[{Polyansky {et~al.}(2018)Polyansky, Kyuberis, Zobov, Tennyson,
  Yurchenko, \& Lodi}]{polyansky_exomol_2018}
Polyansky, O.~L., Kyuberis, A.~A., Zobov, N.~F., {et~al.} 2018,
  \href{http://dx.doi.org/10.1093/mnras/sty1877}{\JournalTitle{Monthly Notices
  of the Royal Astronomical Society}, 480, 2597}, publisher: Oxford Academic

\bibitem[{Price-Whelan {et~al.}(2018)Price-Whelan, Sip{\textbackslash}Hocz,
  Günther, Lim, Crawford, Conseil, Shupe, Craig, Dencheva, Ginsburg,
  VanderPlas, Bradley, Pérez-Suárez, Val-Borro, Aldcroft, Cruz, Robitaille,
  Tollerud, Ardelean, Babej, Bach, Bachetti, Bakanov, Bamford, Barentsen,
  Barmby, Baumbach, Berry, Biscani, Boquien, Bostroem, Bouma, Brammer, Bray,
  Breytenbach, Buddelmeijer, Burke, Calderone, Rodríguez, Cara, Cardoso,
  Cheedella, Copin, Corrales, Crichton, D'Avella, Deil, Depagne, Dietrich,
  Donath, Droettboom, Earl, Erben, Fabbro, Ferreira, Finethy, Fox, Garrison,
  Gibbons, Goldstein, Gommers, Greco, Greenfield, Groener, Grollier, Hagen,
  Hirst, Homeier, Horton, Hosseinzadeh, Hu, Hunkeler, Ivezić, Jain, Jenness,
  Kanarek, Kendrew, Kern, Kerzendorf, Khvalko, King, Kirkby, Kulkarni, Kumar,
  Lee, Lenz, Littlefair, Ma, Macleod, Mastropietro, McCully, Montagnac, Morris,
  Mueller, Mumford, Muna, Murphy, Nelson, Nguyen, Ninan, Nöthe, Ogaz, Oh,
  Parejko, Parley, Pascual, Patil, Patil, Plunkett, Prochaska, Rastogi, Janga,
  Sabater, Sakurikar, Seifert, Sherbert, Sherwood-Taylor, Shih, Sick, Silbiger,
  Singanamalla, Singer, Sladen, Sooley, Sornarajah, Streicher, Teuben, Thomas,
  Tremblay, Turner, Terrón, Kerkwijk, Vega, Watkins, Weaver, Whitmore,
  Woillez, Zabalza, \& and}]{price-whelan_astropy_2018}
Price-Whelan, a. A.~M., Sip{\textbackslash}Hocz, B.~M., Günther, H.~M.,
  {et~al.} 2018,
  \href{http://dx.doi.org/10.3847/1538-3881/aabc4f}{\JournalTitle{The
  Astronomical Journal}, 156, 123}, publisher: American Astronomical Society

\bibitem[{Robitaille {et~al.}(2013)Robitaille, Tollerud, Greenfield,
  Droettboom, Bray, Aldcroft, Davis, Ginsburg, Price-Whelan, Kerzendorf,
  Conley, Crighton, Barbary, Muna, Ferguson, Grollier, Parikh, Nair, Günther,
  Deil, Woillez, Conseil, Kramer, Turner, Singer, Fox, Weaver, Zabalza,
  Edwards, Bostroem, Burke, Casey, Crawford, Dencheva, Ely, Jenness, Labrie,
  Lim, Pierfederici, Pontzen, Ptak, Refsdal, Servillat, \&
  Streicher}]{robitaille_astropy_2013}
Robitaille, T.~P., Tollerud, E.~J., Greenfield, P., {et~al.} 2013,
  \href{http://dx.doi.org/10.1051/0004-6361/201322068}{\JournalTitle{Astronomy
  \& Astrophysics}, 558, A33}, publisher: EDP Sciences

\bibitem[{Rodler {et~al.}(2013)Rodler, Kürster, \&
  Barnes}]{rodler_detection_2013}
Rodler, F., Kürster, M., \& Barnes, J.~R. 2013,
  \href{http://dx.doi.org/10.1093/mnras/stt462}{\JournalTitle{Monthly Notices
  of the Royal Astronomical Society}, 432, 1980}

\bibitem[{Rodler {et~al.}(2012)Rodler, Lopez-Morales, \&
  Ribas}]{rodler_weighing_2012}
Rodler, F., Lopez-Morales, M., \& Ribas, I. 2012,
  \href{http://dx.doi.org/10.1088/2041-8205/753/1/L25}{\JournalTitle{The
  Astrophysical Journal}, 753, L25}

\bibitem[{Rodmann {et~al.}(2006)Rodmann, Henning, Chandler, Mundy, \&
  Wilner}]{rodmann_large_2006}
Rodmann, J., Henning, T., Chandler, C.~J., Mundy, L.~G., \& Wilner, D.~J. 2006,
  \href{http://dx.doi.org/10.1051/0004-6361:20054038}{\JournalTitle{Astronomy
  \& Astrophysics}, 446, 211}, number: 1 Publisher: EDP Sciences

\bibitem[{Rothman {et~al.}(2010)Rothman, Gordon, Barber, Dothe, Gamache,
  Goldman, Perevalov, Tashkun, \& Tennyson}]{rothman_hitemp_2010}
Rothman, L.~S., Gordon, I.~E., Barber, R.~J., {et~al.} 2010,
  \href{http://dx.doi.org/10.1016/j.jqsrt.2010.05.001}{\JournalTitle{Journal of
  Quantitative Spectroscopy and Radiative Transfer}, 111, 2139}

\bibitem[{Schwarz {et~al.}(2015)Schwarz, Brogi, Kok, Birkby, \&
  Snellen}]{schwarz_evidence_2015}
Schwarz, H., Brogi, M., Kok, R.~d., Birkby, J., \& Snellen, I. 2015,
  \href{http://dx.doi.org/10.1051/0004-6361/201425170}{\JournalTitle{Astronomy
  \& Astrophysics}, 576, A111}

\bibitem[{Sing {et~al.}(2016)Sing, Fortney, Nikolov, Wakeford, Kataria, Evans,
  Aigrain, Ballester, Burrows, Deming, Désert, Gibson, Henry, Huitson,
  Knutson, Etangs, Pont, Showman, Vidal-Madjar, Williamson, \&
  Wilson}]{sing_continuum_2016}
Sing, D.~K., Fortney, J.~J., Nikolov, N., {et~al.} 2016,
  \href{http://dx.doi.org/10.1038/nature16068}{\JournalTitle{Nature}, 529, 59},
  number: 7584 Publisher: Nature Publishing Group

\bibitem[{Snellen {et~al.}(2010)Snellen, Kok, Mooij, Brogi, Nefs, \&
  Albrecht}]{snellen_exoplanet_2010}
Snellen, I., Kok, R.~d., Mooij, E.~d., {et~al.} 2010,
  \href{http://dx.doi.org/10.1017/S1743921311020199}{\JournalTitle{Proceedings
  of the International Astronomical Union}, 6, 208}

\bibitem[{Spake {et~al.}(2020)Spake, Sing, Wakeford, Nikolov, Mikal-Evans,
  Deming, Barstow, Anderson, Carter, Gillon, Goyal, Hebrard, Hellier, Kataria,
  Lam, Triaud, \& Wheatley}]{spake_abundance_2020}
Spake, J.~J., Sing, D.~K., Wakeford, H.~R., {et~al.} 2020,
  \href{http://dx.doi.org/10.1093/mnras/staa3116}{\JournalTitle{Monthly Notices
  of the Royal Astronomical Society}}

\bibitem[{Stevenson {et~al.}(2014)Stevenson, Bean, Madhusudhan, \&
  Harrington}]{stevenson_deciphering_2014}
Stevenson, K.~B., Bean, J.~L., Madhusudhan, N., \& Harrington, J. 2014,
  \href{http://dx.doi.org/10.1088/0004-637X/791/1/36}{\JournalTitle{The
  Astrophysical Journal}, 791, 36}, publisher: IOP Publishing

\bibitem[{Swain {et~al.}(2013)Swain, Deroo, Tinetti, Hollis, Tessenyi, Line,
  Kawahara, Fujii, Showman, \& Yurchenko}]{swain_probing_2013}
Swain, M., Deroo, P., Tinetti, G., {et~al.} 2013,
  \href{http://dx.doi.org/10.1016/j.icarus.2013.04.003}{\JournalTitle{Icarus},
  225, 432}

\bibitem[{Sánchez-López {et~al.}(2019)Sánchez-López, Alonso-Floriano,
  López-Puertas, Snellen, Funke, Nagel, Bauer, Amado, Caballero, Czesla,
  Nortmann, Pallé, Salz, Reiners, Ribas, Quirrenbach, Anglada-Escudé, Béjar,
  Casasayas-Barris, Galadí-Enríquez, Guenther, Henning, Kaminski, Kürster,
  Lampón, Lara, Montes, Morales, Stangret, Tal-Or, Sanz-Forcada, Schmitt,
  Osorio, \& Zechmeister}]{sanchez-lopez_water_2019}
Sánchez-López, A., Alonso-Floriano, F.~J., López-Puertas, M., {et~al.} 2019,
  \href{http://dx.doi.org/10.1051/0004-6361/201936084}{\JournalTitle{Astronomy
  \& Astrophysics}, 630, A53}, publisher: EDP Sciences

\bibitem[{Sánchez-López {et~al.}(2020)Sánchez-López, López-Puertas,
  Snellen, Nagel, Bauer, Pallé, Tal-Or, Amado, Caballero, Czesla, Nortmann,
  Reiners, Ribas, Quirrenbach, Aceituno, Béjar, Casasayas-Barris, Henning,
  Molaverdikhani, Montes, Stangret, Osorio, \&
  Zechmeister}]{sanchez-lopez_discriminating_2020}
Sánchez-López, A., López-Puertas, M., Snellen, I. a.~G., {et~al.} 2020,
  \href{http://dx.doi.org/10.1051/0004-6361/202038629}{\JournalTitle{Astronomy
  \& Astrophysics}, 643, A24}, publisher: EDP Sciences

\bibitem[{Todorov {et~al.}(2016)Todorov, Line, Pineda, Meyer, Quanz, Hinkley,
  \& Fortney}]{todorov_water_2016}
Todorov, K.~O., Line, M.~R., Pineda, J.~E., {et~al.} 2016,
  \href{http://dx.doi.org/10.3847/0004-637X/823/1/14}{\JournalTitle{The
  Astrophysical Journal}, 823, 14}, publisher: American Astronomical Society

\bibitem[{Vazan {et~al.}(2018)Vazan, Helled, \& Guillot}]{vazan_jupiters_2018}
Vazan, A., Helled, R., \& Guillot, T. 2018,
  \href{http://dx.doi.org/10.1051/0004-6361/201732522}{\JournalTitle{Astronomy
  \& Astrophysics}, 610, L14}, publisher: EDP Sciences

\bibitem[{Virtanen {et~al.}(2020)Virtanen, Gommers, Oliphant, Haberland, Reddy,
  Cournapeau, Burovski, Peterson, Weckesser, Bright, van~der Walt, Brett,
  Wilson, Millman, Mayorov, Nelson, Jones, Kern, Larson, Carey, Polat, Feng,
  Moore, VanderPlas, Laxalde, Perktold, Cimrman, Henriksen, Quintero, Harris,
  Archibald, Ribeiro, Pedregosa, \& van Mulbregt}]{virtanen_scipy_2020}
Virtanen, P., Gommers, R., Oliphant, T.~E., {et~al.} 2020,
  \href{http://dx.doi.org/10.1038/s41592-019-0686-2}{\JournalTitle{Nature
  Methods}, 17, 261}, number: 3 Publisher: Nature Publishing Group

\bibitem[{Visscher {et~al.}(2010)Visscher, Lodders, \&
  Fegley}]{visscher_atmospheric_2010}
Visscher, C., Lodders, K., \& Fegley, B. 2010,
  \href{http://dx.doi.org/10.1088/0004-637X/716/2/1060}{\JournalTitle{The
  Astrophysical Journal}, 716, 1060}, publisher: American Astronomical Society

\bibitem[{Wakeford \& Sing(2015)}]{wakeford_transmission_2015}
Wakeford, H.~R. \& Sing, D.~K. 2015,
  \href{http://dx.doi.org/10.1051/0004-6361/201424207}{\JournalTitle{Astronomy
  \& Astrophysics}, 573, A122}

\bibitem[{Wakeford {et~al.}(2017)Wakeford, Visscher, Lewis, Kataria, Marley,
  Fortney, \& Mandell}]{wakeford_high-temperature_2017}
Wakeford, H.~R., Visscher, C., Lewis, N.~K., {et~al.} 2017,
  \href{http://dx.doi.org/10.1093/mnras/stw2639}{\JournalTitle{Monthly Notices
  of the Royal Astronomical Society}, 464, 4247}

\bibitem[{Wang {et~al.}(2020)Wang, Wang, Ma, Chilcote, Ertel, Guyon, Ilyin,
  Jovanovic, Kalas, Lozi, Macintosh, Strassmeier, \&
  Stone}]{wang_chemical_2020}
Wang, J., Wang, J.~J., Ma, B., {et~al.} 2020,
  \href{http://dx.doi.org/10.3847/1538-3881/ababa7}{\JournalTitle{The
  Astronomical Journal}, 160, 150}, publisher: American Astronomical Society

\bibitem[{Watson {et~al.}(2019)Watson, de~Mooij, Steeghs, Marsh, Brogi, Gibson,
  \& Matthews}]{watson_doppler_2019}
Watson, C.~A., de~Mooij, E. J.~W., Steeghs, D., {et~al.} 2019,
  \href{http://dx.doi.org/10.1093/mnras/stz2679}{\JournalTitle{Monthly Notices
  of the Royal Astronomical Society}, 490, 1991}, publisher: Oxford Academic

\bibitem[{Webb {et~al.}(2020)Webb, Brogi, Gandhi, Line, Birkby, Chubb, Snellen,
  \& Yurchenko}]{webb_weak_2020}
Webb, R.~K., Brogi, M., Gandhi, S., {et~al.} 2020,
  \href{http://dx.doi.org/10.1093/mnras/staa715}{\JournalTitle{Monthly Notices
  of the Royal Astronomical Society}, 494, 108}, publisher: Oxford Academic

\bibitem[{Welbanks {et~al.}(2019)Welbanks, Madhusudhan, Allard, Hubeny,
  Spiegelman, \& Leininger}]{welbanks_massmetallicity_2019}
Welbanks, L., Madhusudhan, N., Allard, N.~F., {et~al.} 2019,
  \href{http://dx.doi.org/10.3847/2041-8213/ab5a89}{\JournalTitle{The
  Astrophysical Journal}, 887, L20}, publisher: American Astronomical Society

\bibitem[{Yurchenko {et~al.}(2011)Yurchenko, Barber, \&
  Tennyson}]{yurchenko_variationally_2011}
Yurchenko, S.~N., Barber, R.~J., \& Tennyson, J. 2011,
  \href{http://dx.doi.org/10.1111/j.1365-2966.2011.18261.x}{\JournalTitle{Monthly
  Notices of the Royal Astronomical Society}, 413, 1828}

\bibitem[{Yurchenko \& Tennyson(2014)}]{yurchenko_exomol_2014}
Yurchenko, S.~N. \& Tennyson, J. 2014,
  \href{http://dx.doi.org/10.1093/mnras/stu326}{\JournalTitle{Monthly Notices
  of the Royal Astronomical Society}, 440, 1649}

\bibitem[{Zhang {et~al.}(2020{\natexlab{a}})Zhang, Bosman, \&
  Bergin}]{zhang_excess_2020}
Zhang, K., Bosman, A.~D., \& Bergin, E.~A. 2020{\natexlab{a}},
  \href{http://dx.doi.org/10.3847/2041-8213/ab77ca}{\JournalTitle{The
  Astrophysical Journal}, 891, L16}, publisher: American Astronomical Society

\bibitem[{Zhang {et~al.}(2020{\natexlab{b}})Zhang, Chachan, Kempton, Knutson,
  \& Chang}]{zhang_platon_2020}
Zhang, M., Chachan, Y., Kempton, E. M.-R., Knutson, H.~A., \& Chang, W.~H.
  2020{\natexlab{b}},
  \href{http://dx.doi.org/10.3847/1538-4357/aba1e6}{\JournalTitle{The
  Astrophysical Journal}, 899, 27}, publisher: American Astronomical Society

\bibitem[{Zucker(2003)}]{zucker_cross-correlation_2003}
Zucker, S. 2003,
  \href{http://dx.doi.org/10.1046/j.1365-8711.2003.06633.x}{\JournalTitle{Monthly
  Notices of the Royal Astronomical Society}, 342, 1291}, publisher: Oxford
  Academic

\bibitem[{Öberg \& Bergin(2016)}]{oberg_excess_2016}
Öberg, K.~I. \& Bergin, E.~A. 2016,
  \href{http://dx.doi.org/10.3847/2041-8205/831/2/L19}{\JournalTitle{The
  Astrophysical Journal}, 831, L19}

\bibitem[{Öberg {et~al.}(2011)Öberg, Murray-Clay, \&
  Bergin}]{oberg_effects_2011}
Öberg, K.~I., Murray-Clay, R., \& Bergin, E.~A. 2011,
  \href{http://dx.doi.org/10.1088/2041-8205/743/1/L16}{\JournalTitle{The
  Astrophysical Journal Letters}, 743, L16}

\bibitem[{Öberg \& Wordsworth(2019)}]{oberg_jupitertextquotesingles_2019}
Öberg, K.~I. \& Wordsworth, R. 2019,
  \href{http://dx.doi.org/10.3847/1538-3881/ab46a8}{\JournalTitle{The
  Astronomical Journal}, 158, 194}, publisher: American Astronomical Society

\end{thebibliography}
\bibliographystyle{yahapj}



\end{document}